\pdfoutput=1
\documentclass[11pt,letterpaper]{article}

\usepackage[margin=1in]{geometry}
\usepackage[T1]{fontenc}
\usepackage{microtype}
\usepackage{amsmath,amssymb,amsthm,mathtools}
\usepackage[sc,osf]{mathpazo}
\usepackage{bm}
\usepackage{booktabs}
\usepackage{array}
\usepackage{multirow}
\usepackage{enumitem}
\usepackage{placeins}
\usepackage[round,authoryear]{natbib}
\usepackage{tikz}
\usepackage{pgfplots}
\pgfplotsset{compat=1.18}
\usetikzlibrary{arrows.meta,positioning,shapes.geometric,calc,backgrounds,fit}

\usepackage[hidelinks]{hyperref}

\newcommand{\UseLiningFigures}{\renewcommand{\rmdefault}{pplx}\fontfamily{pplx}\selectfont}
\AddToHook{env/table/begin}{\UseLiningFigures}
\AddToHook{env/tabular/begin}{\UseLiningFigures}
\AddToHook{env/figure/begin}{\UseLiningFigures}
\AddToHook{env/tikzpicture/begin}{\UseLiningFigures}

\newcommand{\PaperTitle}{Spillover Inference from Partial Aggregated Relational Data:\texorpdfstring{\\}{ }Identification Boundaries, Polynomial-Time Certification,\texorpdfstring{\\}{ }Survey Design, and Reporting Asymmetry}
\newcommand{\PaperAuthor}{Yen-hsuan Tseng}
\hypersetup{
  pdftitle={\PaperTitle},
  pdfauthor={\PaperAuthor}
}

\theoremstyle{plain}
\newtheorem{theorem}{Theorem}[section]
\newtheorem{proposition}[theorem]{Proposition}
\newtheorem{lemma}[theorem]{Lemma}

\theoremstyle{definition}
\newtheorem{definition}[theorem]{Definition}
\newtheorem{assumption}{Assumption}
\newtheorem{example}[theorem]{Example}

\theoremstyle{remark}
\newtheorem{remark}[theorem]{Remark}
\numberwithin{equation}{section}

\newcommand{\R}{\mathbb{R}}
\newcommand{\E}{\mathbb{E}}
\newcommand{\Prob}{\mathbb{P}}
\newcommand{\Var}{\operatorname{Var}}
\newcommand{\Cov}{\operatorname{Cov}}
\newcommand{\Corr}{\operatorname{Corr}}
\newcommand{\tr}{\operatorname{tr}}
\newcommand{\diag}{\operatorname{diag}}
\newcommand{\blkdiag}{\operatorname{blkdiag}}
\newcommand{\logit}{\operatorname{logit}}
\newcommand{\conv}{\operatorname{conv}}
\newcommand{\spanop}{\operatorname{span}}
\newcommand{\ind}[1]{\mathbf{1}\{#1\}}

\newcommand{\Rset}{\mathcal{R}}
\newcommand{\Nset}{\mathcal{N}}
\newcommand{\PiN}{\Pi_{\mathcal{N}}}
\newcommand{\ThetaI}{\Theta_{I}}
\newcommand{\ThetaIhat}{\widehat{\Theta}_{I}}
\newcommand{\Tstrat}{T_{\mathrm{strat}}}
\newcommand{\Budget}{\mathsf{B}}
\newcommand{\Qitems}{\mathsf{Q}}
\newcommand{\Birk}{\mathfrak{B}}
\newcommand{\MZ}{M_{Z}}

\title{\Large\textbf{\PaperTitle}}

\author{\PaperAuthor}
\date{November 26, 2026}

\begin{document}
\maketitle

\begin{abstract}
Aggregated Relational Data (ARD) lowers survey costs by asking sampled respondents how many contacts they hold in known trait groups rather than eliciting a full network census. Under \emph{partial participation}---when only a subset of the population is surveyed---non-respondents supply no counts. Swapping the link profiles of two non-respondents in the same trait group who differ in treatment preserves the exact joint law of all observed edges and counts, yet holding conditional outcome means fixed can flip a peer-spillover slope from $+2$ to $-2$.

We map the resulting identification, inference, and survey-design boundaries. Relative to a column-profile oracle, the linear-projection spillover estimand is partially identified on a finite non-respondent permutation orbit. We prove an exact near-linear sorting test for sign identification, build polynomial-time outer certifiers for the orbit span, decompose plug-in exposure errors and latent-homophily control-function limits, and analyze model-free errors-in-variables brackets alongside prospective survey designs---adding a single treated-contact question or stratifying each trait group by treatment---that measure realized exposure directly subject to a fixed-budget trade-off.

On the Karnataka microfinance network, two associational empirical findings emerge. First, under the fitted degree-conditioned logistic model, non-respondent permutations selected on one half of respondents yield opposite-signed held-out leader-exposure slopes across all ten 50/50 respondent splits (mean $-0.0089$ vs.\ $+0.0198$). Second, an exact decomposition shows that the drop from the released union-graph slope ($+0.0141$) to hybrid ARD ($+0.0057$) is not mechanical attenuation (which predicts $+0.0147$), but a large gap between ties recorded only from the respondent's report and ties that either side could have reported ($+0.0351$, wild cluster bootstrap $p = 0.0074$).
\end{abstract}

\vspace{-4pt}

\newpage
\setcounter{tocdepth}{2}
\tableofcontents
\newpage

\section{Introduction}
\label{sec:intro}

Measuring a complete social network in a field study is expensive. Eliciting all bilateral links across dozens of villages or schools requires interviewing most of the population and resolving hundreds of names per respondent, as in the Indian village census of \citet{banerjee2013diffusion}. To lower this cost, applied researchers increasingly collect \emph{Aggregated Relational Data} (ARD), replacing full link rosters with simple count questions of the form ``How many people with trait $k$ do you know?'' Originating in statistical network scale-up and overdispersed count modeling \citep{zheng2006how,mccormick2010how,mccormick2015latent}, ARD has entered economics and causal inference as a basis for reconstructing unobserved network structure \citep{breza2020using,alidaee2020recovering,breza2023consistently} and evaluating downstream spillover regressions \citep{wilkinsreeves2026single}, alongside adjacent econometric methods for estimating peer effects from partial network data \citep{boucher2025estimating}.

Yet the feature that makes ARD economical in the field creates a sharp statistical asymmetry. In practice, ARD is rarely collected from every individual on the population roster. A subset $\Rset \subset [r]$ of respondents (participation rate $\varphi = |\Rset|/r < 1$) completes the survey, while the remaining non-respondents $\Nset = [r]\setminus\Rset$ appear on the roster with observed trait groups and treatment indicators $z_j$ (for example, whether a household is a community leader or assigned to an intervention), but supply no ARD counts of their own. This partial-participation regime is the standard empirical setting---roughly $47\%$ of households completed the detailed network module in the Karnataka microfinance villages of \citet{banerjee2013diffusion}.

Every result in this paper stems from a single structural consequence of partial participation: all respondents share the same unknown within-group assignment of non-respondents. Consider two unsurveyed villagers who belong to the same trait group, only one of whom is treated (Example~\ref{ex:toy} and Figure~\ref{fig:toy}). Swapping their link probabilities leaves the joint distribution of every observed respondent edge and ARD count identically unchanged, yet holding outcome means fixed can flip the spillover regression slope from $+2$ to $-2$. Whereas \citet{breza2020using,breza2023consistently} establish recovery of network structure and group-invariant network statistics from ARD, pairing non-respondent links with node-specific treatments $z_j$ in a spillover regression exposes this labeling symmetry. Two natural questions follow immediately: (i) given an existing $K$-group ARD survey, can we test whether the sign of the spillover is identified and bound its feasible range? (ii) before fieldwork, what minimal change to the questionnaire removes the ambiguity at collection?

\subsection{Situating the Problem Across Three Intersecting Literatures}

Partial-participation ARD spillover analysis sits at the intersection of three active literatures that have developed largely in parallel:

\begin{enumerate}[leftmargin=1.5em]
  \item \textbf{Network reconstruction from ARD, sampled networks, and dyadic reporting.}
  In statistics, \citet{zheng2006how}, \citet{mccormick2010how}, and \citet{mccormick2015latent} developed overdispersed count, scale-up, and spherical latent-surface models for personal network size and mixing rates from ARD (reviewed comprehensively by \citealp{laga2021thirty}), alongside foundational likelihood work on sampled networks \citep{handcock2010modeling}. In economics, \citet{breza2020using} demonstrated that ARD can feasibly recover network structure without full censuses, \citet{breza2023consistently} established consistency of network statistics from ARD under latent-space assumptions, and \citet{alidaee2020recovering} studied nuclear-norm penalized recovery of low-rank networks from ARD. Recent spectral theory further characterizes full-roster ($\varphi=1$) community recovery from ARD \citep{davison2026community}, low-rank network smoothing for peer effects \citep{hayes2026smoothing}, and bounded-degree minimax rates for randomized peer treatments \citep{hayes2025minimax}. A parallel econometric and statistical literature documents how incomplete sampling, top-coding, unobserved or misclassified links, proxy networks, and asymmetric bilateral reporting distort network regressions \citep{manski1993identification,comola2014testing,chandrasekhar2016econometrics,comola2017missing,ready2021measuring,griffith2022name,debacco2023latent,lewbel2023social,hardy2024estimating,zhang2024spillovers,marray2025estimating,lewbel2026estimating,sun2026flexible,weinstein2026bayesian}. More recently, \citet{boucher2025estimating} developed simulated GMM (SGMM) for simultaneous linear-in-means peer effects across $M\to\infty$ bounded-size groups with full-group outcome and covariate censuses (Assumptions~1--2) given a consistently estimable network distribution (Assumption~5); they note that ARD violates their bounded-group asymptotics (Appendix~B.1.6), direct ARD applications to a Bayesian MCMC estimator driven by the outcome likelihood (Section~4), and show in Appendix~H.3 (Tables~H.1--H.2) that ARD SGMM degrades sharply whenever contextual peer covariates ($GX$ in their notation) are unobserved. In machine learning and causal inference, \citet{wilkinsreeves2026single} established single-network asymptotics for causal interference with partial network data using marginalized proxy exposures, proving dense full-roster SBM recovery ($\varphi=1$) while noting (in Appendix~A.2.3.2 of the arXiv preprint \texttt{2406.11940v1}) that under degree-heterogeneous $\beta$-models, ARD estimation rates are too slow to ignore first-stage graph estimation error.

  \item \textbf{Causal inference under interference and partial identification.}
  The modern potential-outcomes theory of interference defines direct and spillover estimands via exposure mappings \citep{hudgens2008toward,aronow2017estimating}, with recent work handling unknown or misspecified exposure mappings \citep{savje2021average,savje2024causal}, single-network approximate neighborhood interference \citep{leung2022causal}, exogenous-sensitivity partial identification on known graphs \citep{schroder2026causal}, and minimax impossibility of specification testing for exposure mappings \citep{gao2026impossibility}. Meanwhile, the partial-identification and data-combination literatures \citep{manski1990nonparametric,imbens2004confidence,stoye2009more,dhaultfoeuille2025linear,fan2025partial} provide principled optimal-transport tools when data cannot point-identify a structural parameter. Under partial-participation ARD, classical per-observation Fr\'echet--Hoeffding or rearrangement boxes are substantially looser and often blow up to $(-\infty,+\infty)$, whereas enforcing the shared non-respondent permutation across respondents yields much tighter certified intervals.

  \item \textbf{Measurement error, degree heterogeneity, and unobserved homophily.}
  Downstream regressions using reconstructed exposures inherit classical and Berkson measurement error \citep{fuller1987measurement} as well as generated-regressor uncertainty \citep{pagan1984econometric}. Under sparse, degree-heterogeneous undirected networks \citep{chatterjee2011random,yan2013central,graham2017econometric,dzemski2019empirical,ma2020universal}, node-level fixed effects $\alpha_i = \Theta(1)$ induce heteroskedastic, heavy-tailed estimation errors and correlated inverse-Fisher covariances. Moreover, when unobserved network-forming traits also enter outcomes (latent homophily), control-function methods \citep{goldsmithpinkham2013social,johnsson2021estimation,auerbach2022identification} face sharp collinearity and identifiability limits under partial ARD.
\end{enumerate}

These three literatures meet at one point: because every respondent shares the same unknown within-group assignment of non-respondents, network-reconstruction uncertainty, interference exposures, and measurement-error regressions are coupled across respondents rather than independent. Our work also connects cleanly to, while remaining distinct from, three companion theoretical studies of full-roster and count-constrained ARD: \citet{tseng2026degree} develops degree-conditioned finite-sample association tests under arbitrary activity heterogeneity; \citet{tseng2026alignment} establishes alignment-plus-resolution recovery thresholds under full-roster ($\varphi=1$) dense homogeneous ARD; and \citet{tseng2026countfibres} characterizes conditional uniformity on count fibres via $k$-beta additive log-odds. Here we study \emph{downstream spillover inference} under \emph{partial participation} ($\varphi < 1$), sparse degree heterogeneity, and field-survey data.

\subsection{Summary of Main Contributions}

Organized around the retrospective and prospective questions above, the paper makes four contributions:

\begin{enumerate}[leftmargin=1.5em]
  \item \textbf{Non-respondent orbit partial identification, exact sorting sign test, and outer certification (Section~\ref{sec:orbit}).}
  For any symmetric link kernel and any sparsity regime, the joint distribution of observed respondent edges $A_{\Rset,\Rset}$ and ARD counts $Y_{\Rset,\cdot}$ is strictly invariant to within-group permutations $\pi \in \PiN$ of non-respondents (Proposition~\ref{prop:swap_invariance}). Relative to a column-profile oracle (or a structural model transferring the non-respondent activity multiset), the linear-projection spillover estimand $\theta^\star$ is sharply identified under unrestricted conditional outcome means only up to the finite \emph{Orbit Identified Set} $\ThetaI$ (Proposition~\ref{prop:orbit_set}), collapsing to a point across all outcomes if and only if differently treated non-respondents in each cell share identical projected link profiles. Because all respondents share one permutation and the regression denominator is positive whenever the slope is defined, checking whether sign reversal is excluded reduces over the permutohedron \citep{birkhoff1946tres,rado1952inequality,hardy1952inequalities} to within-group sorting in $O(r_{\mathrm{r}} r_{\mathrm{n}} + \sum_k m_{\mathrm{n},k}\log m_{\mathrm{n},k})$ time (Theorem~\ref{thm:sign_test}). To certify the full feasible span when sign reversal is excluded or when pooling multiple villages, Section~\ref{sec:orbit} further constructs nested polynomial-time outer certifiers based on Frank--Wolfe duality \citep{frank1956algorithm} and Shor semidefinite relaxation \citep{shor1987quadratic,luo2010semidefinite} (Theorem~\ref{thm:certifiers}).

  \item \textbf{Error taxonomy, feasible plug-in certification, and homophily limits (Section~\ref{sec:plugin_homophily}).}
  We establish an exact three-term decomposition of mean-hybrid exposure errors (Proposition~\ref{prop:three_term}) isolating four failure modes (FM1--FM4): classical raw-count attenuation, swap-induced Berkson variance under randomized treatment (clarifying the exact scope boundary of \citealp{wilkinsreeves2026single}), asymptotic bias from within-cell treatment--activity covariance, and nonlinear exposure-mapping smearing. To build feasible bounds without oracle link probabilities, we estimate non-respondent expected counts via degree-conditioned logistic maximum likelihood and derive a perturbation certifier (Lemma~\ref{lem:perturbation}, Theorem~\ref{thm:plugin_asymp}) incorporating rank-1 cell-error projection and hub-kurtosis effective degrees of freedom \citep{satterthwaite1946approximate,rudelson2013hanson}, paired with a five-metric coverage ledger that discloses when sparse single-village uncertainty forces $(-\infty,+\infty)$ outputs. Section~\ref{sec:homophily} separately establishes three structural limits (Propositions~\ref{prop:S1}--\ref{prop:S3}) on control functions for unobserved latent homophily.

  \item \textbf{Model-free remedies: raw-ARD EIV brackets and $(K{+}1)$ vs.\ $2K$ survey designs (Section~\ref{sec:remedies}).}
  We derive exact finite-sample second-moment and variance identities for pure and hybrid raw-ARD exposure proxies, yielding model-free Cauchy--Schwarz errors-in-variables (EIV) brackets under high-SNR expected exposure (Proposition~\ref{prop:raw_eiv}). For prospective fieldwork when binary treatment is salient and accurately reported, adding a single question---``How many of your contacts are treated?'' ($(K{+}1)$ Augmented ARD)---or splitting all $K$ questions by treatment status ($2K$ $\Tstrat$, Proposition~\ref{prop:tstrat}) measures realized exposure directly and eliminates swap ambiguity. Under a fixed question budget, Proposition~\ref{prop:budget} shows that $(K{+}1)$ improves over $2K$ on realized exposure, standard $K$-group ARD + EIV retains an edge under sparse single-graph expected exposure with rare treatments, and $2K$ $\Tstrat$ uniquely supplies disjoint untreated-neighbor counts needed for clean EIV control functions under latent homophily and group-heterogeneous spillovers.

  \item \textbf{Karnataka microfinance empirical analysis (Section~\ref{sec:empirical}).}
  On the 75-village archive of \citet{banerjee2013diffusion}, we benchmark our framework on strictly non-circular and real-SNR semi-synthetic designs and examine associational regressions for real microfinance adoption. First, under the fitted degree-conditioned logistic model, held-out 50/50 respondent cross-fitting produces opposite-signed leader-exposure slopes across all ten splits (while placebo cross-fits center near zero), complementing the in-sample pooled orbit bounds and heterogeneity-shrinkage checks. Second, an exact algebraic decomposition shows that the gap between the released union-graph benchmark and hybrid ARD is not mechanical attenuation ($\kappa \in [1.03, 1.08] > 1$), but a large contrast between ego-only ties to non-respondent leaders and union-recorded ties among respondent leaders that persists under non-leader placebos, four-regressor horseraces, and wild cluster bootstrap inference. In paired directional relation matrices, point estimates of the $\Rset\Nset$--$\Rset\Rset$ contrast are larger for advice sought (\texttt{helpdecision}) than for advice given (\texttt{giveadvice}), a pattern consistent with respondents reporting more salient ties (the cross-relation difference is not formally tested).
\end{enumerate}

\subsection{Reading Guide}

Sections~\ref{sec:setup}--\ref{sec:discussion} form a self-contained narrative covering the observation model, a six-node rational toy example, core proof mechanisms with literature comparisons, survey-design trade-offs, and the Karnataka empirical findings. Mechanical derivations, plug-in calibration formulas, synthetic stress grids, and complete Karnataka diagnostic tables appear in Appendices~\ref{app:orbit}--\ref{app:tables}.

\section{Problem Setup, Scope, and an Illustrative Toy Example}
\label{sec:setup}

We begin by formalizing partial-participation ARD, distinguishing causal and associational target estimands, and illustrating the non-respondent symmetry obstruction on a six-node rational network where every quantity can be verified by hand.

\subsection{Partial-Participation ARD Observation Model}

Consider a finite population roster $[r] := \{1,\dots,r\}$ partitioned into $K \ge 1$ disjoint, publicly observed trait groups $G_1, \dots, G_K$ with sizes $m_k := |G_k|$ and membership matrix $T \in \{0,1\}^{r\times K}$ ($T_{ik} = \ind{i \in G_k}$). A subset $\Rset \subseteq [r]$ of size $r_{\mathrm{r}} := |\Rset|$ participates in the detailed network and outcome survey, while $\Nset := [r]\setminus\Rset$ of size $r_{\mathrm{n}} := r - r_{\mathrm{r}}$ does not. We denote the survey participation rate by $\varphi := r_{\mathrm{r}}/r \in (0,1]$, and partition each trait group into its respondent and non-respondent cells:
\[
  G_k^{\mathrm{r}} := G_k \cap \Rset \quad (m_{\mathrm{r},k} := |G_k^{\mathrm{r}}|),
  \qquad
  G_k^{\mathrm{n}} := G_k \cap \Nset \quad (m_{\mathrm{n},k} := |G_k^{\mathrm{n}}|).
\]

Let $A = (A_{ij})_{i,j\in[r]} \in \{0,1\}^{r\times r}$ be an undirected binary adjacency matrix with $A_{ii}=0$ and $A_{ij}=A_{ji}$. Conditional on latent node attributes $\xi = (\xi_1,\dots,\xi_r)$ taking values in a measurable space $\Xi^r$, edges $\{A_{ij} : 1 \le i < j \le r\}$ are drawn independently as
\begin{equation}
  \label{eq:graph_model}
  A_{ij} \mid \xi \;\sim\; \operatorname{Bernoulli}(p_{ij}),
  \qquad
  p_{ij} = \psi(\xi_i, \xi_j),
\end{equation}
where $\psi : \Xi \times \Xi \to [0,1]$ is an arbitrary symmetric link-formation kernel. When we specialize to parametric feasible estimation (Section~\ref{sec:plugin_homophily}), we work with the sparse degree-heterogeneous logistic latent-sign model
\begin{equation}
  \label{eq:logistic_model}
  \logit(p_{ij}) = \rho_r + \alpha_i + \alpha_j + \tau \zeta_i \zeta_j,
  \qquad
  \xi_i = (\alpha_i, \zeta_i) \in \R \times \{-1, +1\},
\end{equation}
where $\rho_r \le 0$ is a sparsity offset allowing average degree $d_{\mathrm{avg}} \asymp r e^{\rho_r} \ll r$ (such as the sparse village degrees $d_{\mathrm{avg}} \approx 4$--$14$ in Karnataka), $\alpha_i$ captures unrestricted $O(1)$ node-level degree heterogeneity (centered around zero, with multiplicative activity factor $a_i := e^{\alpha_i}$), and $\tau \ge 0$ governs latent community homophily. Our identification results in Section~\ref{sec:orbit} hold for completely general $\psi$ in \eqref{eq:graph_model}.

Each individual $j \in [r]$ has an observed roster treatment (or exposure-source indicator) $z_j \in [0,1]$---binary $z_j \in \{0,1\}$ in our primary experiments and the Karnataka application---and each respondent $i \in \Rset$ has an observed outcome $y_i \in \R$ and observed control vector $W_i \in \R^q$ (which may include village fixed effects and household covariates). Under \emph{partial-participation ARD}, the researcher observes:
\begin{equation}
  \label{eq:observed_data}
  \mathcal{D}_{\mathrm{ARD}}
  \;=\;
  \Big(
    A_{\Rset,\Rset},\;
    Y_{\Rset,\cdot},\;
    T,\;
    z,\;
    y_{\Rset},\;
    W_{\Rset}
  \Big),
\end{equation}
where for each respondent $i \in \Rset$ and group $k \in \{1,\dots,K\}$,
\[
  Y_{ik} := \sum_{j \in G_k \setminus \{i\}} A_{ij},
  \qquad
  Y_{ik}^{\mathrm{r}} := \sum_{j \in G_k^{\mathrm{r}} \setminus \{i\}} A_{ij},
  \qquad
  Y_{ik}^{\mathrm{n}} := Y_{ik} - Y_{ik}^{\mathrm{r}} = \sum_{j \in G_k^{\mathrm{n}}} A_{ij},
  \qquad
  D_i := \sum_{k=1}^K Y_{ik}.
\]
The total degree $D_i$ is a deterministic summary of $Y_{\Rset,\cdot}$, and the expected non-respondent cell count is $\mu_{ik}^{\mathrm{n}} := \E[Y_{ik}^{\mathrm{n}} \mid \xi] = \sum_{j \in G_k^{\mathrm{n}}} p_{ij}$. Individual links $A_{ij}$ between respondents $i \in \Rset$ and non-respondents $j \in \Nset$ are observed only through their cell totals $Y_{ik}^{\mathrm{n}}$, and no data are observed on non-respondent outcomes or $\Nset \times \Nset$ ties. (When $A_{\Rset,\Rset}$ is also unavailable, the setting reduces to \emph{pure ARD}, observing only $Y_{\Rset,\cdot}$; we analyze both hybrid and pure ARD.)

\subsection{Exposures, Linear Projection Target, and Causal vs.\ Associational Scope}

For each respondent $i \in \Rset$, the network literature considers two canonical linear peer-exposure mappings:
\begin{align}
  E_i^{\mathrm{real}} &\;:=\; (Az)_i = \sum_{j \in [r]\setminus\{i\}} A_{ij} z_j
  &&\text{(realized exposure)}, \notag\\
  E_i^{\mathrm{exp}} &\;:=\; (Pz)_i = \sum_{j \in [r]\setminus\{i\}} p_{ij} z_j
  &&\text{(expected exposure)}. \label{eq:exposures}
\end{align}
Both decompose additively into a respondent component and a non-respondent component: for instance, $E_i^{\mathrm{exp}} = E_{i,\mathrm{base}} + \sum_{k=1}^K \sum_{j \in G_k^{\mathrm{n}}} p_{ij} z_j$, where $E_{i,\mathrm{base}} := \sum_{j \in \Rset\setminus\{i\}} p_{ij} z_j$. Let $\MZ \in \R^{r_{\mathrm{r}}\times r_{\mathrm{r}}}$ denote the symmetric idempotent Frisch--Waugh--Lovell (FWL) orthogonal projector onto $\spanop(\mathbf{1}_{\Rset}, z_{\Rset}, W_{\Rset})^\perp$, with diagonal leverages $h_{ii} := 1 - (\MZ)_{ii}$. For any respondent vector $v \in \R^{r_{\mathrm{r}}}$, write $\tilde{v} := \MZ v$; in particular, let $E_{\Rset} := (E_i^{\mathrm{exp}})_{i\in\Rset}$ denote the respondent expected-exposure vector, with $\tilde{y} := \MZ y_{\Rset}$ and $\widetilde{E} := \MZ E_{\Rset}$.

Let $m_i := \E[y_i \mid \xi, z, W_{\Rset}]$ denote the unrestricted conditional mean outcome of respondent $i$. Our primary population target throughout the paper is the FWL linear-projection spillover slope onto expected exposure:
\begin{equation}
  \label{eq:target_theta}
  \theta^\star \;:=\; \frac{\langle \MZ m,\, \widetilde{E} \rangle}{\|\widetilde{E}\|_2^2},
  \qquad\text{assuming } \widetilde{E} \neq 0.
\end{equation}
To keep causal and descriptive claims strictly separated \citep{savje2024causal}, we state explicitly when $\theta^\star$ admits a causal interpretation:

\begin{assumption}[Causal Spillover Scope]
\label{assump:causal}
Let $Y_i(z, A)$ denote respondent $i$'s potential outcome under roster treatment $z$ and network $A$.
\begin{enumerate}[label=(C\arabic*),leftmargin=2.4em]
  \item \emph{Design exogeneity and conditional link independence:} Conditional on $\xi$, the realized graph $A$ is independent of $(z, W_{\Rset})$ so that $\E[A_{ij} \mid \xi, z, W_{\Rset}] = p_{ij}$; moreover, treatment $z$ is conditionally unconfounded with potential outcomes and latent network traits given $(T, W_{\Rset})$ (or assigned via a known randomized design independent of $\xi$ and the outcome shocks in~(C2) conditional on $W_{\Rset}$).
  \item \emph{Linear exposure mapping:} $Y_i(z, A) = \beta_0 + \beta_1 z_i + \beta_{\mathrm{spill}} E_i + W_i^\top \vartheta_W + \varepsilon_i$, where $E_i \in \{E_i^{\mathrm{real}}, E_i^{\mathrm{exp}}\}$ and $\E[\varepsilon_i \mid A, \xi, z, W_{\Rset}] = 0$ with $\Var(\varepsilon_i \mid \xi, z, W_{\Rset}) = \sigma_\varepsilon^2$.
\end{enumerate}
\end{assumption}

Under Assumption~\ref{assump:causal}, if the structural exposure is $E_i^{\mathrm{exp}}$ then $m_i = \beta_0 + \beta_1 z_i + \beta_{\mathrm{spill}} E_i^{\mathrm{exp}} + W_i^\top \vartheta_W$ directly; if the structural exposure is $E_i^{\mathrm{real}}$, averaging over $A$ via $\E[A_{ij}\mid\xi,z,W_{\Rset}]=p_{ij}$ yields the identical conditional mean $m_i$. In either case, $\MZ m = \beta_{\mathrm{spill}} \widetilde{E}$ and therefore $\theta^\star = \beta_{\mathrm{spill}}$ coincides identically with the causal spillover effect \citep{hudgens2008toward,aronow2017estimating}. Without Assumption~\ref{assump:causal}---in particular, in our observational Karnataka microfinance analysis of Section~\ref{sec:empirical}, where village leaders are not randomly assigned---$\theta^\star$ remains a well-defined \emph{associational} linear-projection estimand capturing the partial regression slope of microfinance adoption on leader exposure controlling for own leader status, village fixed effects, and demographics. Crucially, the sign non-identification results of Section~\ref{sec:orbit} concern the projection estimand $\theta^\star$ with unrestricted $m$ (avoiding outcome-based specification filtering of exposure mappings, which faces minimax testing impossibility under general nested mappings by Theorem~3.1 of \citealp{gao2026impossibility}; see Proposition~\ref{prop:orbit_set}). This partial-identification architecture is also strictly orthogonal to \citet{schroder2026causal}, who assume a completely known graph ($\varphi=1$) and bound exposure misspecification via exogenous propensity-ratio sensitivity parameters, whereas here the non-identification orbit arises endogenously with zero tuning parameters from the exact $\PiN$ symmetry of partial-participation ARD.

\subsection{An Illustrative Six-Node Toy Example}
\label{sec:toy_example}

Before introducing general theorems, a six-node network illustrates why partial-participation ARD fails to identify even the \emph{sign} of the projection estimand $\theta^\star$, and why a simple survey redesign resolves the symmetry immediately.

\begin{example}[Exact Rational Sign Flip under Non-Respondent Swapping]
\label{ex:toy}
Consider a roster of $r = 6$ individuals partitioned into respondents $\Rset = \{1,2,3\}$ and non-respondents $\Nset = \{4,5,6\}$, with $K=2$ non-respondent trait cells $G_1^{\mathrm{n}} = \{4,5\}$ and $G_2^{\mathrm{n}} = \{6\}$, and binary treatment $z = (1, 0, 0,\; 1, 0,\; 0)^\top$ (so respondent~1 and non-respondent~4 are treated, while all others are untreated). There are no additional controls ($W_{\Rset} = \emptyset$).

Suppose respondent-block link probabilities are symmetric across respondents~2 and~3 ($p_{12} = p_{13} = 1/4$, $p_{23} = 1/2$), links to the singleton cell $G_2^{\mathrm{n}}=\{6\}$ are uniform ($p_{i6}=1/2$), and links from respondents to $G_1^{\mathrm{n}}=\{4,5\}$ are polarized across two latent communities:
\[
  (p_{14}, p_{15}) = \Bigl(\tfrac{1}{2}, \tfrac{1}{2}\Bigr),
  \qquad
  (p_{24}, p_{25}) = \Bigl(\tfrac{4}{5}, \tfrac{1}{5}\Bigr),
  \qquad
  (p_{34}, p_{35}) = \Bigl(\tfrac{1}{5}, \tfrac{4}{5}\Bigr).
\]
Notice that respondents~2 and~3 have the \emph{exact same} expected ARD count to $G_1^{\mathrm{n}}$:
\[
  \mu_{21}^{\mathrm{n}} = \tfrac{4}{5} + \tfrac{1}{5} = 1,
  \qquad
  \mu_{31}^{\mathrm{n}} = \tfrac{1}{5} + \tfrac{4}{5} = 1.
\]
Suppose the true conditional mean outcome satisfies $m_i = 1 + z_i + 2 E_i^{\mathrm{exp}}$ ($\beta_{\mathrm{spill}} = +2$). Because $z_{\Rset} = (1,0,0)^\top$, the FWL projector $\MZ$ onto $\spanop(\mathbf{1}_3, z_{\Rset})^\perp$ maps any vector $v = (v_1,v_2,v_3)^\top$ to $\big(0,\; (v_2-v_3)/2,\; -(v_2-v_3)/2\big)^\top$, so the projection slope simplifies to the single ratio
\[
  \theta(\pi) \;=\; \frac{m_2 - m_3}{E_2(\pi) - E_3(\pi)}.
\]
Under the true labeling $\pi^{\mathrm{id}}$ (node~4 treated, node~5 control), we have $E_2(\pi^{\mathrm{id}}) = \frac{1}{4} + \frac{4}{5} = \frac{21}{20}$ and $E_3(\pi^{\mathrm{id}}) = \frac{1}{4} + \frac{1}{5} = \frac{9}{20}$, so $E_2 - E_3 = +\frac{3}{5}$, $m_2 - m_3 = 2 \cdot \frac{3}{5} = +\frac{6}{5}$, and $\theta(\pi^{\mathrm{id}}) = +2$.

Now swap the latent identities of non-respondents~4 and~5 ($\pi^{\mathrm{swap}} = (4 \leftrightarrow 5)$). Every observed edge probability in $P_{\Rset,\Rset}$ is unchanged, and every ARD count distribution $Y_{i1}^{\mathrm{n}} \sim \operatorname{Poisson\text{-}Binomial}(p_{i4}, p_{i5})$ is \emph{identically unchanged} (because summing over $G_1^{\mathrm{n}}=\{4,5\}$ erases which member is which). Yet under $\pi^{\mathrm{swap}}$, respondent~2's expected treated exposure becomes $\frac{1}{4} + \frac{1}{5} = \frac{9}{20}$ while respondent~3's becomes $\frac{1}{4} + \frac{4}{5} = \frac{21}{20}$, reversing $E_2(\pi^{\mathrm{swap}}) - E_3(\pi^{\mathrm{swap}}) = -\frac{3}{5}$ and flipping the spillover projection slope to
\[
  \theta(\pi^{\mathrm{swap}}) \;=\; \frac{+6/5}{-3/5} \;=\; -2.
\]
Thus, for unrestricted $m$---even given an oracle knowing $P$ up to non-respondent relabeling---standard $K$-group ARD yields the orbit identified set $\ThetaI = \{-2, +2\}$: the sign of the spillover projection is strictly unidentified. (Notice that because $r_{\mathrm{r}}=3$ here, $\dim(\operatorname{Im}\MZ)=1$, so even the exact linear structural condition $\tilde{m} = \beta\widetilde{E}(\pi)$ is satisfied at both $\pm 2$; for $r_{\mathrm{r}} > 3$, exact linear collinearity $\tilde{m} \in \spanop\{\widetilde{E}(\pi)\}$ generically filters $\pi$ to a point when $m$ is noiseless and strictly linear, whereas the projection estimand with unrestricted $m$ remains partially identified on $\ThetaI$ at every $r_{\mathrm{r}}$.) In an asymmetric three-non-respondent variant with $G_1^{\mathrm{n}}=\{4,5,6\}$, $(p_{2j})_{j=4}^6 = (\frac{2}{5}, \frac{3}{5}, 0)$ and $(p_{3j})_{j=4}^6 = (\frac{1}{5}, \frac{1}{5}, \frac{3}{5})$---again with identical $\mu_{21}^{\mathrm{n}}=\mu_{31}^{\mathrm{n}}=1$---the orbit takes three distinct values $\ThetaI = \{-\frac{2}{3}, +1, +2\}$, still straddling zero.

\emph{How prospective treatment-aware ARD breaks the symmetry:} Under $2K$ treatment-stratified ARD ($\Tstrat$), splitting each trait group into treated and control cells ($G_{k,1}, G_{k,0}$) places nodes~4 and~5 into distinct cells ($G_{1,1}^{\mathrm{n}} = \{4\}$, $G_{1,0}^{\mathrm{n}} = \{5\}$), forbidding the swap $(4 \leftrightarrow 5)$ and collapsing $\ThetaI$ immediately to the singleton $\{+2\}$. Under $(K{+}1)$ Augmented ARD---asking one extra question about total treated contacts---the $K$ trait cells remain unchanged, but directly observing $Y_{i,\mathrm{tr}} = E_i^{\mathrm{real}}$ bypasses non-respondent relabeling altogether (Proposition~\ref{prop:tstrat}(i)).
\end{example}

\begin{figure}[t]
\centering
\resizebox{\textwidth}{!}{%
\begin{tikzpicture}[
  font=\small,
  rnode/.style={circle, draw=blue!70!black, fill=blue!10, thick, minimum size=7.5mm, inner sep=0pt},
  ntreat/.style={circle, draw=red!75!black, fill=red!14, thick, minimum size=7.5mm, inner sep=0pt},
  nctrl/.style={circle, draw=gray!65!black, fill=gray!12, thick, minimum size=7.5mm, inner sep=0pt},
  strong/.style={-{}, very thick, blue!75!black},
  weak/.style={-{}, dashed, gray!70!black, thin},
  box/.style={draw=black!35, rounded corners=4pt, inner sep=6pt, dashed}
]

\begin{scope}[xshift=0cm]
  \node[font=\bfseries\small, anchor=west] at (-1.1, 3.25) {(a) Standard $K$-group ARD: swap $(4\leftrightarrow 5)$ preserves $\mathcal{D}_{\mathrm{ARD}}$};

  \node[rnode] (r2) at (0, 1.8) {$2$};
  \node[rnode] (r3) at (0, 0.2) {$3$};
  \node[font=\scriptsize, anchor=east, text=blue!70!black] at (-0.5, 1.8) {$z_2=0$};
  \node[font=\scriptsize, anchor=east, text=blue!70!black] at (-0.5, 0.2) {$z_3=0$};

  \node[ntreat] (n4) at (2.3, 1.8) {$4$};
  \node[nctrl]  (n5) at (2.3, 0.2) {$5$};
  \node[font=\scriptsize, anchor=west, text=red!75!black]  at (2.8, 1.8) {$z_4=1$};
  \node[font=\scriptsize, anchor=west, text=gray!65!black] at (2.8, 0.2) {$z_5=0$};

  \draw[strong] (r2) -- node[above, font=\scriptsize, sloped]{$4/5$} (n4);
  \draw[weak]   (r2) -- node[pos=0.28, above, font=\scriptsize]{$1/5$} (n5);
  \draw[weak]   (r3) -- node[pos=0.28, below, font=\scriptsize]{$1/5$} (n4);
  \draw[strong] (r3) -- node[below, font=\scriptsize, sloped]{$4/5$} (n5);

  \node[box, fit=(n4)(n5), label={[font=\scriptsize]above:$G_1^{\mathrm{n}}=\{4,5\}$}] (g1) {};

  \node[font=\scriptsize, align=center] at (1.15, -0.75) {
    $\pi^{\mathrm{id}}$: $\mu_{21}^{\mathrm{n}}=\mu_{31}^{\mathrm{n}}=1$\\
    $E_2 - E_3 = +\frac{3}{5} \;\Rightarrow\; \theta(\pi^{\mathrm{id}}) = \mathbf{+2}$
  };

  \draw[{Latex[length=2.2mm]}-{Latex[length=2.2mm]}, thick, orange!80!black]
    (3.95, 1.0) -- node[above, font=\scriptsize, align=center]{swap\\$(4\leftrightarrow 5)$} (5.15, 1.0);

  \begin{scope}[xshift=6.0cm]
    \node[rnode] (r2b) at (0, 1.8) {$2$};
    \node[rnode] (r3b) at (0, 0.2) {$3$};
    \node[nctrl]  (n4b) at (2.3, 1.8) {$5$};
    \node[ntreat] (n5b) at (2.3, 0.2) {$4$};
    \node[font=\scriptsize, anchor=west, text=gray!65!black] at (2.8, 1.8) {$z_5=0$};
    \node[font=\scriptsize, anchor=west, text=red!75!black]  at (2.8, 0.2) {$z_4=1$};

    \draw[strong] (r2b) -- node[above, font=\scriptsize]{$4/5$} (n4b);
    \draw[weak]   (r2b) -- node[pos=0.28, above, font=\scriptsize]{$1/5$} (n5b);
    \draw[weak]   (r3b) -- node[pos=0.28, below, font=\scriptsize]{$1/5$} (n4b);
    \draw[strong] (r3b) -- node[below, font=\scriptsize]{$4/5$} (n5b);

    \node[box, fit=(n4b)(n5b), label={[font=\scriptsize]above:$G_1^{\mathrm{n}}=\{4,5\}$}] (g1b) {};

    \node[font=\scriptsize, align=center, text=red!70!black] at (1.15, -0.75) {
      $\pi^{\mathrm{swap}}$: \emph{identical} $\mu_{21}^{\mathrm{n}}=\mu_{31}^{\mathrm{n}}=1$\\
      $E_2 - E_3 = -\frac{3}{5} \;\Rightarrow\; \theta(\pi^{\mathrm{swap}}) = \mathbf{-2}$
    };
  \end{scope}
\end{scope}

\begin{scope}[xshift=11.0cm]
  \node[font=\bfseries\small, anchor=west] at (-0.6, 3.25) {(b) $2K$ $\Tstrat$: symmetry broken};

  \node[rnode] (r2c) at (0, 1.8) {$2$};
  \node[rnode] (r3c) at (0, 0.2) {$3$};
  \node[ntreat] (n4c) at (2.2, 1.8) {$4$};
  \node[nctrl]  (n5c) at (2.2, 0.2) {$5$};

  \draw[strong] (r2c) -- node[above, font=\scriptsize]{$4/5$} (n4c);
  \draw[weak]   (r2c) -- (n5c);
  \draw[weak]   (r3c) -- (n4c);
  \draw[strong] (r3c) -- node[below, font=\scriptsize]{$4/5$} (n5c);

  \node[draw=red!65!black, rounded corners=3pt, inner sep=4pt, dashed, fit=(n4c),
        label={[font=\scriptsize, text=red!75!black]right:$G_{1,1}^{\mathrm{n}}$}] {};
  \node[draw=gray!65!black, rounded corners=3pt, inner sep=4pt, dashed, fit=(n5c),
        label={[font=\scriptsize, text=gray!65!black]right:$G_{1,0}^{\mathrm{n}}$}] {};

  \node[font=\scriptsize, align=center, text=green!45!black] at (1.2, -0.75) {
    $\mu_{2,(1,1)}^{\mathrm{n}}=\frac{4}{5} \neq \mu_{3,(1,1)}^{\mathrm{n}}=\frac{1}{5}$\\
    Swap forbidden $\;\Rightarrow\; \ThetaI = \{\mathbf{+2}\}$
  };
\end{scope}

\end{tikzpicture}}
\caption{Visualizing Example~\ref{ex:toy} ($r=6$, $\Rset=\{1,2,3\}$, $\Nset=\{4,5,6\}$, displaying the pivotal block between control respondents $\{2,3\}$ and non-respondent cell $G_1^{\mathrm{n}}=\{4,5\}$; respondent~$1$ and non-respondent~$6$ have symmetric links $p_{12}=p_{13}=1/4$ and $p_{i6}=1/2$). \textbf{(a)}~Under standard $K$-group ARD, permuting non-respondents $(4\leftrightarrow 5)$ preserves every observed respondent edge and every ARD count distribution identically, yet reverses the sign of the spillover projection slope from $+2$ to $-2$. \textbf{(b)}~Splitting the ARD count by treatment status into $2K$ cells ($G_{1,1}^{\mathrm{n}}$ vs.\ $G_{1,0}^{\mathrm{n}}$) breaks the swap invariance and point-identifies $\theta^\star = +2$.}
\label{fig:toy}
\end{figure}
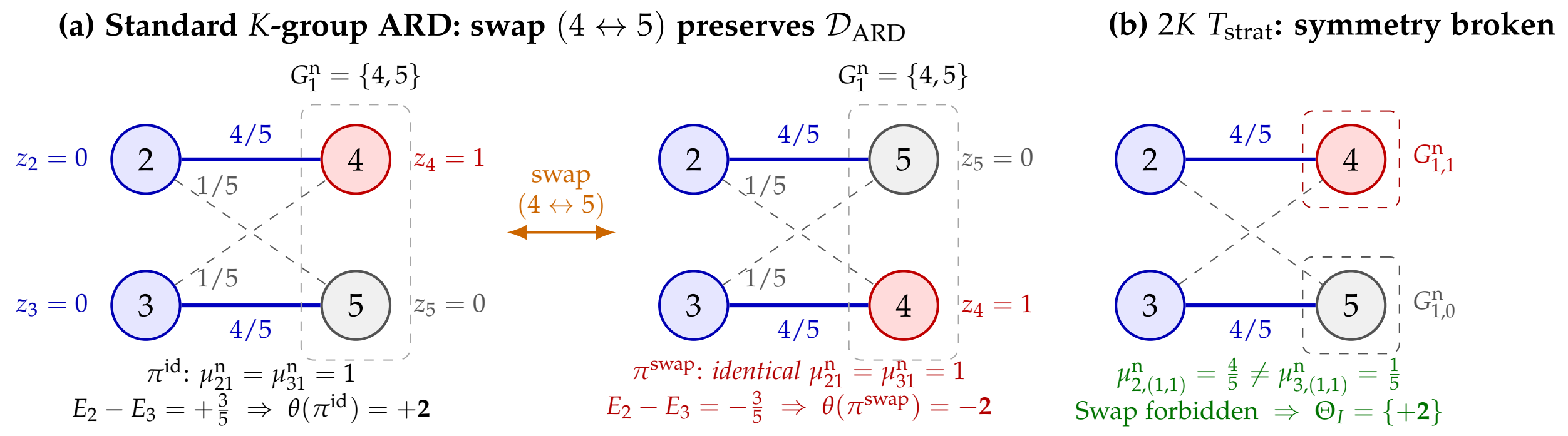

Figure~\ref{fig:toy} visualizes Example~\ref{ex:toy}. The remainder of this paper generalizes this mechanism to large sparse networks, polynomial-time optimization bounds, feasible plug-in estimation, and the 75-village Karnataka field data.

\subsection{Summary of Notation}

To eliminate symbol overload across the theoretical, optimization, and empirical sections, Table~\ref{tab:notation} records the fixed notation used throughout the paper. In particular, $\varphi$ always denotes the survey participation rate while $\pi$ is reserved exclusively for non-respondent permutations; $a_i = e^{\alpha_i}$ denotes node activity while $\theta(\pi)$ denotes the orbit projection estimand; $\eta_{\mathrm{plug}}$ and $(\varrho_{\mathrm{WC}}, \varrho_{\mathrm{stoch}})$ denote plug-in perturbation quantities while $(\rho_\alpha, \rho_\zeta)$ denote unobserved homophily confounding strengths and $\rho_r$ denotes the network sparsity offset; $e_i^{\mathrm{raw}}$ denotes raw-ARD classical edge error; and $\Budget = r_{\mathrm{r}} \Qitems$ ($\Qitems \in \{K, K{+}1, 2K\}$ items per respondent) denotes survey budget while $\Birk_m$ denotes the Birkhoff polytope.

\begin{table}[t]
\centering
\caption{Fixed notation summary used throughout the paper.}
\label{tab:notation}
\resizebox{\textwidth}{!}{%
\begin{tabular}{@{}ll@{\qquad}ll@{}}
\toprule
Symbol & Meaning & Symbol & Meaning \\
\midrule
$[r], \Rset, \Nset$ & Roster, respondents ($r_{\mathrm{r}}$), non-resp.\ ($r_{\mathrm{n}}$) & $\varphi := r_{\mathrm{r}}/r$ & Survey participation rate \\
$G_k, G_k^{\mathrm{r}}, G_k^{\mathrm{n}}$ & Trait group $k$ and $\Rset$/$\Nset$ cells ($m_{\mathrm{r},k}, m_{\mathrm{n},k}$) & $\pi \in \PiN$ & Non-respondent cell permutation \\
$A, P, \psi(\xi_i,\xi_j)$ & Adjacency (union matrix in \S\ref{sec:empirical}), prob., kernel & $\xi_i=(\alpha_i,\zeta_i), \rho_r$ & Latent activity ($a_i=e^{\alpha_i}$), sign ($\pm 1$), sparsity offset \\
$Y_{ik}, Y_{ik}^{\mathrm{r}}, Y_{ik}^{\mathrm{n}}, D_i$ & Total, $\Rset$-cell, $\Nset$-cell ARD counts, total degree & $\mu_{ik}^{\mathrm{n}}$ & Expected $\Nset$-cell count $\sum_{j\in G_k^{\mathrm{n}}} p_{ij}$ \\
$z, \bar{z}_{\Nset,k}, \bar{z}_{k,-i}$ & Treatment, $\Nset$-cell mean, LOO group mean & $\MZ, h_{ii}$ & FWL projector onto $(\mathbf{1},z_{\Rset},W_{\Rset})^\perp$, leverage \\
$E^{\mathrm{real}}, E^{\mathrm{exp}}, E^{\mathrm{obs}}$ & Realized $(Az)$, expected $(Pz)$, union-graph exposure & $\widetilde{E}, \tilde{y}, m, r^\star$ & Residualized $\MZ E_{\Rset}, \MZ y_{\Rset}$; cond.\ mean; proj.\ residual \\
$\theta^\star, \theta(\pi), \ThetaI, \ThetaIhat$ & Target slope, orbit slope, population/sample orbit sets & $\beta_{\mathrm{spill}}, \mathcal{O}, \mathcal{P}$ & Structural slope; discrete orbit; permutohedron hull \\
$Q_\beta(x), \underline{L}, \overline{L}$ & Level-set quadratic form and linear extrema & $S_k, Q_k, N_{\mathrm{LB}}$ & Cell treatment sum, norm squared, denom.\ floor \\
$\Delta^{\mathrm{est}}, \Delta^{\mathrm{swap}}, \Delta^{\mathrm{real}}$ & 3-term exposure error decomposition & $\omega_{ik}$ & Within-cell cov.\ $\Cov_{G_k^{\mathrm{n}}}(p_{i\cdot}, z)$ \\
$\eta_{\mathrm{plug}}, \underline{\sigma}_{\widehat{E}}, \mathrm{se}_{\max}$ & Plug-in norm radius, denom.\ floor, noise bound & $\varrho_{\mathrm{WC}}, \varrho_{\mathrm{stoch}}$ & Worst-case norm ratio; stochastic-EIV heuristic ratio \\
$\mu^{\mathrm{raw}}, e_i^{\mathrm{raw}}, \hat{v}^{\mathrm{lo}/\mathrm{up}}$ & Raw-ARD mean, edge error, C--S variance bounds & $r_{\mathrm{eff}}, \alpha^{(1)}, \alpha^{(2)}$ & Satterthwaite--Welch df; union-bound tail levels \\
$\rho_\alpha, \rho_\zeta$ & Unobserved homophily confounding slopes & $Y_{i,\mathrm{tr}}, Y_{i,(k,z)}, D_{i,0}$ & $(K{+}1)$ / $2K$ $\Tstrat$ treated counts and control degree \\
$\Budget := r_{\mathrm{r}} \Qitems, \Birk_m$ & Survey budget ($\Qitems$ items/resp.); Birkhoff polytope & $G, n_{\Rset}$ & Pooled village clusters and total respondents \\
$\kappa, \Delta_{\mathrm{res}}$ & Exact $\hat{\beta}_{\mathrm{RawHyb}} = \hat{\beta}_{\mathrm{obs}}\kappa + \Delta_{\mathrm{res}}$ & $\beta_{\Rset\Rset}, \beta_{\Rset\Nset}, \lambda^\star_{\mathrm{OOS}}$ & $\Rset\Rset$/$\Rset\Nset$ slopes; held-out shrinkage threshold \\
\bottomrule
\end{tabular}}
\end{table}

\section[Non-Respondent Symmetry, Orbit Partial Identification, and Polynomial-Time Certification]{Non-Respondent Symmetry, Orbit Partial Identification,\\and Polynomial-Time Certification}
\label{sec:orbit}

Example~\ref{ex:toy} is not an artifact of a six-node graph; it reflects an exact group symmetry inherent to partial-participation ARD under any symmetric network model. This section establishes that invariance, specifies carefully what the network law identifies versus what requires a column-profile oracle, characterizes when the resulting \emph{Orbit Identified Set} $\ThetaI$ collapses to a point, proves an exact near-linear test for sign identification, and develops nested polynomial-time outer certifiers exploiting permutohedron geometry.

\subsection{Unconditional Non-Respondent Swap Invariance}

Let $\PiN := \prod_{k=1}^K \mathfrak{S}(G_k^{\mathrm{n}})$ denote the group of permutations $\pi = (\pi_1,\dots,\pi_K)$ that permute non-respondents arbitrarily within each trait cell $G_k^{\mathrm{n}}$ while leaving every respondent $i \in \Rset$ fixed. For any latent profile $\xi = (\xi_1,\dots,\xi_r) \in \Xi^r$ and $\pi \in \PiN$, write $\xi^\pi$ for the permuted profile with $(\xi^\pi)_i = \xi_i$ for $i \in \Rset$ and $(\xi^\pi)_j = \xi_{\pi_k(j)}$ for $j \in G_k^{\mathrm{n}}$.

\begin{proposition}[Unconditional Non-Respondent Swap Invariance]
\label{prop:swap_invariance}
Under the independent-edge model \eqref{eq:graph_model} with any symmetric link kernel $\psi : \Xi \times \Xi \to [0,1]$, any roster size $r$, any sparsity regime, and any participation rate $\varphi \in (0,1)$, the joint law of the observed network data $(A_{\Rset,\Rset}, Y_{\Rset,\cdot})$ satisfies
\[
  \mathcal{L}_{\xi}\!\big(A_{\Rset,\Rset},\, Y_{\Rset,\cdot}\big)
  \;=\;
  \mathcal{L}_{\xi^\pi}\!\big(A_{\Rset,\Rset},\, Y_{\Rset,\cdot}\big)
  \qquad \text{for every } \pi \in \PiN.
\]
Consequently, the Hellinger distance between $\mathcal{L}_{\xi}$ and $\mathcal{L}_{\xi^\pi}$ is identically zero ($H^2 \equiv 0$). More strongly, because conditional on $\xi$ the non-respondent counts $\{Y_{ik}^{\mathrm{n}}\}_{i\in\Rset}$ sum over disjoint edge sets across distinct respondents $i \in \Rset$, for an unrestricted kernel $\psi$ the network law $\mathcal{L}_\xi$ is invariant even to \emph{respondent-specific} permutations of $\{p_{ij} : j \in G_k^{\mathrm{n}}\}$ within each row $(i,k)$, identifying only $P_{\Rset,\Rset}$ and the row-wise multisets $\{p_{ij} : j \in G_k^{\mathrm{n}}\}$ (not the cross-row column alignment, which must be supplied by a structural network model such as \eqref{eq:logistic_model} or a column-profile oracle).
\end{proposition}

\begin{proof}
Fix $\pi \in \PiN$. Because $\pi$ acts as the identity on $\Rset$, every respondent-block probability $p_{ij}(\xi^\pi) = \psi(\xi_i, \xi_j) = p_{ij}(\xi)$ for $i,j \in \Rset$ is unchanged, so $A_{\Rset,\Rset}$ has the identical distribution under $\xi$ and $\xi^\pi$. Independently of $A_{\Rset,\Rset}$, the non-respondent cell counts $Y_{ik}^{\mathrm{n}} = \sum_{j \in G_k^{\mathrm{n}}} A_{ij}$ depend on disjoint $\Rset \times G_k^{\mathrm{n}}$ edge sets across distinct $(i,k)$. Because edges $\{A_{ij}\}$ are conditionally independent given $\xi$, each $Y_{ik}^{\mathrm{n}} \sim \operatorname{Poisson\text{-}Binomial}(\{p_{ij} : j \in G_k^{\mathrm{n}}\})$ depends only on the unordered multiset of probabilities in row $(i,k)$, which is preserved under both shared permutations $\pi \in \PiN$ and arbitrary row-specific permutations of $G_k^{\mathrm{n}}$. Finally, $Y_{ik} = Y_{ik}^{\mathrm{r}} + Y_{ik}^{\mathrm{n}}$ is a deterministic function of $(A_{\Rset,\Rset}, Y_{\Rset,\cdot}^{\mathrm{n}})$.
\end{proof}

\begin{remark}[Scope Boundary Relative to Full-Roster ARD and Sampled Networks]
\label{rem:swap_scope}
Proposition~\ref{prop:swap_invariance} delineates an exact scope boundary with existing ARD and missing-link recovery theory. Full-roster recovery theorems---including Lemma~3.4 of \citet{wilkinsreeves2026single} (numbering throughout refers to preprint \texttt{arXiv:2406.11940v1}), the alignment-plus-resolution analysis of \citet{tseng2026alignment}, the row-compression spectral recovery of \citet{davison2026community} (where Remark~4.7 notes second moments remain informative even at mean collapse), and the low-rank ARD network smoothing of \citet{hayes2026smoothing} (which requires $\varphi=1$, dense $\min_i \tilde{d}_i = \omega(n^{3/4}\log n)$, and confronts the bounded-degree minimax scissor of \citealp{hayes2025minimax} for randomized peer treatments $Gz$)---all operate at $\varphi = 1$ ($\Nset = \emptyset$), where every individual supplies their own ARD row. Likewise, \citet{breza2020using,breza2023consistently} establish recovery of network structure and group-invariant statistics under latent-surface models, while \citet{boucher2025estimating} (Appendix~B.1.6) note that ARD fails their bounded-group Assumption~5; and in missing-link econometrics, \citet{zhang2024spillovers} shows sign reversal from omitted edges only under \emph{heterogeneous non-monotone} spillovers (Corollary~3.2) while noting (fn.~10) that partial node sampling breaks her Assumption~5. Proposition~\ref{prop:swap_invariance} establishes a sharper barrier under partial participation ($\varphi < 1$) that persists as $r \to \infty$ with zero reporting error and homogeneous linear spillovers whenever non-respondent edges are paired with node-specific treatments $z_j$ without predicting individual $\xi_j$ from $(z_j, W)$.
\end{remark}

\subsection{The Orbit Identified Set \texorpdfstring{$\ThetaI$}{Theta\_I} and Collapse-to-Point Characterization}

To separate the unavoidable non-respondent labeling symmetry $\PiN$ from cross-row alignment uncertainty, consider a \emph{column-profile oracle} that observes $P_{\Rset,\Rset}$ and the unordered multiset of non-respondent column profiles $\{(p_{ij})_{i\in\Rset} : j \in G_k^{\mathrm{n}}\}$ for each group $k$ (equivalently, observes the probability matrix $P_{\Rset,\cdot}$ up to the shared group action $\PiN$, as happens when a structural model such as \eqref{eq:logistic_model} aligns columns across respondents). Each candidate permutation $\pi \in \PiN$ induces a candidate expected exposure vector on respondents:
\begin{equation}
  \label{eq:E_pi}
  E_{\Rset}(\pi) \;=\; E_{\Rset,\mathrm{base}} + \sum_{k=1}^K P_{\Rset, G_k^{\mathrm{n}}} \Pi_k z^{(k)},
  \qquad
  \widetilde{E}(\pi) \;:=\; \MZ E_{\Rset}(\pi),
\end{equation}
where $E_{i,\mathrm{base}} := \sum_{j \in \Rset\setminus\{i\}} p_{ij} z_j$, $z^{(k)} := (z_j)_{j\in G_k^{\mathrm{n}}} \in [0,1]^{m_{\mathrm{n},k}}$, and $\Pi_k \in \mathcal{P}_{m_{\mathrm{n},k}}$ is the permutation matrix representing $\pi_k$.

\begin{definition}[Orbit Identified Set and Sample Analogue]
\label{def:orbit_set}
Let $m = (\E[y_i \mid \xi, z, W_{\Rset}])_{i\in\Rset}$ and $\tilde{m} := \MZ m$. Define the population orbit projection slope and the \emph{Orbit Identified Set} by
\begin{equation}
  \label{eq:orbit_set}
  \theta(\pi) \;:=\; \frac{\langle \tilde{m},\, \widetilde{E}(\pi) \rangle}{\|\widetilde{E}(\pi)\|_2^2},
  \qquad
  \ThetaI \;:=\; \big\{ \theta(\pi) : \pi \in \PiN,\; \widetilde{E}(\pi) \neq 0 \big\},
\end{equation}
with the convention that $\ThetaI$ is reported as unbounded ($\R$) whenever $\widetilde{E}(\pi)=0$ for some $\pi\in\PiN$ with $\tilde{m}\neq 0$. Replacing $\tilde{m}$ with observed residualized outcomes $\tilde{y} = \MZ y_{\Rset}$ yields the \emph{sample orbit set} $\ThetaIhat := \{ \hat{\theta}(\pi) := \langle \tilde{y}, \widetilde{E}(\pi)\rangle / \|\widetilde{E}(\pi)\|_2^2 : \pi \in \PiN, \widetilde{E}(\pi) \neq 0 \}$.
\end{definition}

Separating the column-profile oracle set $\ThetaI$ from both the row-wise general-$\psi$ set and the sample analogue $\ThetaIhat$---and specifying the model class for $m$---is essential for identification clarity \citep{manski1990nonparametric,imbens2004confidence}:

\begin{proposition}[Sharpness Relative to the Column-Profile Oracle, Outer Validity, and Point-ID Collapse]
\label{prop:orbit_set}
Consider the population replication thought experiment in which $m = (\E[y_i \mid \xi, z, W_{\Rset}])_{i\in\Rset}$ and the column-profile oracle $(P_{\Rset,\Rset}, \{P_{\Rset, G_k^{\mathrm{n}}}\}_{k=1}^K \bmod \PiN)$ are known, and assume $\widetilde{E}(\pi) \neq 0$ for all $\pi \in \PiN$.
\begin{enumerate}[label=(\roman*),leftmargin=2em]
  \item \emph{Sharpness for the column-profile oracle under unrestricted $m$:} If $m \in \R^{r_{\mathrm{r}}}$ is unrestricted (allowing arbitrary nonlinear exposure responses or realized-graph deviations), then $\ThetaI$ is the \emph{sharp} identified set for the projection estimand $\theta^\star = \theta(\pi^\star)$ given $(m, P_{\Rset,\Rset}, \{P_{\Rset,G_k^{\mathrm{n}}}\}\bmod\PiN)$: every $\theta \in \ThetaI$ is attained by a relabeled column-profile structure $(\xi^\pi, m)$ inducing the identical observable law.
  \item \emph{Misspecification-robust outer validity under linear structural models:} Under Assumption~\ref{assump:causal}, $\theta^\star = \beta_{\mathrm{spill}} \in \ThetaI$, so $\ThetaI$ is a valid outer identified set for $\beta_{\mathrm{spill}}$. For $r_{\mathrm{r}} > \operatorname{rank}(\mathbf{1}, z_{\Rset}, W_{\Rset}) + 1$, $\ThetaI$ is generically strictly wider than the structural conditional-mean identified set $\{\beta : \exists \pi \in \PiN,\, \tilde{m} = \beta \widetilde{E}(\pi)\}$, because $\ThetaI$ purposefully avoids relying on exact zero-residual collinearity across respondents.
  \item \emph{Exact characterization of point-identification collapse:}
  \begin{enumerate}[label=(\alph*),leftmargin=1.6em]
    \item The raw exposure vector $E_{\Rset}(\pi)$ is constant across all $\pi \in \PiN$ if and only if, for every group $k \in \{1,\dots,K\}$, either $z^{(k)}$ is constant on $G_k^{\mathrm{n}}$ or every respondent's probability vector $(p_{ij})_{j\in G_k^{\mathrm{n}}}$ is constant in $j \in G_k^{\mathrm{n}}$. Under strictly interior probabilities $p_{ij}\in(0,1)$, the latter condition is the \emph{degenerate special case} $\logit(p_{ij}) = \gamma_{i,k}$ (zero node-$j$ effect) of the $k$-beta additive log-odds class of \citet{tseng2026countfibres}.
    \item The projected exposure vector $\widetilde{E}(\pi) = \MZ E_{\Rset}(\pi)$ is constant across $\pi \in \PiN$ (so $\ThetaI = \{\theta^\star\}$ collapses to a singleton for every $m$) if and only if
    \begin{equation}
      \label{eq:exact_collapse_iff}
      \MZ P_{\Rset, G_k^{\mathrm{n}}}(e_j - e_{j'}) \;=\; 0
      \qquad \text{for every } k \in \{1,\dots,K\} \text{ and every } j, j' \in G_k^{\mathrm{n}} \text{ with } z_j \neq z_{j'}.
    \end{equation}
  \end{enumerate}
\end{enumerate}
\end{proposition}

\begin{proof}[Proof sketch (complete proof in Appendix~\ref{app:orbit})]
Part~(i) follows directly from Proposition~\ref{prop:swap_invariance}: holding $m$ fixed and relabeling columns by any $\pi \in \PiN$ leaves the observable law of $(A_{\Rset,\Rset}, Y_{\Rset,\cdot})$ and the column-profile oracle invariant while mapping $\theta(\pi^\star)$ to $\theta(\pi)$. Part~(ii) holds because $\tilde{m} = \beta_{\mathrm{spill}}\widetilde{E}(\pi^\star)$ yields $\theta(\pi^\star)=\beta_{\mathrm{spill}} \in \ThetaI$. For Part~(iii), comparing any transposition $\sigma_{j,j'} = (j \leftrightarrow j')$ within $G_k^{\mathrm{n}}$ gives $E_{\Rset}(\mathrm{id}) - E_{\Rset}(\sigma_{j,j'}) = (z_j - z_{j'}) P_{\Rset, G_k^{\mathrm{n}}}(e_j - e_{j'})$, and applying $\MZ$ gives $\widetilde{E}(\mathrm{id}) - \widetilde{E}(\sigma_{j,j'}) = (z_j - z_{j'}) \MZ P_{\Rset, G_k^{\mathrm{n}}}(e_j - e_{j'})$, which vanishes across all transpositions if and only if \eqref{eq:exact_collapse_iff} holds.
\end{proof}

\begin{remark}[Column Alignment vs.\ Unrestricted $\psi$ and Point-ID Collapse]
\label{rem:collapse_and_alignment}
Under a completely unrestricted link kernel $\psi$ without cross-row column alignment, row-wise permutations produce a strictly larger identified set containing $\ThetaI$; the certified bounds in Theorem~\ref{thm:certifiers} below are valid outer envelopes specifically for the column-aligned orbit set $\ThetaI$. Notice also why condition~(b) in \eqref{eq:exact_collapse_iff} is strictly weaker than~(a): if $p_{ij} = c_j$ depends only on $j$ (not on $i \in \Rset$) within cell $k$, then $P_{\Rset, G_k^{\mathrm{n}}}(e_j - e_{j'}) = (c_j - c_{j'}) \mathbf{1}_{\Rset} \in \ker(\MZ)$ is annihilated by the intercept projection, so $\ThetaI$ collapses to a point even though $p_{ij}$ and $z_j$ both vary in $j$. Whenever non-respondents carry heterogeneous latent traits whose link profiles \emph{interact} with respondent attributes (so $\MZ P_{\Rset, G_k^{\mathrm{n}}}(e_j - e_{j'}) \neq 0$), $\ThetaI$ is non-degenerate. (Appendix~\ref{app:orbit} records a verified numerical contrast of Part~(ii), where the discrete orbit $\ThetaI$ has endpoint range $[1.1918, 3.8372]$ under unrestricted $m$ while exact zero-residual structural filtering collapses to $\{2.0000\}$. This clarifies the trade-off behind likelihood-based proxy-network estimators---such as the Bayesian MCMC estimator of \citealp{boucher2025estimating}, Section~4, Algorithm~1, Step~1 and fn.~30, the full-graph Bayesian latent-network feedback examined in Appendix~C.3 of \citealp{weinstein2026bayesian}, and the mixture/EM deconvolution of \citealp{hardy2024estimating}, whose Section~4.2 notes non-identification for binary outcomes: borrowing identification from a parametric outcome likelihood $P(y \mid \theta, A)$ can shrink the orbit, though on a sparse $\varphi < 1$ graph a single non-respondent swap shifts only $O(d_{\mathrm{avg}})$ respondents' outcomes and changes the outcome log-likelihood by only $O(d_{\mathrm{avg}}\beta_{\mathrm{spill}}^2/\sigma_\varepsilon^2) = O(1)$, so $\ThetaI$ deliberately avoids outcome-model filtering to preserve misspecification robustness.)
\end{remark}

\subsection{Permutohedron Reduction and Exact Linear Sign-Reversal Equivalence}

Because $|\PiN| = \prod_{k=1}^K m_{\mathrm{n},k}!$ grows super-exponentially even in a single village ($m_{\mathrm{n},k} \approx 20$ gives $20! > 2\times 10^{18}$), enumerating $\ThetaI$ or $\ThetaIhat$ is intractable, and minimizing a non-convex quadratic form over permutations is NP-hard in general (reducing to quadratic assignment). To obtain exact sign tests and tight certified outer bounds in polynomial time, we lift $\PiN$ into Euclidean space via classical majorization geometry \citep{birkhoff1946tres,rado1952inequality,hardy1952inequalities}.

Let $x^{(k)} := \Pi_k z^{(k)} \in \R^{m_{\mathrm{n},k}}$, stack $x := (x^{(1)}; \dots; x^{(K)}) \in \R^{r_{\mathrm{n}}}$, and define the discrete orbit $\mathcal{O} := \prod_{k=1}^K \mathcal{O}_k$ with $\mathcal{O}_k := \{\Pi_k z^{(k)} : \Pi_k \in \mathcal{P}_{m_{\mathrm{n},k}}\}$. Set $\tilde{e}_0 := \MZ E_{\Rset,\mathrm{base}}$, $\widetilde{P}^{(k)} := \MZ P_{\Rset, G_k^{\mathrm{n}}}$, and $\widetilde{P} := [\widetilde{P}^{(1)}, \dots, \widetilde{P}^{(K)}] \in \R^{r_{\mathrm{r}} \times r_{\mathrm{n}}}$. Then $\widetilde{E}(\pi) = \widetilde{E}(x) := \tilde{e}_0 + \widetilde{P} x$ is affine in $x$. By the Birkhoff--von Neumann theorem \citep{birkhoff1946tres} and Rado's theorem \citep{rado1952inequality}, the convex hull of $\mathcal{O}$ is the Cartesian product of permutohedra
\[
  \mathcal{P} \;:=\; \conv(\mathcal{O}) \;=\; \prod_{k=1}^K \mathcal{P}_k,
  \qquad
  \mathcal{P}_k = \Birk_{m_{\mathrm{n},k}} z^{(k)} = \{ u \in \R^{m_{\mathrm{n},k}} : u \prec z^{(k)} \},
\]
which simplifies for binary treatments $z^{(k)} \in \{0,1\}^{m_{\mathrm{n},k}}$ with $S_k := \sum_{j \in G_k^{\mathrm{n}}} z_j$ treated units to the hypersimplex $\mathcal{P}_k = \{u \in [0,1]^{m_{\mathrm{n},k}} : \mathbf{1}^\top u = S_k\}$. By the rearrangement inequality \citep{hardy1952inequalities}, any linear objective $c^\top x = \sum_{k=1}^K c^{(k)\top} x^{(k)}$ is minimized over $\mathcal{P}$ at an extreme point in $\mathcal{O}$ by sorting each $c^{(k)}$ ascending against $z^{(k)}$ descending, in $O(r_{\mathrm{n}} + \sum_k m_{\mathrm{n},k}\log m_{\mathrm{n},k})$ time.

For any candidate threshold $\beta \in \R$ and outcome vector $\tilde{y}$ (replacing $\tilde{y}$ by $\tilde{m}$ gives the population version), define the level-set quadratic form
\begin{equation}
  \label{eq:Q_beta}
  Q_\beta(x) \;:=\; \tilde{y}^\top \widetilde{E}(x) - \beta \|\widetilde{E}(x)\|_2^2
  \;=\;
  x^\top (-\beta M_0) x + (y_P - 2\beta e_{0,P})^\top x + (\tilde{y}^\top \tilde{e}_0 - \beta \|\tilde{e}_0\|_2^2),
\end{equation}
where $M_0 := \widetilde{P}^\top \widetilde{P} \succeq 0$, $y_P := \widetilde{P}^\top \tilde{y}$, and $e_{0,P} := \widetilde{P}^\top \tilde{e}_0$. Notice a sharp geometric asymmetry:

\begin{lemma}[Level-Set Certification and Convexity Asymmetry]
\label{lem:level_set}
\begin{enumerate}[label=(\roman*),leftmargin=2em]
  \item If a computable lower bound satisfies $L(\beta) \le \min_{x\in\mathcal{O}} Q_\beta(x)$ and $L(\beta) > 0$, then $\widetilde{E}(x) \neq 0$ and $\hat{\theta}(x) > \beta$ for all $x \in \mathcal{O}$. Symmetrically, if $U(\beta) \ge \max_{x\in\mathcal{O}} Q_\beta(x)$ satisfies $U(\beta) < 0$, then $\widetilde{E}(x) \neq 0$ and $\hat{\theta}(x) < \beta$ for all $x \in \mathcal{O}$.
  \item Because $M_0 \succeq 0$, $Q_\beta(x)$ is \emph{concave} on $\mathcal{P}$ whenever $\beta > 0$ (so certifying upper bounds $\overline{\beta} > 0$ via $\max_{\mathcal{P}} Q_\beta$ is convex, whereas certifying positive lower bounds $\underline{\beta} > 0$ via $\min_{\mathcal{O}} Q_\beta$ is non-convex concave minimization) and \emph{convex} whenever $\beta < 0$ (reversing which endpoint is convex).
\end{enumerate}
\end{lemma}

Setting $\beta = 0$ eliminates the quadratic term $-\beta M_0$ identically and reduces sign identification to exact linear optimization over $\mathcal{P}$:

\begin{theorem}[Exact Near-Linear Sign-Reversal Equivalence]
\label{thm:sign_test}
Define the linear numerator extrema
\[
  \underline{L} \;:=\; \min_{x\in\mathcal{P}} \tilde{y}^\top \widetilde{E}(x) \;=\; \tilde{y}^\top \tilde{e}_0 + \min_{x\in\mathcal{P}} y_P^\top x,
  \qquad
  \overline{L} \;:=\; \max_{x\in\mathcal{P}} \tilde{y}^\top \widetilde{E}(x),
\]
computable in $O(r_{\mathrm{r}} r_{\mathrm{n}} + \sum_{k=1}^K m_{\mathrm{n},k} \log m_{\mathrm{n},k})$ time with attaining extreme points $x_{\min}, x_{\max} \in \mathcal{O}$ for $\underline{L}$ and $\overline{L}$, respectively.
\begin{enumerate}[label=(\roman*),leftmargin=2em]
  \item If $\widetilde{E}(x_{\min}) \neq 0$, then $\underline{L} > 0 \Longleftrightarrow \inf \ThetaIhat > 0$ (sign reversal $\hat{\theta}(\pi) \le 0$ is excluded over $\PiN$). Symmetrically, if $\widetilde{E}(x_{\max}) \neq 0$, then $\overline{L} < 0 \Longleftrightarrow \sup \ThetaIhat < 0$.
  \item Whenever $\underline{L} \le 0 < \overline{L}$ and $0 \notin \widetilde{E}(\mathcal{P})$, \emph{both} endpoints of the convex-hull relaxation $\inf_{\mathcal{P}}\hat{\theta}(x) \le 0 < \sup_{\mathcal{P}}\hat{\theta}(x)$ lie on the convex side of Lemma~\ref{lem:level_set}(ii).
\end{enumerate}
\end{theorem}

\begin{proof}
Because $\mathcal{P} = \conv(\mathcal{O})$ and $x \mapsto \tilde{y}^\top \widetilde{E}(x)$ is affine, its minimum and maximum over $\mathcal{P}$ equal those over $\mathcal{O}$ and are attained at extreme points $x_{\min}, x_{\max} \in \mathcal{O}$ via cell-wise sorting. If $\underline{L} > 0$, then for every $x \in \mathcal{O}$ we have $\tilde{y}^\top \widetilde{E}(x) \ge \underline{L} > 0$, forcing $\|\widetilde{E}(x)\|_2^2 > 0$ and $\hat{\theta}(x) = \tilde{y}^\top \widetilde{E}(x)/\|\widetilde{E}(x)\|_2^2 > 0$; finiteness of $\mathcal{O}$ gives $\inf \ThetaIhat = \min \ThetaIhat > 0$. Conversely, if $\underline{L} \le 0$ and $\widetilde{E}(x_{\min}) \neq 0$, then $x_{\min}\in\mathcal{O}$ witnesses $\hat{\theta}(x_{\min}) = \underline{L}/\|\widetilde{E}(x_{\min})\|_2^2 \le 0$. The upper-bound equivalence follows identically by replacing $\tilde{y}$ with $-\tilde{y}$ and invoking $\widetilde{E}(x_{\max})\neq 0$. Part~(ii) follows immediately from Lemma~\ref{lem:level_set}(ii), since certifying $\underline{\beta} \le 0$ uses $\beta < 0$ (convex minimization) and certifying $\overline{\beta} > 0$ uses $\beta > 0$ (concave maximization).
\end{proof}

Theorem~\ref{thm:sign_test} explains two phenomena that govern our empirical analysis: (a) testing whether non-respondent relabeling can flip the sign of a spillover projection costs only sorting, with zero relaxation gap; and (b) whenever the orbit straddles zero (as occurs for the model-implied sample orbit $\ThetaIhat(\widehat{P})$ in every pooled Karnataka specification, examined out of sample in Section~\ref{sec:real_mf}), \emph{both} outer endpoints lie on the tractable convex side. Notice the structural contrast with unlinked-data optimal-transport bounds \citep{dhaultfoeuille2025linear,fan2025partial}: in two-sample linear regressions and conditional optimal transport, the regression denominator is point-identified and moments are \emph{linear} in the unknown coupling, whereas network mixing $P_{\Rset,\Nset}\Pi z$ makes the denominator $\|\widetilde{E}(x)\|_2^2$ itself quadratic in $\Pi$ (a non-convex quadratic assignment problem), collapsing to exact linear sorting only at $\beta = 0$.

\subsection{Polynomial-Time Outer Certification: From Uncoupled Boxes to Cholesky-Certified Shor-SDP}

When one seeks not merely the sign but a tight certified interval $[\underline{\beta}_{\mathrm{out}}, \overline{\beta}_{\mathrm{out}}] \supseteq \ThetaIhat$---especially when sign reversal is excluded ($\underline{L} > 0$) so that certifying the positive lower endpoint $\underline{\beta} > 0$ requires non-convex minimization of $Q_\beta(x)$ over $\mathcal{O}$---per-respondent bounds break down. Specifically, computing each respondent's individual rearrangement interval $[\ell_i, u_i] = [\min_{\mathcal{O}} \widetilde{E}_i(x), \max_{\mathcal{O}} \widetilde{E}_i(x)]$ ignores the fact that all $r_{\mathrm{r}}$ respondents must share the \emph{same} permutation $\pi \in \PiN$. Because the uncoupled Cartesian box $\prod_{i=1}^{r_{\mathrm{r}}} [\ell_i, u_i]$ readily contains the origin even when $\min_{x\in\mathcal{O}} \|\widetilde{E}(x)\|_2^2 \gg 0$, uncoupled boxes frequently blow up to $(-\infty, +\infty)$ or stay $2\times$--$10\times$ too wide.

To preserve graph coupling across all respondents in polynomial time, we combine classical convex-optimization tools \citep{frank1956algorithm,shor1987quadratic,luo2010semidefinite} with exact orbit invariants satisfied by every $x \in \mathcal{O}$ (group sums $\mathbf{1}^\top x^{(k)} = S_k$, squared norms $\|x^{(k)}\|_2^2 = Q_k := \sum_{j \in G_k^{\mathrm{n}}} z_j^2$, and coordinate box bounds; full formulas in Appendix~\ref{app:orbit}), yielding a nested three-tier outer certification hierarchy:
\begin{itemize}[leftmargin=1.5em]
  \item \textbf{Tier~1 (\texttt{LP-Secant}):} Certifies a denominator floor $N_{\mathrm{LB}} \le \min_{x\in\mathcal{P}} \|\widetilde{E}(x)\|_2^2$ via Pairwise Frank--Wolfe \citep{frank1956algorithm} duality gaps (returning $(-\infty,+\infty)$ if $N_{\mathrm{LB}} \le 0$) and linearizes each $\widetilde{E}_i(x)^2$ on the non-convex side by its chord secant over exact row extrema $[\ell_i, u_i]$.
  \item \textbf{Tier~2 (\texttt{Perm-QP} and Pooled Matrix-Free \texttt{Perm-QP}):} Intersects Tier~1 with exact convex-side optimization of $\pm Q_\beta(x)$ over $\mathcal{P}$ via Dinkelbach iteration and Pairwise Frank--Wolfe dual bisection. Because each step uses only matrix--vector products with $\widetilde{P}$ and cell-wise sorting, Tier~2 scales \emph{matrix-free} to pooled multi-village fixed-effects regressions across $G$ villages ($n_{\Rset} = \sum_{v=1}^G r_{\mathrm{r},v} = 4{,}891$ respondents across $G = 49$ villages).
  \item \textbf{Tier~3 (\texttt{FastShorSDP+PermRefine}):} Tightens the non-convex side ($\min_{x\in\mathcal{O}} Q_\beta(x)$ for $\beta > 0$) by attaching dual multipliers $\lambda \in \R^K$ to the sphere invariants $\|x^{(k)}\|_2^2 = Q_k$ and $\nu \in \R^{r_{\mathrm{n}}}$ to the coordinate box invariants ($\nu_\ell \ge 0$ on non-binary cells where $z^{(k(\ell))} \notin \{0,1\}^{m_{\mathrm{n},k(\ell)}}$) to form the Shor quadratic Lagrangian $\mathcal{L}_\beta(x;\lambda,\nu) \le Q_\beta(x)$ \citep{shor1987quadratic,luo2010semidefinite}. Eliminating all $K$ group-sum equalities via $x = \bar{x} + Q_V w$---where $\bar{x}$ is the cell-mean vector and $Q_V \in \R^{r_{\mathrm{n}}\times (r_{\mathrm{n}}-K)}$ is an orthonormal Helmert basis of the group-sum kernel---reduces $\mathcal{L}_\beta$ to the unconstrained quadratic $\mathcal{L}_\beta(\bar{x}+Q_V w; \lambda,\nu) = w^\top H(\lambda,\nu) w + 2 h(\lambda,\nu)^\top w + c_0(\lambda,\nu)$ (explicit forms in Appendix~\ref{app:orbit}, \eqref{eq:orbit_invariants}--\eqref{eq:shor_lagrangian}). Whenever $(\lambda,\nu)$ passes Cholesky factorization $H(\lambda,\nu) = R^\top R \succ 0$ on the reduced subspace, $\mathcal{L}_\beta$ is strictly convex on the affine hull of $\mathcal{P}$, so both its closed-form Helmert minimum and any Pairwise Frank--Wolfe dual lower bound over $\mathcal{P}$ (\texttt{PermRefine}) unconditionally lower-bound $\min_{x\in\mathcal{O}} Q_\beta(x)$.
\end{itemize}

\begin{theorem}[Nested Polynomial-Time Certified Outer Envelopes]
\label{thm:certifiers}
For any $(\widetilde{P}, \tilde{e}_0, \tilde{y}, z)$ with $N_{\mathrm{LB}} > 0$, intersecting each tier with the previous tier yields the unconditional nesting
\[
  \ThetaIhat
  \;\subseteq\;
  \big[\underline{\beta}_{\mathrm{SDP}},\, \overline{\beta}_{\mathrm{SDP}}\big]
  \;\subseteq\;
  \big[\underline{\beta}_{\mathrm{QP}},\, \overline{\beta}_{\mathrm{QP}}\big]
  \;\subseteq\;
  \big[\underline{\beta}_{\mathrm{LP}},\, \overline{\beta}_{\mathrm{LP}}\big],
\]
regardless of early stopping of the Frank--Wolfe or dual ascent iterations. In particular, for any dual multipliers $(\lambda,\nu)$ with $\nu_\ell \ge 0$ on non-binary cells such that $H(\lambda,\nu) \succ 0$ is verified by Cholesky factorization \citep{neumaier2004safe,rump2006verification}, and any iterate $\hat{x} \in \mathcal{P}$, both the reduced quadratic minimum $c_0(\lambda,\nu) - h(\lambda,\nu)^\top H(\lambda,\nu)^{-1} h(\lambda,\nu)$ and the Frank--Wolfe linear dual bound $\mathcal{L}_\beta(\hat{x};\lambda,\nu) + \min_{s\in\mathcal{P}} \nabla_x \mathcal{L}_\beta(\hat{x};\lambda,\nu)^\top (s - \hat{x})$ lower-bound $\min_{x\in\mathcal{O}} Q_\beta(x)$ (Appendix~\ref{app:orbit}, \eqref{eq:cholesky_cert}). Under rescaling $y \mapsto \kappa_y y$ ($\kappa_y > 0$) and $z \mapsto \kappa_z z$ ($\kappa_z > 0$), the exact relaxation intervals are \emph{scale-equivariant}, mapping $[\underline{\beta}, \overline{\beta}] \mapsto (\kappa_y/\kappa_z)[\underline{\beta}, \overline{\beta}]$ (up to finite-step bisection tolerances). Each tier runs in polynomial time in $(r_{\mathrm{r}}, r_{\mathrm{n}})$.
\end{theorem}

\begin{proof}[Proof mechanism and positioning (Appendix~\ref{app:orbit})]
The underlying optimization primitives---Birkhoff--Rado sorting \citep{birkhoff1946tres,rado1952inequality}, Frank--Wolfe duality \citep{frank1956algorithm}, and Shor Lagrangian relaxation \citep{shor1987quadratic,luo2010semidefinite}---are classical. What is new here is combining them for the graph-coupled ARD orbit problem: (i) eliminating the $K$ group-sum constraints via the explicit orthonormal Helmert basis $x = \bar{x} + Q_V w$ absorbs all $K r_{\mathrm{n}}$ row-sum RLT equalities without primal $O(r_{\mathrm{n}}^2)$ lifting; (ii) checking $H(\lambda,\nu) = R^\top R \succ 0$ directly on the reduced subspace simultaneously certifies the closed-form quadratic bound and certifies convexity of $\mathcal{L}_\beta(\cdot;\lambda,\nu)$ on $\mathcal{P}$, enabling the Pairwise Frank--Wolfe \texttt{PermRefine} linear dual lower bound to tighten the relaxation onto the permutohedron unconditionally; and (iii) the matrix-free Tier~2 extension scales directly to pooled multi-village fixed-effects regressions. When paired with multi-restart 2-opt swaps over $\mathcal{O}$ to generate constructive inner witnesses $\{\underline{\beta}_{\mathrm{in}}, \overline{\beta}_{\mathrm{in}}\} \subseteq \ThetaIhat$, the endpoint span $\overline{\beta}_{\mathrm{in}} - \underline{\beta}_{\mathrm{in}}$ lower-bounds $\operatorname{diam}(\ThetaIhat)$ (noting that $\ThetaIhat$ itself is a finite discrete set, as in Example~\ref{ex:toy}), and the outer-to-inner span ratio certifies practical tightness.
\end{proof}

\begin{remark}[Constructive Inner Witnesses vs.\ Outer Bounds]
\label{rem:inner_vs_outer}
We emphasize an elementary logical asymmetry: while a certified outer bound excluding zero ($\underline{\beta}_{\mathrm{SDP}} > 0$) proves sign identification of $\ThetaIhat$, an outer bound \emph{containing} zero does \emph{not} prove sign non-identification (it could merely reflect relaxation slack). Demonstrating that $\ThetaIhat$ straddles zero requires constructive \emph{inner feasible witnesses} $\pi_-, \pi_+ \in \PiN$ attaining $\hat{\theta}(\pi_-) \le 0 < \hat{\theta}(\pi_+)$ (or $\underline{L} \le 0 < \overline{L}$ in Theorem~\ref{thm:sign_test}). Even then, as we examine carefully in Section~\ref{sec:real_mf}, sample extremum bounds over a large permutation group $\PiN$ can widen outward due to outcome noise $\tilde{\varepsilon}$ unless checked against null-outcome placebos and out-of-sample cross-fitting.
\end{remark}

\section{Proxy Exposure Errors, Feasible Plug-In Certification, and Unobserved Homophily}
\label{sec:plugin_homophily}

Section~\ref{sec:orbit} analyzed identification given a column-profile oracle knowing $P_{\Rset,\cdot}$ up to $\PiN$. In practice, applied researchers observe neither $P$ nor the true non-respondent labeling $\pi^\star$, and must instead construct exposure proxies or plug-in probability estimates $\widehat{P}$ from $\mathcal{D}_{\mathrm{ARD}}$. This section gives a clean three-term error taxonomy for proxy exposures, develops a feasible degree-conditioned plug-in certifier under sparse degree heterogeneity, and characterizes the limits of control functions under unobserved latent homophily.

\subsection{Exact Three-Term Proxy Error Identity and Four Failure Modes}

Consider the canonical mean-hybrid exposure proxy constructed from an estimated respondent-block probability matrix $\widehat{P}_{\Rset,\Rset}$ and estimated non-respondent cell expectations $\widehat{\mu}_{ik}^{\mathrm{n}}$:
\begin{equation}
  \label{eq:mean_hyb_proxy}
  \widehat{E}_i^{\mathrm{mean\text{-}hyb}}
  \;:=\;
  \sum_{j \in \Rset \setminus \{i\}} \hat{p}_{ij} z_j
  + \sum_{k=1}^K \bar{z}_{\Nset,k}\, \widehat{\mu}_{ik}^{\mathrm{n}},
  \qquad
  \bar{z}_{\Nset,k} := \frac{1}{m_{\mathrm{n},k}} \sum_{j \in G_k^{\mathrm{n}}} z_j.
\end{equation}
Comparing $\widehat{E}_i^{\mathrm{mean\text{-}hyb}}$ against realized exposure $E_i^{\mathrm{real}} = (Az)_i$ or expected exposure $E_i^{\mathrm{exp}} = (Pz)_i$ yields an elementary but revealing algebraic decomposition:

\begin{proposition}[Three-Term Exposure Proxy Error Decomposition]
\label{prop:three_term}
For every respondent $i \in \Rset$, define the within-cell treatment--propensity covariance $\omega_{ik} := \frac{1}{m_{\mathrm{n},k}} \sum_{j \in G_k^{\mathrm{n}}} (p_{ij} - \bar{p}_{ik}^{\mathrm{n}})(z_j - \bar{z}_{\Nset,k})$, where $\bar{p}_{ik}^{\mathrm{n}} := \mu_{ik}^{\mathrm{n}} / m_{\mathrm{n},k}$. Then
\begin{align}
  \widehat{E}_i^{\mathrm{mean\text{-}hyb}} - E_i^{\mathrm{real}}
  &\;=\;
  \underbrace{\sum_{j \in \Rset\setminus\{i\}} (\hat{p}_{ij} - p_{ij}) z_j + \sum_{k=1}^K \bar{z}_{\Nset,k}(\widehat{\mu}_{ik}^{\mathrm{n}} - \mu_{ik}^{\mathrm{n}})}_{\Delta_i^{\mathrm{est}}\text{ (first-stage estimation error)}} \notag\\
  &\qquad-\;
  \underbrace{\sum_{k=1}^K m_{\mathrm{n},k}\, \omega_{ik}}_{\Delta_i^{\mathrm{swap}}\text{ (within-cell swap error)}}
  \;-\;
  \underbrace{\sum_{j \in [r]\setminus\{i\}} (A_{ij} - p_{ij}) z_j}_{\Delta_i^{\mathrm{real}}\text{ (realized-edge noise)}}.
  \label{eq:three_term}
\end{align}
For expected exposure $E_i^{\mathrm{exp}} = (Pz)_i$, the final term $\Delta_i^{\mathrm{real}}$ is identically zero.
\end{proposition}

\begin{proof}
Write $\sum_{j \in G_k^{\mathrm{n}}} p_{ij} z_j = \bar{z}_{\Nset,k} \sum_{j \in G_k^{\mathrm{n}}} p_{ij} + \sum_{j \in G_k^{\mathrm{n}}} (p_{ij} - \bar{p}_{ik}^{\mathrm{n}})(z_j - \bar{z}_{\Nset,k}) = \bar{z}_{\Nset,k} \mu_{ik}^{\mathrm{n}} + m_{\mathrm{n},k}\omega_{ik}$. Subtracting $E_i^{\mathrm{real}} = \sum_{j \in \Rset\setminus\{i\}} p_{ij} z_j + \sum_{k=1}^K \sum_{j \in G_k^{\mathrm{n}}} p_{ij} z_j + \Delta_i^{\mathrm{real}}$ from \eqref{eq:mean_hyb_proxy} yields \eqref{eq:three_term} identically.
\end{proof}

Proposition~\ref{prop:three_term} connects directly to classical versus Berkson measurement error \citep{fuller1987measurement,pagan1984econometric} and isolates four distinct failure modes (FM1--FM4) when partial-participation ARD proxies are plugged into spillover regressions:

\begin{enumerate}[label=\textbf{FM\arabic*.},leftmargin=2.8em]
  \item \textbf{Raw-count classical attenuation vs.\ Berkson orthogonality.}
  If one plugs raw observed counts $Y_{ik}^{\mathrm{n}}$ (and raw $A_{ij}$ on $\Rset\times\Rset$) into $\widehat{E}_i^{\mathrm{raw\text{-}hyb}} = \sum_{j\in\Rset\setminus\{i\}} A_{ij} z_j + \sum_k \bar{z}_{\Nset,k} Y_{ik}^{\mathrm{n}}$, then under expected exposure $y \sim E^{\mathrm{exp}}$ the count fluctuation $Y_{ik}^{\mathrm{n}} - \mu_{ik}^{\mathrm{n}}$ is \emph{classical} measurement error correlated with the proxy, attenuating OLS toward zero. By contrast, when $y \sim E^{\mathrm{real}}$, using $\widehat{E}^{\mathrm{mean\text{-}hyb}}$ renders $\Delta_i^{\mathrm{real}}$ mean-zero Berkson error orthogonal to the oracle conditional expectation $\E[E_i^{\mathrm{real}}\mid\xi, z]$, avoiding classical attenuation \emph{provided} $\Delta_i^{\mathrm{est}}$ and $\Delta_i^{\mathrm{swap}}$ are controlled.

  \item \textbf{Swap-induced Berkson variance floor under randomized treatment.}
  Under randomized treatment $z_j \stackrel{\mathrm{i.i.d.}}{\sim} \operatorname{Ber}(p_z)$ independent of $\xi$, within-cell exchangeability makes $\Delta_i^{\mathrm{swap}} = \sum_k m_{\mathrm{n},k}\omega_{ik}$ mean-zero Berkson-type error conditional on the cell-averaged proxy, so it does \emph{not} induce asymptotic OLS bias. However, conditional on the realized treatment vector $z$, the swap residual has non-vanishing magnitude whenever non-respondents exhibit within-cell link heterogeneity and the number of trait cells $K$ is bounded---inflating estimation variance by a composite residual of size $\beta_{\mathrm{spill}}(\Delta^{\mathrm{swap}} + \Delta^{\mathrm{real}})$. In single-network asymptotics, this regime ($\varphi < 1$, non-vanishing within-cell heterogeneity) is not covered by the affinity-set concentration conditions (Condition~F1 / Theorem~3.3, \texttt{arXiv:2406.11940v1} numbering) of \citet{wilkinsreeves2026single}.

  \item \textbf{Invisible within-cell treatment--activity correlation ($\omega_{ik} \neq 0$).}
  Whenever treatment is correlated with unobserved network activity or homophily inside a trait cell (for instance, village leaders having higher degree than non-leaders of the same caste and roof type), $\E[\omega_{ik}] \neq 0$ produces systematic $O(1)$ asymptotic bias in $\widehat{E}_i^{\mathrm{mean\text{-}hyb}}$ that is strictly invisible to standard $K$-group ARD by Proposition~\ref{prop:swap_invariance}.

  \item \textbf{Nonlinear exposure-mapping smearing.}
  For nonlinear exposure responses $y_i = g(z_i, E_i^{\mathrm{real}}) + \varepsilon_i$ such as threshold adoption $g(z_i, E_i) = \ind{E_i > c}$, Berkson orthogonality fails: replacing $E_i^{\mathrm{real}}$ with its conditional expectation smears the step discontinuity via Jensen's inequality (motivating mixture/EM estimators in \citealp{hardy2024estimating}---whose Section~4.2 notes non-identification for binary outcomes---and Appendix~A.2.4 of \citealp{wilkinsreeves2026single}, \texttt{arXiv:2406.11940v1}).
\end{enumerate}

\subsection{Feasible Degree-Conditioned Plug-In Certification}
\label{sec:feasible_plugin}

Because $\ThetaI$ depends on unknown link probabilities $P$, constructing feasible confidence intervals for $\theta^\star$ requires (i) estimating $P_{\Rset,\Rset}$ and the unordered non-respondent multiset $\{p_{ij} : j \in G_k^{\mathrm{n}}\}$ under the sparse degree-heterogeneous model \eqref{eq:logistic_model}, and (ii) propagating first-stage estimation error without producing misleadingly narrow or vacuous intervals.

\paragraph{Joint degree-conditioned estimation.}
Building on undirected $\beta$-model and latent-space asymptotics \citep{chatterjee2011random,yan2013central,graham2017econometric,dzemski2019empirical,ma2020universal} and the total-degree conditioning principle of \citet{tseng2026degree}, we fit \eqref{eq:logistic_model} jointly on $(A_{\Rset,\Rset}, Y_{\Rset,\cdot}^{\mathrm{n}})$ by alternating coordinate ascent over respondent activities $\hat{\alpha}_{\Rset}$, signs $\hat{\zeta}_{\Rset} \in \{\pm 1\}^{r_{\mathrm{r}}}$, homophily strength $\hat{\tau} \ge 0$, and sign-specific non-respondent cell rates $(\hat{b}_k^+, \hat{b}_k^-)$. Incorporating $Y_{\Rset,\cdot}^{\mathrm{n}}$ alongside $A_{\Rset,\Rset}$ anchors each respondent's total degree $D_i = D_i^{\mathrm{r}} + D_i^{\mathrm{n}}$ ($D_i^{\mathrm{r}} := \sum_k Y_{ik}^{\mathrm{r}}$, $D_i^{\mathrm{n}} := \sum_k Y_{ik}^{\mathrm{n}}$), lowering $P_{\Rset,\Rset}$ RMSE relative to $A_{\Rset,\Rset}$-only fits in sparse graphs. To populate the unordered non-respondent multiset $\widehat{P}_{\Rset, G_k^{\mathrm{n}}}$, we transfer the empirical respondent activity-and-sign distribution inside cell $k$ under an MCAR-spread working assumption (which we stress-test under non-MCAR shifts and misspecified kernels in Section~\ref{sec:empirical}). Given $\widehat{P}$, we compute the uninflated feasible outer envelope $[\hat{b}_{\min}, \hat{b}_{\max}] \supseteq \ThetaIhat(\widehat{P})$ and certified denominator floor $\underline{\sigma}_{\widehat{E}} := r_{\mathrm{r}}^{-1/2} \min_{x \in \mathcal{P}} \|\MZ \widehat{E}(x)\|_2$ via an exact 2D Minkowski-polygon certifier (exploiting the two-sign-class rank structure) or \texttt{FastShorSDP+PermRefine}.

\paragraph{Deterministic perturbation envelope.}
Because the transferred non-respondent multiset $\widehat{P}_{\Rset, G_k^{\mathrm{n}}}$ carries no canonical node labels, we measure estimation accuracy up to the best permutation alignment $\pi \in \PiN$. Separating deterministic algebra from probabilistic calibration yields a clean envelope valid both under Assumption~\ref{assump:causal} ($r^\star = 0$) and under general unrestricted $m$ ($r^\star \neq 0$):

\begin{lemma}[Deterministic Perturbation Envelope]
\label{lem:perturbation}
Let $\widehat{E}(\pi)$ be any estimated exposure orbit with uninflated certified outer interval $[\hat{b}_{\min}, \hat{b}_{\max}] \supseteq \ThetaIhat(\widehat{P})$ and certified denominator floor $\underline{\sigma}_{\widehat{E}} > 0$, and set $b_{\max}^{\mathrm{abs}} := \max(|\hat{b}_{\min}|, |\hat{b}_{\max}|)$. Let $r^\star := \tilde{m} - \theta^\star \widetilde{E}(\pi^\star)$ be the oracle projection residual with $\sigma_{r^\star} := \|r^\star\|_2 / \sqrt{r_{\mathrm{r}}}$ (noting $\sigma_{r^\star} = 0$ identically under Assumption~\ref{assump:causal}). Suppose there exists an aligning permutation $\tilde{\pi} \in \PiN$ such that
\[
  \frac{1}{\sqrt{r_{\mathrm{r}}}} \|\MZ(\widehat{E}(\tilde{\pi}) - E(\pi^\star))\|_2 \;\le\; \eta_{\mathrm{plug}},
  \qquad
  \varrho := \frac{\eta_{\mathrm{plug}}}{\underline{\sigma}_{\widehat{E}}} < 1,
  \qquad
  \frac{|\langle \MZ \varepsilon,\, \MZ \widehat{E}(\tilde{\pi}) \rangle|}{r_{\mathrm{r}} \underline{\sigma}_{\widehat{E}}^2} \;\le\; \mathrm{se}_{\max}.
\]
Then the true target $\theta^\star = \theta(\pi^\star)$ lies unconditionally in
\begin{equation}
  \label{eq:perturbation_iv}
  \begin{aligned}
    \widehat{\mathcal{I}}(\varrho, \mathrm{se}_{\max}, \sigma_{r^\star})
    \;:=\;
    \Bigg[
      &\hat{b}_{\min} - \frac{\varrho\, b_{\max}^{\mathrm{abs}} + \varrho(\sigma_{r^\star}/\underline{\sigma}_{\widehat{E}}) + \mathrm{se}_{\max}}{1 - \varrho},\\
      &\hat{b}_{\max} + \frac{\varrho\, b_{\max}^{\mathrm{abs}} + \varrho(\sigma_{r^\star}/\underline{\sigma}_{\widehat{E}}) + \mathrm{se}_{\max}}{1 - \varrho}
    \Bigg].
  \end{aligned}
\end{equation}
In particular, under the linear structural model (Assumption~\ref{assump:causal}), $\sigma_{r^\star} = 0$ and the radius simplifies to $(\varrho\, b_{\max}^{\mathrm{abs}} + \mathrm{se}_{\max})/(1-\varrho)$.
\end{lemma}

\paragraph{Worst-case perturbation theory vs.\ finite-sample heuristic calibration.}
To make Lemma~\ref{lem:perturbation} operational under sparse degree heterogeneity, the first-stage error $\MZ(\widehat{E}(\tilde{\pi}) - E(\pi^\star))$ decomposes into respondent activity errors $\Delta_{\mathrm{node}}$, sign-class cell-profile errors $\Delta_{\mathrm{cell}}$, and within-cell spread mismatch $\Delta_{\mathrm{spr}}$. Two finite-sample geometric refinements---documented with exact formulas, unit calibration benchmarks, and inverse-Fisher diagnostics in Appendix~\ref{app:plugin}---prevent miscalibration: (a) exact rank-1 FWL projection $\|\MZ v_s\|_2^2$ of each sign class's effective activity vector $v_s$ rather than diagonal leverage traces, and (b) Satterthwaite--Welch effective degrees of freedom $r_{\mathrm{eff}}$ \citep{satterthwaite1946approximate} to capture heavy-tailed hub kurtosis in $\|\MZ \Delta_{\mathrm{node}}\|_2^2$.

The next theorem states the formal asymptotic coverage guarantee for the worst-case perturbation ratio $\varrho_{\mathrm{WC}}$ under Assumption~\ref{assump:causal}. Alongside it, Appendix~\ref{app:plugin} specifies a tighter \emph{stochastic-EIV heuristic ratio} $\varrho_{\mathrm{stoch}}$ \eqref{eq:rho_stoch} (a simulation-calibrated slope heuristic rather than a consequence of Lemma~\ref{lem:perturbation}). Whenever either $\varrho_{\mathrm{WC}}$ or $\varrho_{\mathrm{stoch}}$ reaches the stability cutoff $\varrho \ge 0.88$, the procedure returns $(-\infty, +\infty)$ rather than an unstable bound.

\begin{theorem}[Asymptotic Coverage of Worst-Case Plug-In Certification]
\label{thm:plugin_asymp}
Consider a sequence of sparse networks generated by \eqref{eq:logistic_model} with sparsity offset $\rho_r$, bounded centered activities $\max_i |\alpha_i| \le M_\alpha$, diverging average degree $d_{\mathrm{avg}} \asymp r e^{\rho_r} \gg \log r$, fixed $K$, participation rate $\varphi \in (c_\varphi, 1)$, and MCAR non-respondent activity transfer, under Assumption~\ref{assump:causal} ($r^\star = 0$). Suppose the joint degree-conditioned MLE satisfies $\max_{i\in\Rset} |\hat{\alpha}_i - \alpha_i| = O_p(\sqrt{\log r / d_{\mathrm{avg}}})$ with asymptotically Gaussian score expansion and well-separated population denominator floor $\underline{\sigma}_{E} := r_{\mathrm{r}}^{-1/2}\min_{x\in\mathcal{P}} \|\MZ E(x)\|_2 \ge c_E > 0$. Then:
\begin{enumerate}[label=(\roman*),leftmargin=2em]
  \item Under bounded exposure-Jacobian regularity conditions (Appendix~\ref{app:plugin}), off-diagonal inverse-Fisher covariances satisfy $\Cov(\hat{\alpha}_i, \hat{\alpha}_{i'}) = O((r d_{\mathrm{avg}})^{-1})$ and the diagonal Fisher trace approximates the full inverse-Fisher quadratic-form variance up to relative error $O(r^{-1})$.
  \item Let $\varrho_{\mathrm{WC}}(\alpha^{(1)})$ be the worst-case perturbation ratio in Appendix~\ref{app:plugin} \eqref{eq:rho_wc} calibrated with one-sided upper quantile $z_{1-\alpha^{(1)}}$ via Satterthwaite--Welch / Hanson--Wright concentration \citep{satterthwaite1946approximate,rudelson2013hanson} plus the within-cell spread RMS $\sqrt{S_{\mathrm{spr}}/r_{\mathrm{r}}}/\underline{\sigma}_{\widehat{E}}$ (where $S_{\mathrm{spr}}$ is the sum of within-cell spread variances defined in Appendix~\ref{app:plugin}), and let $\mathrm{se}_{\max}(\alpha^{(2)}) := z_{1-\alpha^{(2)}/2} \hat{\sigma}_{\varepsilon}/(\sqrt{r_{\mathrm{r}}}\underline{\sigma}_{\widehat{E}})$. If $S_{\mathrm{spr}}/r_{\mathrm{r}} = o_p(1)$, then by a union bound $\widehat{\mathcal{I}}(\varrho_{\mathrm{WC}}(\alpha^{(1)}), \mathrm{se}_{\max}(\alpha^{(2)}), 0)$ achieves asymptotic pointwise coverage at least $1 - (\alpha^{(1)} + \alpha^{(2)})$ for $\beta_{\mathrm{spill}} \in \ThetaI$ while converging to the population outer relaxation envelope at rate $O_p(d_{\mathrm{avg}}^{-1/2} + r_{\mathrm{r}}^{-1/2} + \sqrt{S_{\mathrm{spr}}/r_{\mathrm{r}}})$.
\end{enumerate}
\end{theorem}

\begin{remark}[Coverage Levels, Inverse-Fisher Sign, and Five-Metric Reporting]
\label{rem:plugin_diagnostics}
In Theorem~\ref{thm:plugin_asymp}(ii), choosing $\alpha^{(1)} = \alpha^{(2)} = \alpha/2$ yields nominal level $1-\alpha$, whereas the worst-case \texttt{WC} columns in Appendix~\ref{app:tables} evaluate $\alpha^{(1)} = \alpha^{(2)} = \alpha$ (nominal union-bound level $\ge 1-2\alpha$), and the feasible \texttt{Stoch} tables in Section~\ref{sec:empirical} evaluate the heuristic \eqref{eq:rho_stoch} at nominal $\alpha = 0.05$. (Note also that requiring $d_{\mathrm{avg}} \gg \log r$ here does not collide with the bounded-degree minimax lower bound of Theorem~2 in \citealp{hayes2025minimax}, because our target uses unnormalized total treated contacts $(Az)_i$---whose variance grows with degree---rather than row-normalized averages $(D^{-1}Az)_i$.) Exact inverse-Fisher inversion at $r \in \{60, 120\}$ ($d_{\mathrm{avg}}=15$; Table~\ref{tab:block1_diag} in Appendix~\ref{app:plugin}) confirms that $88.6\%$--$98.4\%$ of off-diagonal entries $\Cov(\hat{\alpha}_i, \hat{\alpha}_{i'})$ are negative (with positive off-diagonal entries arising from degree heterogeneity and logistic hub saturation rather than any homogeneous sign threshold, increasing smoothly with $\tau$), and dropping off-diagonal entries matches the full respondent-exposure Jacobian variance within $\pm 1\%$ ($0.998\times$--$1.009\times$) rather than acting as a deterministic upper bound. In low-degree subgraphs ($d_{\Rset,\Rset} \approx 4.4$), $\varrho$ frequently exceeds $0.88$ and outputs $(-\infty,+\infty)$; reporting overall coverage \texttt{1.000} without disclosing the finite rate would be vacuous. Throughout Section~\ref{sec:empirical}, we therefore report a \emph{five-metric coverage ledger}: overall coverage, finite rate, \emph{finite-and-covering rate}, median finite width, and right- vs.\ wrong-sign zero-exclusion rates.
\end{remark}

\subsection{Unobserved Latent Homophily and Control-Function Boundaries}
\label{sec:homophily}

Assumption~\ref{assump:causal}---like the strict network-exogeneity assumption in partial-network peer-effects estimators (e.g., Assumption~4 of \citealp{boucher2025estimating})---fails when unobserved traits that drive link formation also enter outcomes directly (latent homophily; \citealp{goldsmithpinkham2013social,johnsson2021estimation,auerbach2022identification}):
\begin{equation}
  \label{eq:homophily_model}
  y_i \;=\; \beta_0 + \beta_1 z_i + \beta_{\mathrm{spill}} E_i + \rho_\alpha \alpha_i + \rho_\zeta \zeta_i + \varepsilon_i^{\mathrm{hom}},
  \qquad \E[\varepsilon_i^{\mathrm{hom}} \mid \xi, z] = 0.
\end{equation}
Under partial-participation ARD, adding observed ARD summaries $(D_i, Y_{i,\cdot})$ or estimated latent control functions $(\hat{\alpha}_i, \hat{\zeta}_i)$ to $W_{\Rset}$ faces three concrete boundaries (proofs in Appendix~\ref{app:plugin}):

\begin{proposition}[S1: Group-Balance Blindness and Raw-Count Collinearity]
\label{prop:S1}
Under \eqref{eq:logistic_model} and \eqref{eq:homophily_model} with zero-mean outcome noise:
\begin{enumerate}[label=(\roman*),leftmargin=2em]
  \item \emph{First-moment blindness at symmetric group balance:} Let $b_k := \sum_{j \in G_k} \zeta_j$ be the latent community imbalance in group $k$. Suppose $b_k = 0$ and the activity multiset $\{\alpha_j : j \in G_k, \zeta_j = +1\} = \{\alpha_j : j \in G_k, \zeta_j = -1\}$ is sign-symmetric for every group $k \in \{1,\dots,K\}$. Then for every $k$ not containing $i$ (and up to the $O(1/m_k)$ self-exclusion term on $i$'s own group), $\E[Y_{ik} \mid \alpha_i, \zeta_i] = \E[Y_{ik} \mid \alpha_i, -\zeta_i]$, so raw ARD counts carry zero first-moment signal for $\zeta_i$ and linear controls on $(D_i, Y_{i,\cdot})$ fail to absorb $\rho_\zeta \zeta_i$ (whereas nonlinear ratio controls $(\log(D_i+1), Y_{i,\cdot}/(D_i+1))$ recover $\zeta_i$ when $b_k \neq 0$).
  \item \emph{Span collinearity of raw linear ARD controls:} Controlling linearly for $(D_i, Y_{i,1},\dots,Y_{i,K-1})$ absorbs $\spanop\{Y_{i,1},\dots,Y_{i,K}\}$, which contains the pure raw-ARD exposure proxy $\sum_k \bar{z}_k Y_{ik}$ identically (up to leave-one-out self-exclusion) and severely attenuates residual exposure variation.
\end{enumerate}
\end{proposition}

\begin{proposition}[S2: Rank-2 Span Collapse vs.\ Logistic Curvature Survival]
\label{prop:S2}
Suppose link probabilities follow an exact rank-2 multiplicative kernel $p_{ij} = a_i a_j (c_0 + c_1 \zeta_i \zeta_j)$ without saturation clipping. Then for every respondent $i \in \Rset$, expected exposure $E_i^{\mathrm{exp}} = a_i \big(c_0 \sum_{j\neq i} a_j z_j + c_1 \zeta_i \sum_{j\neq i} a_j \zeta_j z_j\big)$ lies (up to the $O(1/r)$ self-exclusion term) inside $\spanop\{a_i, a_i \zeta_i\}$. Controlling for $(a_i, \zeta_i, a_i \zeta_i)$ therefore collapses residual exposure variation $\|M_{W^\star} E_{\Rset}\|_2^2 / \|\MZ E_{\Rset}\|_2^2 = O(r^{-2})$ and destroys spillover identification. Under the nonlinear logistic link \eqref{eq:logistic_model}, saturation curvature breaks exact rank-2 collinearity and preserves nonzero identifying variation, though finite-sample variance inflation remains substantial.
\end{proposition}

\begin{proposition}[S3: Non-Collinearity Bound for Noisy Controls and Control-Function EIV Overshoot]
\label{prop:S3}
Under latent homophily with noisy proxy controls:
\begin{enumerate}[label=(\roman*),leftmargin=2em]
  \item \emph{Non-collinearity requirement for noisy proxy controls:} Applying standard partitioned-regression measurement-error algebra \citep{fuller1987measurement} to a scalar latent confounder $w_i$ (with $\rho w_i$ entering $y_i$) proxied by a noisy network estimate $\hat{w}_i = w_i + \nu_i$ ($\E[\nu_i \mid E, w] = 0$, $\Var(\nu_i) = \sigma_\nu^2 = O(d_{\mathrm{avg}}^{-1})$, normalizing residualized variables to unit variance) with $r_{Ew} := \Corr(\widetilde{E}, \tilde{w})$ gives asymptotic OLS bias
  \begin{equation}
    \label{eq:S3_bias}
    \operatorname{Bias}\!\big(\hat{\beta}_{\mathrm{CF}}\big)
    \;=\;
    \frac{\rho\, r_{Ew}\, \sigma_\nu^2}{(1 - r_{Ew}^2) + \sigma_\nu^2}.
  \end{equation}
  Hence $O(d_{\mathrm{avg}}^{-1})$ bias decay requires a uniform non-collinearity margin $|r_{Ew}| \le 1 - c < 1$; as $|r_{Ew}| \to 1$, the bias converges to $\rho \operatorname{sign}(r_{Ew}) \neq 0$ for every $d_{\mathrm{avg}}$.
  \item \emph{EIV variance overshoot under control functions:} When applying $2K$ treatment-stratified EIV variance subtraction (Section~\ref{sec:remedies}, Definition~\ref{def:tstrat}) jointly with degree or latent control functions $W$, the residualized denominator $\|M_{Z,W} \widehat{E}^{\mathrm{strat}}\|_2^2$ shrinks while the conservative count variance bound $\sum_k \frac{m_{k,1}^{(-i)}}{m_{k,1}^{(-i)}-1} Y_{i,(k,1)}(1 - Y_{i,(k,1)}/m_{k,1}^{(-i)})$ ($m_{k,1}^{(-i)} := |\{j \in G_k\setminus\{i\} : z_j = 1\}|$) does not condition out $W$, causing upward overshoot unless the tighter oracle Bernoulli variance $\sum_j p_{ij}(1-p_{ij})z_j$ is known.
\end{enumerate}
\end{proposition}

\begin{remark}[Numerical Illustration of S1--S3]
\label{rem:homophily_num}
Synthetic diagnostics at $\beta_{\mathrm{spill}} = 2.0$ confirm each boundary cleanly: in unclipped rank-2 M1, residual exposure share drops to $0.00069$ and oracle $\mathrm{CF}(a^\star,\zeta^\star,a^\star\zeta^\star)$ collapses to $+0.018$, whereas in nonlinear logistic M2 curvature preserves $0.04974$ share ($\mathrm{CF}(\hat{a},\hat{\zeta},\hat{a}\hat{\zeta}) = +1.542$ vs.\ raw $\mathrm{CF}(D,Y) = +0.896$). For \eqref{eq:S3_bias}, bias decays from $0.1739$ to $0.0035$ across $d_{\mathrm{avg}} \in \{10, 40, 160, 640\}$ at $|r_{Ew}| = 0.80$, yet stays identically $1.0000$ at $|r_{Ew}| = 1.00$ (Figure~\ref{fig:budget_homophily}(b)). For S3(ii), adding oracle control functions pushes stratified EIV from $2.363$ up to $2.985$ under the conservative variance bound, compared to $2.162$ under the oracle Bernoulli variance.
\end{remark}

\section{Model-Free Remedies: Raw-ARD Errors-in-Variables and Treatment-Stratified Surveys}
\label{sec:remedies}

Parametric plug-in certification (Section~\ref{sec:feasible_plugin}) requires fitting a network model $\widehat{P}$ and transferring the non-respondent activity distribution, and its single-village intervals can become wide in sparse subgraphs. This section develops two complementary \emph{model-free} remedies: (i) finite-sample errors-in-variables (EIV) variance bracketing directly from raw ARD counts without estimating $\widehat{P}$ (complementing parametric two-draw simulated-network EIV proxies such as Appendix~D.1, Proposition~2 of \citealp{boucher2025estimating}), and (ii) minimal prospective redesigns of the ARD questionnaire---specifically a $(K{+}1)$-question augmentation and $2K$ treatment-stratified ARD ($\Tstrat$)---that eliminate non-respondent swap invariance at the data-collection stage.

\subsection{Exact Raw-ARD Variance Identities and Model-Free EIV Brackets}

Consider the two raw-ARD linear exposure proxies that require no fitted network model $\widehat{P}$:
\begin{align}
  &\begin{aligned}
    \widehat{E}_i^{\mathrm{raw\text{-}pure}}
    &\;:=\; \sum_{k=1}^K \bar{z}_{k,-i}\, Y_{ik}
    \;=\; \sum_{j \in [r]\setminus\{i\}} c_{ij}^{\mathrm{pure}} A_{ij}, \\
    &\qquad c_{ij}^{\mathrm{pure}} := \bar{z}_{k(j),-i}
    := \frac{1}{|G_{k(j)}\setminus\{i\}|}\sum_{u \in G_{k(j)}\setminus\{i\}} z_u,
  \end{aligned}
  \label{eq:raw_pure}\\[6pt]
  &\begin{aligned}
    \widehat{E}_i^{\mathrm{raw\text{-}hyb}}
    &\;:=\; \sum_{j \in \Rset\setminus\{i\}} A_{ij} z_j + \sum_{k=1}^K \bar{z}_{\Nset,k}\, Y_{ik}^{\mathrm{n}}
    \;=\; \sum_{j \in [r]\setminus\{i\}} c_{ij}^{\mathrm{hyb}} A_{ij}, \\
    &\qquad c_{ij}^{\mathrm{hyb}} :=
    \begin{cases}
      z_j & j \in \Rset\setminus\{i\},\\
      \bar{z}_{\Nset, k(j)} & j \in \Nset.
    \end{cases}
  \end{aligned}
  \label{eq:raw_hyb}
\end{align}
Notice the sharp distinction between the two: \texttt{Raw-Pure} \eqref{eq:raw_pure} uses only total ARD counts $Y_{ik}$ (smoothing \emph{both} respondent and non-respondent blocks by leave-one-out cell means $\bar{z}_{k,-i}$), whereas \texttt{Raw-Hyb} \eqref{eq:raw_hyb} keeps the exact observed respondent-block ties $A_{ij}z_j$ and smooths only the non-respondent block.

Under expected exposure $E_i^{\mathrm{exp}} = (Pz)_i$, any linear proxy $\widehat{E}_i = \sum_{j\neq i} c_{ij} A_{ij}$ has conditional mean $\mu_i^{\mathrm{raw}} := \E[\widehat{E}_i \mid \xi] = \sum_{j\neq i} c_{ij} p_{ij}$ and classical independent-edge measurement error $e_i^{\mathrm{raw}} := \widehat{E}_i - \mu_i^{\mathrm{raw}} = \sum_{j\neq i} c_{ij}(A_{ij} - p_{ij})$. Crucially, because the graph $A$ is \emph{undirected} ($A_{ij} = A_{ji}$), errors $e_i^{\mathrm{raw}}$ and $e_{i'}^{\mathrm{raw}}$ for two distinct respondents $i \neq i' \in \Rset$ share exactly one undirected edge $A_{ii'}$, producing an exact off-diagonal covariance $\Cov(e_i^{\mathrm{raw}}, e_{i'}^{\mathrm{raw}} \mid \xi) = c_{i i'} c_{i' i} p_{ii'}(1 - p_{ii'})$.

\begin{proposition}[Exact Finite-Sample Second-Moment Identity and Model-Free EIV Bracket]
\label{prop:raw_eiv}
Let $\widehat{E}_i = \sum_{j\neq i} c_{ij} A_{ij}$ be either \texttt{Raw-Pure} \eqref{eq:raw_pure} or \texttt{Raw-Hyb} \eqref{eq:raw_hyb}, with conditional mean $\mu^{\mathrm{raw}} = (\mu_i^{\mathrm{raw}})_{i\in\Rset}$ and per-node variance $\sigma_{e,i}^2 := \Var(\widehat{E}_i \mid \xi) = \sum_{j\neq i} c_{ij}^2 p_{ij}(1-p_{ij})$.
\begin{enumerate}[label=(\roman*),leftmargin=2em]
  \item \emph{Exact projected second moment:}
  \begin{equation}
    \label{eq:exact_m2}
    \E\!\Big[\|\MZ \widehat{E}\|_2^2 \,\Big|\, \xi\Big]
    \;=\;
    \|\MZ \mu^{\mathrm{raw}}\|_2^2
    \;+\; \sum_{i \in \Rset} (1 - h_{ii}) \sigma_{e,i}^2
    \;+\; \sum_{\substack{i, i' \in \Rset \\ i \neq i'}} (\MZ)_{ii'}\, c_{i i'} c_{i' i}\, p_{ii'}(1 - p_{ii'}).
  \end{equation}
  In particular, writing $\mu^{\mathrm{pure}}$ and $\mu^{\mathrm{hyb}}$ for $\mu^{\mathrm{raw}}$ under \eqref{eq:raw_pure} and \eqref{eq:raw_hyb}, replacing $\|\MZ \mu^{\mathrm{pure}}\|_2^2$ by $\|\MZ(E_{\Rset}^{\mathrm{exp}} - \Delta^{\mathrm{swap}})\|_2^2$ (which equals $\|\MZ \mu^{\mathrm{hyb}}\|_2^2$, not $\|\MZ \mu^{\mathrm{pure}}\|_2^2$) introduces an $O(1)$ error whenever respondent-block smoothing differs from $E_{i,\mathrm{base}}$.
  \item \emph{Model-free Cauchy--Schwarz variance bracket:} For any cell $C$ of size $m \ge 1$ with count $Y_C = \sum_{j\in C} A_{ij}$ and constant weight $c$, define the conservative upper estimator $\hat{v}_C^{\mathrm{up}} := c^2 \frac{m}{m-1} Y_C (1 - Y_C / m)$ for $m \ge 2$ (and $\hat{v}_C^{\mathrm{up}} := c^2 Y_C$ for $m=1$) and the lower estimator $\hat{v}_C^{\mathrm{lo}}(p_{\max}) := \min\{c^2 (1 - p_{\max}) Y_C,\; \hat{v}_C^{\mathrm{up}}\}$ where $p_{\max} \ge \max_{j \in C} p_{ij}$. Then
  \begin{equation}
    \label{eq:cs_bracket}
    \E[\hat{v}_C^{\mathrm{lo}}(p_{\max}) \mid \xi]
    \;\le\;
    \sum_{j \in C} c^2 p_{ij}(1 - p_{ij})
    \;\le\;
    \E[\hat{v}_C^{\mathrm{up}} \mid \xi].
  \end{equation}
  For $m \ge 2$, exact equality holds on the upper bound if and only if $p_{ij}$ is constant across $j \in C$; for $m=1$, $\E[c^2 Y_C \mid \xi] = c^2 p_{ij} \ge c^2 p_{ij}(1-p_{ij})$, with equality iff $p_{ij}=0$. Summing over cells (plus $\sum_{j\in\Rset\setminus\{i\}} A_{ij} z_j^2$ and $(1-p_{\max})\sum_{j\in\Rset\setminus\{i\}} A_{ij} z_j^2$ for \texttt{Raw-Hyb}) yields plug-in variance bounds $(\hat{v}_i^{\mathrm{lo}}, \hat{v}_i^{\mathrm{up}})$ and the FWL errors-in-variables \citep{fuller1987measurement} bracket pair
  \begin{equation}
    \label{eq:eiv_bracket}
    \hat{\beta}_{\mathrm{EIV}}(v)
    \;:=\;
    \frac{\langle \tilde{y},\, \MZ \widehat{E} \rangle}{\|\MZ \widehat{E}\|_2^2 - \sum_{i\in\Rset}(1-h_{ii}) v_i},
    \qquad v \in \{\hat{v}^{\mathrm{lo}},\, \hat{v}^{\mathrm{up}}\},
  \end{equation}
  defined whenever the debiased denominator exceeds a positive safety threshold. Because \eqref{eq:cs_bracket} bounds the \emph{expected} variance rather than each realized sample denominator, substituting $\hat{v}^{\mathrm{lo}}$ and $\hat{v}^{\mathrm{up}}$ into \eqref{eq:eiv_bracket} orders the ratio of expectations (and hence the large-$r_{\mathrm{r}}$ probability limit, in magnitude, when the debiased denominator is uniformly bounded away from zero and the off-diagonal sum in \eqref{eq:exact_m2} is negligible) rather than guaranteeing finite-sample realization-by-realization coverage.
\end{enumerate}
\end{proposition}

\begin{proof}
Expanding $\MZ \widehat{E} = \MZ \mu^{\mathrm{raw}} + \MZ e^{\mathrm{raw}}$ and using $\E[e^{\mathrm{raw}} \mid \xi] = 0$ gives $\E[\|\MZ \widehat{E}\|_2^2 \mid \xi] = \|\MZ \mu^{\mathrm{raw}}\|_2^2 + \tr(\MZ \Cov(e^{\mathrm{raw}} \mid \xi))$ because $\MZ^2 = \MZ$. Since $(\MZ)_{ii} = 1 - h_{ii}$ and $\Cov(e_i^{\mathrm{raw}}, e_{i'}^{\mathrm{raw}} \mid \xi) = c_{ii'} c_{i'i} p_{ii'}(1-p_{ii'})$ for $i \neq i'$, \eqref{eq:exact_m2} follows identically. For \eqref{eq:cs_bracket} with $m \ge 2$, let $\mu_C = \sum_{j\in C} p_{ij}$ and $V_C = \sum_{j\in C} p_{ij}(1-p_{ij}) = \mu_C - \sum_{j\in C} p_{ij}^2$. Using $\E[Y_C^2 \mid \xi] = \mu_C^2 + V_C$, direct algebra gives
\[
  \E\!\Big[\frac{m}{m-1} Y_C \Big(1 - \frac{Y_C}{m}\Big) \,\Big|\, \xi\Big]
  \;=\; \frac{m}{m-1}\Big(\mu_C - \frac{\mu_C^2 + V_C}{m}\Big)
  \;=\; V_C + \frac{1}{m-1}\Big(m \sum_{j\in C} p_{ij}^2 - \mu_C^2\Big) \;\ge\; V_C
\]
by Cauchy--Schwarz ($m \sum_{j\in C} p_{ij}^2 \ge \mu_C^2$, with equality iff $p_{ij}$ is constant on $C$). The $m=1$ case is immediate from $p_{ij} \ge p_{ij}(1-p_{ij})$. Meanwhile $p_{ij} \le p_{\max}$ implies $p_{ij}(1-p_{ij}) \ge (1-p_{\max})p_{ij}$, establishing the lower bound in \eqref{eq:cs_bracket}.
\end{proof}

\begin{remark}[Numerical Verification and Scope of Model-Free EIV]
\label{rem:eiv_scope}
Exact $2^{15}$-graph enumeration on $r=6$ confirms \eqref{eq:exact_m2} to machine precision ($1.44\times 10^{-15}$), and a $10{,}000$-draw Monte Carlo study ($r=48, r_{\mathrm{r}}=32$) verifies \eqref{eq:exact_m2} to within $0.13\%$ for \texttt{Raw-Pure} (exact $89.884$ vs.\ MC $89.998$; omitting respondent-block smoothing errs by $+100.2\%$) and $0.21\%$ for \texttt{Raw-Hyb} (exact $216.358$ vs.\ MC $215.894$), while confirming both sides of \eqref{eq:cs_bracket}. Three scope boundaries govern \eqref{eq:eiv_bracket}:
\begin{enumerate}[label=(\alph*),leftmargin=1.8em]
  \item \emph{Estimand and off-diagonal neglect:} Equation~\eqref{eq:eiv_bracket} removes the diagonal classical edge variance around $\mu^{\mathrm{raw}}$, so it targets the linear-projection slope on $\mu^{\mathrm{raw}}$ (which coincides with $E^{\mathrm{exp}}$ up to within-cell treatment--link smoothing error $\mu^{\mathrm{raw}} - E^{\mathrm{exp}}$) and drops the undirected off-diagonal sum in \eqref{eq:exact_m2} (which is $O(r_{\mathrm{r}}^{-1})$ relative to the diagonal term when $(\MZ)_{ii'} = O(r_{\mathrm{r}}^{-1})$, e.g., $-1.84$ out of $89.88$ for \texttt{Raw-Pure} above).
  \item \emph{Expected vs.\ realized exposure:} EIV variance subtraction targets \emph{expected} exposure $y \sim E^{\mathrm{exp}}$ where $e_i^{\mathrm{raw}}$ is classical measurement error; under \emph{realized} exposure $y \sim E^{\mathrm{real}}$, subtracting $\hat{v}_i$ over-corrects Berkson/sampling noise and inflates the slope upward (Section~\ref{sec:empirical}, Table~\ref{tab:semisynth}).
  \item \emph{SNR floor and finite-sample realizations:} In low-SNR, sparse homogeneous graphs where $\|\MZ \mu^{\mathrm{raw}}\|_2^2$ is small relative to $\sum (1-h_{ii}) \hat{v}_i$, the debiased denominator can approach zero (causing dropouts), whereas degree heterogeneity (such as real Karnataka degree dispersion) rescues SNR cleanly (zero dropouts, $0/35$--$0/36$, across all eight Karnataka expected-exposure cells). Moreover, because \eqref{eq:cs_bracket} bounds $\E[\hat{v}_i\mid\xi]$ rather than each draw of the random denominator in \eqref{eq:eiv_bracket}, finite-sample negative correlation between $\|\MZ \widehat{E}\|_2^2$ and $\hat{v}_i^{\mathrm{up}}$ can push $\hat{\beta}_{\mathrm{EIV}}(\hat{v}^{\mathrm{up}})$ below $\beta_{\mathrm{spill}}$ on individual small-$r_{\mathrm{r}}$ draws (e.g., when $r_{\mathrm{r}}=2$ with $m=2$ treated non-respondents, $\hat{v}_i^{\mathrm{up}}=Y_i(2-Y_i)$ vanishes at $Y=(0,2)$ where $\|\MZ \widehat{E}\|_2^2$ is largest). When $\beta_{\mathrm{spill}}<0$, dividing by a smaller positive denominator likewise amplifies $|\hat{\beta}_{\mathrm{EIV}}|$ in magnitude (making it more negative) rather than acting as a signed upper bound.
  \item \emph{Contrast with missing-degree multiplicative corrections:} \citet{marray2025estimating} (Propositions~3--4) proposes a model-free multiplicative correction $\hat{\beta}_{\mathrm{OLS}}/(1+\hat{\eta})$ using total missing degree $D_i^{\mathrm{n}}$ when links are sampled independently of treatment ($z \perp (G, H)$) under equal slopes across observed and missing links. Under partial-participation ARD, knowing $D_i^{\mathrm{n}}$ alone cannot correct within-cell treatment--activity covariance ($\omega_{ik} \neq 0$), dividing by $1+\hat{\eta}>1$ shifts slopes downward (the wrong direction under expected-exposure classical attenuation $y \sim Pz$), and Section~\ref{sec:real_mf} rejects equal slopes across $A_{\Rset,\Rset}$ and $A_{\Rset,\Nset}$ on Karnataka ($p = 0.0074$).
\end{enumerate}
\end{remark}

\subsection{Prospective Survey Design: \texorpdfstring{$(K{+}1)$}{(K+1)} Augmented ARD vs.\ \texorpdfstring{$2K$}{2K} Treatment-Stratified ARD}

When the researcher designs the ARD instrument prospectively---and when binary treatment status $z_j \in \{0,1\}$ is salient to respondents (for example, village leaders, program beneficiaries, or visibly treated households)---asking respondents directly about treated contacts removes the symmetry barrier of Proposition~\ref{prop:swap_invariance} at the source. Two natural designs span the trade-off between questionnaire economy and structural flexibility:

\begin{definition}[$(K{+}1)$ Augmented ARD and $2K$ Treatment-Stratified ARD ($\Tstrat$)]
\label{def:tstrat}
For binary roster treatment $z \in \{0,1\}^r$ and trait partition $\{G_1,\dots,G_K\}$:
\begin{enumerate}[label=(\alph*),leftmargin=1.8em]
  \item \emph{$(K{+}1)$ Augmented ARD:} Each respondent $i \in \Rset$ answers the $K$ standard trait-group count questions $Y_{i1},\dots,Y_{iK}$ plus \emph{one} additional count of total treated contacts,
  \[
    Y_{i,\mathrm{tr}} \;:=\; \sum_{j \in [r]\setminus\{i\}} A_{ij} z_j.
  \]
  \item \emph{$2K$ Treatment-Stratified ARD ($\Tstrat$):} Refine $\{G_1,\dots,G_K\}$ into $2K$ cells $G_{k,1} := \{j \in G_k : z_j = 1\}$ and $G_{k,0} := \{j \in G_k : z_j = 0\}$. Each respondent $i \in \Rset$ reports the $2K$ counts
  \[
    Y_{i,(k,1)} := \sum_{j \in G_{k,1}\setminus\{i\}} A_{ij},
    \qquad
    Y_{i,(k,0)} := \sum_{j \in G_{k,0}\setminus\{i\}} A_{ij},
    \qquad k = 1,\dots,K,
  \]
  yielding $\widehat{E}_i^{\mathrm{strat}} := \sum_{k=1}^K Y_{i,(k,1)}$ and disjoint untreated-neighbor counts $(Y_{i,(k,0)})_{k=1}^K$ with control-neighbor degree $D_{i,0} := \sum_{k=1}^K Y_{i,(k,0)}$.
\end{enumerate}
\end{definition}

In Appendix~B.2 (item~2, p.~A16) of \citet{boucher2025estimating}, the authors informally recommend asking respondents directly for the fraction of friends sharing a covariate so as to observe peer covariate averages ($GX$ in their notation) and reduce reliance on simulated network draws---a recommendation underscored by their Appendix~H.3 simulations (Tables~H.1--H.2), where ARD-based SGMM is accurate when peer covariate averages are observed but degrades severely when they are unobserved. Under partial participation ($\varphi < 1$), incorporating treatment counts into ARD is not merely an efficiency improvement; it resolves the exact swap non-identification barrier of Proposition~\ref{prop:swap_invariance} while introducing a concrete fixed-budget trade-off analyzed below:

\begin{proposition}[Properties of $(K{+}1)$ ARD and $2K$ $\Tstrat$]
\label{prop:tstrat}
Under binary $z \in \{0,1\}^r$ and error-free count reporting on a static graph $A$:
\begin{enumerate}[label=(\roman*),leftmargin=2em]
  \item \emph{Exact elimination of non-respondent swap error:} Both $(K{+}1)$ ARD (via $Y_{i,\mathrm{tr}}$) and $2K$ $\Tstrat$ (via $\widehat{E}_i^{\mathrm{strat}}$) observe $E_i^{\mathrm{real}}$ directly and eliminate non-respondent swap error ($\Delta_i^{\mathrm{swap}} \equiv 0$), removing both FM2 variance floors and FM3 treatment--activity bias. Under $2K$ $\Tstrat$, treatment $z_j$ is constant within every refined cell $G_{k,1}^{\mathrm{n}}$ and $G_{k,0}^{\mathrm{n}}$, so the refined-cell treatment--propensity covariance vanishes identically and the orbit collapses to a singleton by Proposition~\ref{prop:orbit_set}(iii); by contrast, $(K{+}1)$ ARD bypasses swap error directly through $Y_{i,\mathrm{tr}}$ even when the original $K$ trait groups retain nonzero $\omega_{ik}$.
  \item \emph{Exact realized exposure and expected-exposure EIV variance bracket:} On static error-free counts, $Y_{i,\mathrm{tr}} = \widehat{E}_i^{\mathrm{strat}} \equiv \sum_{j\neq i} A_{ij} z_j = E_i^{\mathrm{real}}$ holds identically (without requiring $A_{\Rset,\Rset}$). Under expected exposure $y \sim E^{\mathrm{exp}}$, $\E[E_i^{\mathrm{real}} \mid \xi] = E_i^{\mathrm{exp}}$ is unbiased. The plug-in variance estimator
  \[
    \hat{v}_i^{\mathrm{strat,up}}
    \;:=\;
    \sum_{k: m_{k,1}^{(-i)}\ge 2} \frac{m_{k,1}^{(-i)}}{m_{k,1}^{(-i)}-1} Y_{i,(k,1)}\Big(1 - \frac{Y_{i,(k,1)}}{m_{k,1}^{(-i)}}\Big)
    \;+\; \sum_{k: m_{k,1}^{(-i)}=1} Y_{i,(k,1)}
  \]
  (where $m_{k,1}^{(-i)} := |G_{k,1}\setminus\{i\}|$, or the single-stratum pooled analogue $\hat{v}_i^{(K+1),\mathrm{up}} := \frac{m_{\mathrm{tr}}^{(-i)}}{m_{\mathrm{tr}}^{(-i)}-1} Y_{i,\mathrm{tr}}(1 - Y_{i,\mathrm{tr}}/m_{\mathrm{tr}}^{(-i)})$ for $m_{\mathrm{tr}}^{(-i)} = \sum_{j\neq i} z_j \ge 2$) satisfies $\E[\hat{v}_i^{\mathrm{strat,up}}\mid\xi] \ge \Var(E_i^{\mathrm{real}}\mid\xi)$ with strict inequality whenever $p_{ij}$ varies across treated alters. Substituting $\hat{v}^{\mathrm{strat,up}}$ into \eqref{eq:eiv_bracket} defines the upper-variance EIV plug-in ratio (\texttt{Tstrat-EIV}), paired with $\hat{v}_i^{\mathrm{strat,lo}}(p_{\max}) = (1-p_{\max})E_i^{\mathrm{real}}$ as in Proposition~\ref{prop:raw_eiv}(ii); as in Remark~\ref{rem:eiv_scope}, this variance ordering guarantees an upper-magnitude endpoint in the ratio-of-expectations / large-$r_{\mathrm{r}}$ limit (under positive denominator separation and negligible off-diagonal covariance), not a finite-sample realization-by-realization covering bound.
\end{enumerate}
\end{proposition}

Notice the practical consequence of Proposition~\ref{prop:tstrat}(ii): because $\E[\hat{v}_i^{\mathrm{strat,up}}\mid\xi]$ conservatively overestimates $\Var(E_i^{\mathrm{real}} \mid \xi)$ whenever link probabilities vary across treated alters (by Cauchy--Schwarz), in high-SNR multi-respondent villages with positive $\beta_{\mathrm{spill}}$ the plug-in ratio \texttt{Tstrat-EIV} $= \hat{\beta}_{\mathrm{EIV}}(\hat{v}^{\mathrm{strat,up}})$ subtracts slightly more variance on average than is present and therefore behaves as an asymptotic upper-magnitude bracket endpoint rather than a centered point estimator (explaining why its cross-village medians sit at $2.40$--$3.13$ against true $2.0$ in Table~\ref{tab:semisynth}, while still requiring standard large-$r_{\mathrm{r}}$ concentration rather than offering finite-sample realization coverage).

\begin{remark}[Practical Field Caveats: Transmission, Response Bias, Recall Heaping, and Fatigue]
\label{rem:tstrat_caveats}
Proposition~\ref{prop:tstrat}(ii) is stated as an algebraic identity on error-free static counts, \emph{not} as a claim that real respondents report treated-contact totals without measurement error. In actual network scale-up and ARD field surveys \citep{mccormick2010how,laga2021thirty}, both $(K{+}1)$ ARD and $2K$ $\Tstrat$ (directly answering the call in \citealp{laga2021thirty} for task-targeted ``How many $X$'s do you know?'' items) face four operational constraints: (a) \emph{imperfect transmission/visibility (transmission factor below one)}---respondents may not know which contacts hold $z_j = 1$ even for non-stigmatized traits (plausible for village leaders or public program sites, implausible for private medical or financial treatments); (b) \emph{response and stigma bias}---asking directly about sensitive treatments can trigger under-reporting and forfeit the privacy protection of indirect ARD groups; (c) \emph{size-asymmetric recall and heaping}---respondents tend to enumerate small treated cells ($G_{k,1}$) carefully while under-estimating large control cells ($G_{k,0}$) and rounding counts to multiples of five; and (d) \emph{questionnaire fatigue}---where $(K{+}1)$ adds only a single item (targeting treated contacts rather than total missing degree as in \citealp{marray2025estimating}), whereas $2K$ doubles the count module.
\end{remark}

\subsection{Fixed-Budget Trade-Offs: Comparing \texorpdfstring{$K$}{K}, \texorpdfstring{$(K{+}1)$}{(K+1)}, and \texorpdfstring{$2K$}{2K} Designs}

Under a constrained total question budget $\Budget = r_{\mathrm{r}} \Qitems$ (where $\Qitems \in \{K,\, K{+}1,\, 2K\}$ is the number of ARD items per respondent), comparing the three instruments reveals a sharp three-way boundary:

\begin{proposition}[Fixed Question Budget Trade-Off ($\Budget = r_{\mathrm{r}} \Qitems$)]
\label{prop:budget}
Suppose total ARD questions are budgeted at $\Budget = r_{\mathrm{r}} \Qitems$, so $K$-group ARD surveys $r_{\mathrm{r}}^{(K)} = \lfloor \Budget/K \rfloor$ respondents, $(K{+}1)$ ARD surveys $r_{\mathrm{r}}^{(K+1)} = \lfloor \Budget/(K{+}1) \rfloor$, and $2K$ $\Tstrat$ surveys $r_{\mathrm{r}}^{(2K)} = \lfloor \Budget/(2K) \rfloor$.
\begin{enumerate}[label=(\roman*),leftmargin=2em]
  \item \emph{Realized exposure ($y \sim E^{\mathrm{real}}$) or correlated treatment ($\omega_{ik} \neq 0$):} Under correlated treatment ($\omega_{ik} \neq 0$), standard $K$-group ARD retains an $O(1)$ asymptotic \emph{bias} from $\Delta^{\mathrm{swap}}$; under randomized treatment ($\E[\omega_{ik}]=0$) with realized exposure $y \sim E^{\mathrm{real}}$, $\Delta^{\mathrm{swap}}$ acts as mean-zero Berkson error that induces an irreducible finite-sample estimation-error \emph{variance floor} under FM2. Both $(K{+}1)$ ARD and $2K$ $\Tstrat$ eliminate $\Delta^{\mathrm{swap}}$ identically and achieve RMSE $O(\sigma_\varepsilon / \sqrt{r_{\mathrm{r}}^{(\Qitems)}})$, so $(K{+}1)$ dominates $2K$ by roughly $\sqrt{2K/(K{+}1)}$ whenever a homogeneous scalar spillover slope $\beta_{\mathrm{spill}}$ is targeted without homophily-adjusted EIV.
  \item \emph{Sparse single-graph expected exposure ($y \sim E^{\mathrm{exp}}$) under randomized treatment ($\E[\omega_{ik}]=0$):} Directly applying EIV to $E_i^{\mathrm{real}}$ suffers $p_{\mathrm{treat}}^{-1}$-inflated Bernoulli count noise across treated contacts (especially when prevalence $p_{\mathrm{treat}}$ is small) on top of the reduced sample size $r_{\mathrm{r}}^{(\Qitems)}$. Standard $K$-group \texttt{Raw-Pure + EIV} therefore outperforms direct $E_i^{\mathrm{real}}$-based EIV under small $p_{\mathrm{treat}}$, though $(K{+}1)$ ARD can fall back to its own $K$ standard counts to nearly match standard $K$-group ARD.
  \item \emph{When $2K$ $\Tstrat$ strictly improves over $(K{+}1)$:} Even though $(K{+}1)$ is strictly more budget-efficient for scalar realized-exposure OLS, $2K$ $\Tstrat$ provides three structural capabilities unavailable from $(K{+}1)$: (a) \emph{disjoint control-function counts under expected exposure with latent homophily}---because $(K{+}1)$'s total counts $Y_{ik} = Y_{i,(k,1)} + Y_{i,(k,0)}$ share classical edge noise with $Y_{i,\mathrm{tr}}$, using $Y_{ik}/D_i$ as an EIV control function induces endogenous measurement-error overshoot, whereas $2K$ $\Tstrat$ constructs its control function exclusively from disjoint untreated-neighbor counts $(Y_{i,(k,0)}/D_{i,0})$; (b) \emph{group-heterogeneous spillover identification} ($\beta_1,\dots,\beta_K$ across trait groups); and (c) \emph{tighter Cauchy--Schwarz variance bounds} $\hat{v}_i^{\mathrm{strat,up}}$ than pooling all treated nodes into a single stratum.
\end{enumerate}
\end{proposition}

\begin{remark}[Per-Interview Fixed Costs]
\label{rem:fixed_cost}
Proposition~\ref{prop:budget} models cost as purely proportional to questions ($\Budget = r_{\mathrm{r}} \Qitems$), which is the most conservative benchmark against longer questionnaires. In field surveys with a substantial per-household fixed contact and listing cost $c_0$ alongside per-question marginal cost $c_1$ (total budget $\Budget = r_{\mathrm{r}}(c_0 + c_1 \Qitems)$), the sample-size reduction from $K$ to $(K{+}1)$ or $2K$ is strictly smaller than $K/\Qitems$, shifting the practical break-even further in favor of $(K{+}1)$ and $2K$ $\Tstrat$.
\end{remark}

\begin{figure}[t]
\centering
\resizebox{\textwidth}{!}{%
\begin{tikzpicture}
  \begin{axis}[
    name=axA,
    width=0.49\textwidth,
    height=6.2cm,
    ybar=0.9pt,
    bar width=4.2pt,
    ymode=log,
    log origin=infty,
    ymin=0.0028, ymax=3.0,
    ylabel={RMSE (log scale, true $\beta_{\mathrm{spill}}=2.0$)},
    title={\textbf{(a) Fixed Budget $\Budget=720$: $K$ vs.\ $(K{+}1)$ vs.\ $2K$}},
    title style={font=\small},
    label style={font=\small},
    tick label style={font=\scriptsize},
    symbolic x coords={K6-p50, K6-p15, K12-p50, K12-p15},
    xticklabels={{\shortstack{$K{=}6$\\$p{=}0.50$}}, {\shortstack{$K{=}6$\\$p{=}0.15$}}, {\shortstack{$K{=}12$\\$p{=}0.50$}}, {\shortstack{$K{=}12$\\$p{=}0.15$}}},
    xtick=data,
    enlarge x limits=0.17,
    ymajorgrids=true,
    grid style={dashed, gray!30},
    legend style={at={(0.5,-0.28)}, anchor=north, legend columns=2, font=\scriptsize, draw=black!30, column sep=3pt}
  ]
    \addplot[fill=blue!70!black, draw=blue!80!black] coordinates {(K6-p50, 0.6559) (K6-p15, 0.9805) (K12-p50, 0.6301) (K12-p15, 1.0495)};
    \addlegendentry{Std $K$-ARD ($y{\sim}Az$)}
    \addplot[fill=teal!65!white, draw=teal!80!black] coordinates {(K6-p50, 0.0041) (K6-p15, 0.0077) (K12-p50, 0.0068) (K12-p15, 0.0147)};
    \addlegendentry{$(K{+}1)$ ARD ($y{\sim}Az$)}
    \addplot[fill=cyan!55!white, draw=blue!55!black] coordinates {(K6-p50, 0.0053) (K6-p15, 0.0107) (K12-p50, 0.0097) (K12-p15, 0.0166)};
    \addlegendentry{$2K\ \Tstrat$ ($y{\sim}Az$)}
    \addplot[fill=orange!80!black, draw=orange!90!black] coordinates {(K6-p50, 0.1365) (K6-p15, 0.2982) (K12-p50, 0.1474) (K12-p15, 0.2632)};
    \addlegendentry{Std $K$+EIV ($y{\sim}Pz$)}
    \addplot[fill=magenta!65!white, draw=magenta!80!black] coordinates {(K6-p50, 0.1550) (K6-p15, 0.5652) (K12-p50, 0.2387) (K12-p15, 0.8362)};
    \addlegendentry{$(K{+}1)$ $E_{\mathrm{real}}$+EIV ($y{\sim}Pz$)}
    \addplot[fill=red!65!black, draw=red!75!black] coordinates {(K6-p50, 0.2801) (K6-p15, 0.7584) (K12-p50, 0.2672) (K12-p15, 1.7157)};
    \addlegendentry{$2K\ \Tstrat$+EIV ($y{\sim}Pz$)}
  \end{axis}

  \begin{axis}[
    name=axB,
    at={($(axA.east)+(2.05cm,0)$)},
    anchor=west,
    width=0.47\textwidth,
    height=6.2cm,
    xmode=log,
    log basis x=2,
    xtick={10,40,160,640},
    xticklabels={10,40,160,640},
    ymin=-0.03, ymax=1.08,
    xlabel={Average degree $d_{\mathrm{avg}}$ (log scale)},
    ylabel={Control-function bias ($\rho = 1.0$)},
    title={\textbf{(b) Prop.~\ref{prop:S3}(i): Sharpness of $|r_{Ew}| \le 1-c$}},
    title style={font=\small},
    label style={font=\small},
    tick label style={font=\scriptsize},
    ymajorgrids=true,
    grid style={dashed, gray!30},
    legend style={at={(0.5,-0.28)}, anchor=north, legend columns=3, font=\scriptsize, draw=black!30, column sep=4pt}
  ]
    \addplot[very thick, red!75!black, mark=*, mark size=1.8pt] coordinates {(10, 1.0000) (40, 1.0000) (160, 1.0000) (640, 1.0000)};
    \addlegendentry{$|r_{Ew}|{=}1.00$}
    \addplot[thick, orange!85!black, mark=square*, mark size=1.6pt] coordinates {(10, 0.8257) (40, 0.5512) (160, 0.2366) (640, 0.0721)};
    \addlegendentry{$|r_{Ew}|{=}0.99$}
    \addplot[thick, purple!75!black, mark=triangle*, mark size=1.8pt] coordinates {(10, 0.4810) (40, 0.1939) (160, 0.0572) (640, 0.0150)};
    \addlegendentry{$|r_{Ew}|{=}0.95$}
    \addplot[thick, blue!70!black, mark=diamond*, mark size=1.8pt] coordinates {(10, 0.1739) (40, 0.0519) (160, 0.0137) (640, 0.0035)};
    \addlegendentry{$|r_{Ew}|{=}0.80$}
    \addplot[thick, teal!75!black, mark=o, mark size=1.8pt] coordinates {(10, 0.0588) (40, 0.0161) (160, 0.0041) (640, 0.0010)};
    \addlegendentry{$|r_{Ew}|{=}0.50$}
  \end{axis}
\end{tikzpicture}}
\caption{Survey-design trade-offs and homophily non-collinearity boundary. \textbf{(a)}~Under a fixed question budget $\Budget = r_{\mathrm{r}} \Qitems = 720$ ($r=240$, $40$ trials/cell), both $(K{+}1)$ ARD and $2K$ $\Tstrat$ eliminate swap error under realized exposure $y \sim Az$ with degree-correlated treatment ($63\times$--$123\times$ RMSE reduction for $2K$; $71\times$--$160\times$ for $(K{+}1)$, which beats $2K$ by $1.13\times$--$1.55\times$). Under sparse single-graph expected exposure $y \sim Pz$, standard $K$-group ARD + EIV beats direct $E_{\mathrm{real}}$-based EIV under both $2K$ ($1.81\times$--$6.52\times$) and $(K{+}1)$ at small $p_{\mathrm{treat}}$ (though $(K{+}1)$ can fall back to its own $K$ ARD counts to achieve $0.140$--$0.308$; Appendix Table~\ref{tab:app_budget}). \textbf{(b)}~Exact analytical bias \eqref{eq:S3_bias} (matching Monte Carlo within $\le 0.004$) illustrating Proposition~\ref{prop:S3}(i): bias decays as $O(d_{\mathrm{avg}}^{-1})$ when $|r_{Ew}| \le 1-c < 1$, but remains pinned at $1.0000$ for all $d_{\mathrm{avg}}$ at collinearity $|r_{Ew}|=1$.}
\label{fig:budget_homophily}
\end{figure}
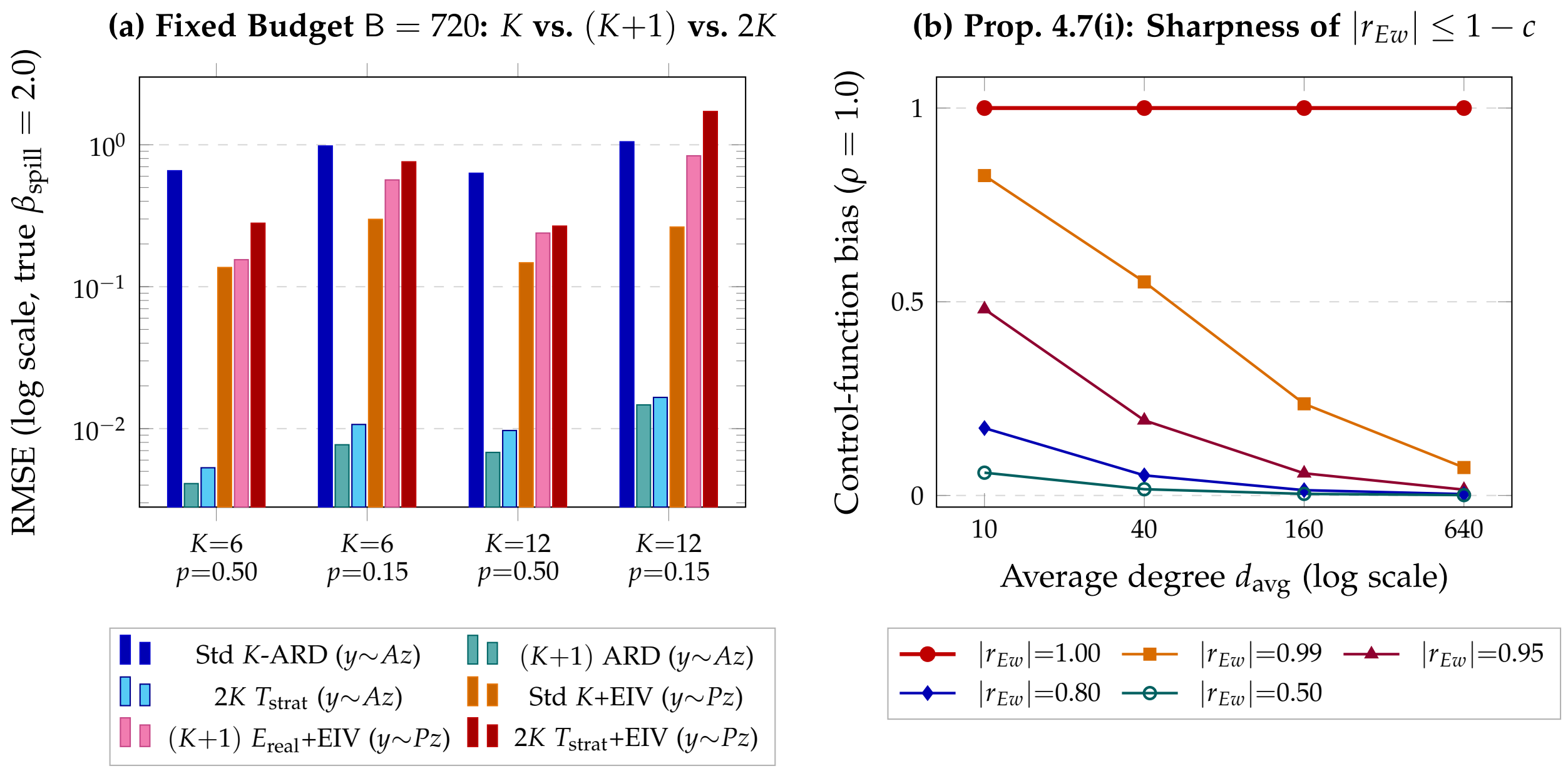

Figure~\ref{fig:budget_homophily}(a) illustrates Proposition~\ref{prop:budget} across $40$ Monte Carlo trials per cell ($r = 240$, $\Budget = 720$, true $\beta_{\mathrm{spill}} = 2.0$; full table in Appendix Table~\ref{tab:app_budget}):
\begin{itemize}[leftmargin=1.5em]
  \item Under realized exposure $y \sim Az$, $(K{+}1)$ ARD beats $2K$ $\Tstrat$ across all eight grid cells by $1.13\times$--$1.55\times$ in RMSE, while $2K$ $\Tstrat$ already cuts RMSE relative to standard $K$-ARD by $13\times$--$24\times$ under i.i.d.\ treatment and $63\times$--$123\times$ under degree-correlated treatment ($0.0053$--$0.0166$ vs.\ $0.630$--$1.050$, and $0.0041$--$0.0147$ for $(K{+}1)$).
  \item Under sparse single-graph expected exposure $y \sim Pz$, standard $K$-group ARD + EIV beats $2K$ \texttt{Tstrat-EIV} across every cell by $1.81\times$--$6.52\times$ ($0.263$ vs.\ $1.716$ at $K=12, p_{\mathrm{treat}}=0.15$) and beats direct $(K{+}1)$ $E_i^{\mathrm{real}}$-EIV by $1.90\times$--$3.18\times$ at $p_{\mathrm{treat}}=0.15$ ($0.298$ vs.\ $0.565$; $0.263$ vs.\ $0.836$), whereas falling back to $(K{+}1)$'s own $K$ ARD counts yields $0.140$--$0.308$, nearly matching standard $K$-ARD.
  \item Confirming Proposition~\ref{prop:budget}(iii), in a homophily simulation with group-clustered $z$ under $y \sim Pz$ (Appendix Table~\ref{tab:app_budget}, Panel~B), $(K{+}1)$ EIV with a total-count control function $(Y_{ik}/D_i)$ suffers endogenous edge-noise overshoot (median $3.019$ vs.\ true $2.0$), whereas $2K$ $\Tstrat$ EIV using disjoint untreated counts $(Y_{i,(k,0)}/D_{i,0})$ remains tightly centered (median $2.196$) and tightens the expected upper variance ratio from $1.096\times$ to $1.064\times$.
\end{itemize}
Figure~\ref{fig:budget_homophily}(b) pairs this design boundary with the sharp non-collinearity decay of Proposition~\ref{prop:S3}(i).

\section{Empirical Analysis on the Karnataka Microfinance Network}
\label{sec:empirical}

We evaluate our framework on the 75-village Karnataka social network and microfinance adoption dataset of \citet{banerjee2013diffusion}, a central empirical benchmark for ARD network reconstruction \citep{breza2020using,breza2023consistently}. Our empirical analysis has two parts: (i) semi-synthetic stress tests under realistic village degree distributions, non-circular ground truths, and real-outcome signal-to-noise ratios; and (ii) real microfinance adoption (\texttt{MF}), where we evaluate held-out sign ambiguity under standard ARD and diagnose a large non-mechanical reporting asymmetry in the released union network.

\subsection{Data Hygiene, Archive Audit, and Design A / Design B Protocol}
\label{sec:data_hygiene}

The public Karnataka archive (\texttt{datav4.0-rezipped.zip}, $75$ villages, $14{,}904$ census households; SHA-256 verified in Appendix~\ref{app:tables}) exhibits three structural features that any valid network validation must respect. Throughout Section~\ref{sec:empirical}, $A$ denotes the released pre-symmetrized OR-union household adjacency matrix, so observed union-graph leader exposure $E_i^{\mathrm{obs}} := (Az)_i$ can differ from true realized exposure $E_i^{\mathrm{real}}$ whenever alter-only or unrecorded ties exist:
\begin{enumerate}[label=(\arabic*),leftmargin=2em]
  \item \emph{Zero unsurveyed$\times$unsurveyed edges by construction:} Only $\approx 47\%$ of households completed the detailed individual network module, and ties in the released household union matrix $A$ exist only when reported by at least one surveyed household. Across all $75$ villages, the unsurveyed$\times$unsurveyed block contains \emph{identically zero} recorded edges. Treating the released matrix as a complete ground-truth network therefore conflates missing survey coverage with genuine absence of ties.
  \item \emph{Exhaustive audit of symmetry in the standard \texttt{datav4.0} bundle:} Inspecting all $2{,}738$ entries in \texttt{datav4.0-rezipped.zip}---including all $2{,}100$ CSV adjacency matrices ($14$ household and $14$ individual relationship types across $75$ villages) and all $4$ Stata \texttt{.dta} files---confirms that $0/2{,}100$ matrices are asymmetric and $0$ raw directed nomination-roster columns exist in this standard pre-symmetrized \texttt{datav4.0} adjacency-matrix bundle. Every adjacency matrix in \texttt{datav4.0} is pre-symmetrized (an undirected union across ego and alter nominations). Raw directed individual nomination rosters from the same $75$-village survey are separately available on Harvard Dataverse (\texttt{doi:10.7910/DVN/U3BIHX}; analyzed by \citealp{ready2021measuring} and \citealp{debacco2023latent}); our empirical analysis specifically audits the standard pre-symmetrized \texttt{datav4.0} release widely used across diffusion and ARD benchmarks, while four paired relation types (\texttt{borrowmoney}/\texttt{lendmoney}, \texttt{helpdecision}/\texttt{giveadvice}, \texttt{visitgo}/\texttt{visitcome}, \texttt{keroricego}/\texttt{keroricecome}) are released as distinct matrices within \texttt{datav4.0} and permit directional mechanism checks on the $\Rset\times\Nset$ block.
  \item \emph{Missing caste and microfinance subsample structure:} Roughly $30\%$ of census households have blank caste ($25$ villages have $>50\%$ blank and $14$ have $1\%$--$50\%$ blank, leaving $36$ zero-blank-caste villages), and $49$ villages have recorded microfinance adoption files: $21$ clean-caste MF villages, their complementary $28$ blank-caste MF villages, and the $43$-village \citet{banerjee2013diffusion} subset of those $49$ MF villages. We never collapse blank caste into ``Other''; instead we analyze \texttt{caste} $\times$ \texttt{roof} on clean-caste villages and \texttt{electricity} $\times$ \texttt{roof} across all $75$ / $49$ / $43$ / $28$ villages (excluding missing \texttt{elec=101} households and zeroing any individual self-loops).
\end{enumerate}

To avoid any $\Nset\times\Nset$ artifact, we split our semi-synthetic benchmarks into two clean designs:
\begin{itemize}[leftmargin=1.5em]
  \item \textbf{Design~A (Surveyed-Subgraph Ground Truth):} Restrict the population to the surveyed-household induced subgraph $\mathcal{S}_{\mathrm{surv}}$ (where every dyad was queried) and hold out artificial non-respondents $\Nset \subset \mathcal{S}_{\mathrm{surv}}$ under MCAR ($\varphi \in \{0.50, 0.75\}$) or degree-selective non-response (\texttt{SelDeg}).
  \item \textbf{Design~B (Native Surveyed vs.\ Unsurveyed Split, Strictly $\Rset\times[r]$):} Take $\Rset = \mathcal{S}_{\mathrm{surv}}$ and $\Nset = [r]\setminus\mathcal{S}_{\mathrm{surv}}$ directly on the full village roster, restricting all exposures and claims strictly to the observed $\Rset\times[r]$ block (zero $\Nset\times\Nset$ ties ever used). To avoid circularity when constructing expected-exposure ground truths $P^\star_{\Rset,\Nset}$ in Design~B, we benchmark \emph{three} distinct generators: (a) a baseline MCAR logistic $P^\star$; (b) \textbf{NonCircular-1}, constructed independently of our imputation routine by drawing non-MCAR shifted, right-skewed lognormal non-respondent activities ($\delta_{\Nset}=+0.40$, $1.25\times$ dispersion) with continuous Fiedler-vector homophily calibrated only to observed village $Y_{\Rset,\cdot}^{\mathrm{n}}$ cell totals; and (c) \textbf{NonCircular-2}, a misspecified full-rank 2D Euclidean latent-distance network with an asymmetric complementary log-log (\texttt{cloglog}) link and non-MCAR non-respondent shift.
\end{itemize}

\subsection{Semi-Synthetic Benchmarks: Tightness, Non-Circular Robustness, and Low-SNR Behavior}

Table~\ref{tab:semisynth} summarizes Design~A and Design~B across the $36$ clean-caste villages at high signal-to-noise ratio ($\beta_{\mathrm{spill}} = 2.0$, real \texttt{leader} indicator plus synthetic i.i.d.\ and caste-clustered treatments), supplemented by a real-MF-calibrated low-SNR benchmark ($\beta_{\mathrm{spill}} = 0.015$, $\sigma_\varepsilon = 0.35$) in Appendix Table~\ref{tab:app_low_snr}. In Table~\ref{tab:semisynth} and Appendix~\ref{app:tables}, \texttt{Raw-Hyb}, \texttt{Raw-Pure}, and \texttt{EIV-Pure} denote the proxies and Cauchy--Schwarz bracket in \eqref{eq:raw_pure}--\eqref{eq:raw_hyb} and \eqref{eq:eiv_bracket}; \texttt{Tstrat-Raw} and \texttt{Tstrat-EIV}$^{\text{up}}$ denote the $2K$ OLS slope and upper-variance EIV ratio (Proposition~\ref{prop:tstrat}); \texttt{Poly} denotes the uninflated plug-in Minkowski-polygon outer bound (Section~\ref{sec:feasible_plugin}); \texttt{Stoch} and \texttt{WC} denote the heuristic \eqref{eq:rho_stoch} and worst-case \eqref{eq:rho_wc} perturbation intervals (Lemma~\ref{lem:perturbation}, Theorem~\ref{thm:plugin_asymp}); and \texttt{FH\_Box} denotes the uncoupled per-respondent rearrangement box (Section~\ref{sec:orbit}). Four empirical takeaways emerge:

\begin{enumerate}[leftmargin=1.5em]
  \item \textbf{Oracle-$P^\star$ \texttt{FastShorSDP+PermRefine} stays near-sharp across non-logistic link kernels.}
  Across all three Design~B expected-exposure generators---including the non-MCAR Fiedler (NonCircular-1) and 2D \texttt{cloglog} (NonCircular-2) kernels---Oracle-$P^\star$ \texttt{FastShorSDP+PermRefine} achieves $1.000$ finite-and-covering rate and $1.000$ right-sign zero exclusion, staying within $1.005\times$--$1.073\times$ of constructive inner spans while tightening exact rearrangement boxes by $2.15\times$--$2.22\times$. Misspecification robustness of the feasible degree-conditioned logistic fit is evaluated in the \texttt{Stoch} and \texttt{WC} columns below.

  \item \textbf{Five-metric reporting exposes the sparse-subgraph limit of single-village plug-in bounds.}
  In Design~A at $\varphi = 0.50$ ($d_{\Rset,\Rset} \approx 4.4$), feasible \texttt{Stoch} maintains $1.000$ overall coverage but its \emph{finite-and-covering rate} drops to $0.028$--$0.306$ because plug-in uncertainty frequently trips the denominator cutoff. Once respondent degree increases (Design~A $\varphi = 0.75$ or Design~B $d_{\mathrm{full}} \approx 13.8$), \texttt{Stoch} finite-and-covering rises to $0.528$--$0.667$ with zero wrong-sign exclusions (Table~\ref{tab:semisynth}, Panel~B).

  \item \textbf{Model-free \texttt{EIV-Pure} tracks expected exposure cleanly with zero dropouts.}
  Under expected exposure $y \sim P^\star z$, \texttt{Raw-Hyb} attenuates heavily ($0.642$--$1.123$ vs.\ $2.000$, illustrating FM1), whereas \texttt{EIV-Pure} achieves zero dropouts ($0/35$--$0/36$) across every Design~A and Design~B expected-exposure cell and brackets $2.000$ closely (with \texttt{Tstrat-EIV}$^{\text{up}}$ sitting conservatively above $2.000$ by Proposition~\ref{prop:tstrat}(ii)). Under \emph{realized} exposure $y \sim Az$ on the unmodified real graph, \texttt{EIV-Pure} overshoots as expected (Berkson error), \texttt{Tstrat-Raw} recovers $2.001$, and \texttt{Raw-Hyb} shifts slightly \emph{above} $2.000$ ($2.092$).

  \item \textbf{Real-MF-calibrated low-SNR benchmark ($\beta_{\mathrm{spill}} = 0.015$, $\sigma_\varepsilon = 0.35$).}
  Because real microfinance slopes are two orders of magnitude smaller than $2.0$, Appendix Table~\ref{tab:app_low_snr} repeats the pooled $G=21$ ($n_{\Rset}=2{,}240$) and single-village benchmarks at $\beta_{\mathrm{spill}} = 0.015$ with continuous ($\sigma_\varepsilon = 0.35$) and binary ($\mathrm{SD} = 0.38$) outcomes. At $\beta_{\mathrm{spill}} = 0.015$, the exact noiseless population orbit $\ThetaI(P^\star)$ is strictly positive and narrow under both Baseline MCAR (outer span $[+0.0099, +0.0179]$) and NonCircular-1 ($[+0.0082, +0.0180]$)---there is no population sign flip. Once realistic outcome noise is added, however, pooled sample witnesses widen $2.1\times$--$3.3\times$ and straddle zero in $25\%$--$75\%$ of draws even when $\ThetaI(P^\star) > 0$, while single-village \texttt{Stoch} achieves $0.73$--$0.83$ finite-and-covering rate with median width $41.5\times$--$56.8\times$ the true slope and zero exclusions. Consequently, in-sample sign straddling on noisy real outcomes cannot stand on its own without held-out validation.
\end{enumerate}

\begin{table}[t]
\centering
\caption{Semi-synthetic Karnataka benchmarks across Design~A (surveyed subgraph $\mathcal{S}_{\mathrm{surv}}$, $36$ clean-caste villages) and Design~B (native surveyed $\Rset$ vs.\ unsurveyed $\Nset$, strictly $\Rset\times[r]$, zero $\Nset\times\Nset$ claims; true $\beta_{\mathrm{spill}}=2.0$). Panel~A reports median point estimates and model-free EIV brackets; Panel~B reports the five-metric coverage ledger for feasible plug-in certification (\texttt{Stoch}, $\alpha=0.05$) alongside uninflated \texttt{Poly} and Oracle-$P^\star$ \texttt{FastShorSDP+PermRefine} tightness.}
\label{tab:semisynth}
\scriptsize
\setlength{\tabcolsep}{4.0pt}
\resizebox{\textwidth}{!}{%
\begin{tabular}{@{}lccccc@{}}
\toprule
\multicolumn{6}{@{}l}{\textbf{Panel A: Median Point Estimators \& Model-Free EIV Brackets (true $\beta_{\mathrm{spill}} = 2.000$)}} \\
\midrule
Regime ($n_V$ villages) & Raw-Hyb & Raw-Pure & EIV-Pure $[\text{lo}, \text{hi}]$ (drops) & Tstrat-Raw & Tstrat-EIV$^{\text{up}}$ (drops) \\
\midrule
\multicolumn{6}{@{}l}{\emph{Design A ($\mathcal{S}_{\mathrm{surv}}$): Expected Exposure $y \sim Pz$ on fresh $A_{\mathrm{sim}}$}} \\
MCAR $\varphi{=}0.50$, leader ($36$) & 0.888 & 1.629 & $[1.871, 2.610]$ ($0/36$) & 0.731 & 3.129 ($5/36$) \\
MCAR $\varphi{=}0.75$, leader ($35$) & 0.769 & 1.540 & $[1.861, 2.393]$ ($0/35$) & 0.694 & 2.780 ($5/35$) \\
SelDeg $\varphi{=}0.50$, leader ($36$) & 0.642 & 1.416 & $[1.771, 2.615]$ ($0/36$) & 0.573 & 2.655 ($10/36$) \\
MCAR $\varphi{=}0.50$, i.i.d.\ $25\%$ ($36$) & 1.038 & 1.409 & $[1.621, 2.111]$ ($0/36$) & 0.857 & 2.472 ($3/36$) \\
MCAR $\varphi{=}0.50$, caste-cl.\ ($36$) & 1.123 & 1.351 & $[1.593, 2.018]$ ($0/36$) & 1.011 & 2.445 ($0/36$) \\
\addlinespace[2pt]
\multicolumn{6}{@{}l}{\emph{Design A ($\mathcal{S}_{\mathrm{surv}}$) \& Design B (Native): Realized Exposure $y \sim Az$ on unmodified real $A$}} \\
Design A MCAR $\varphi{=}0.50$, leader ($36$) & 2.069 & 2.322 & $[2.812, 3.851]$ ($0/36$) & 2.024 & 7.298 ($11/36$) \\
Design B Native $\Rset$ vs.\ $\Nset$, leader ($36$) & 2.092 & 2.746 & $[3.280, 4.510]$ ($0/36$) & 2.001 & --- \\
\addlinespace[2pt]
\multicolumn{6}{@{}l}{\emph{Design B (Native $\Rset\times[r]$, $36$ villages): Expected Exposure $y \sim P^\star z$ (Baseline vs.\ Strictly Non-Circular $P^\star$)}} \\
Baseline (MCAR logistic $P^\star$) & 0.705 & 1.515 & $[1.878, 2.421]$ ($0/36$) & --- & 2.403 ($8/36$) \\
NonCirc-1 (Non-MCAR $+0.40$ + Fiedler) & 0.896 & 1.801 & $[2.082, 2.457]$ ($0/36$) & --- & 2.582 ($2/36$) \\
NonCirc-2 (2D Eucl.\ + Cloglog + Non-MCAR) & 0.751 & 1.609 & $[1.933, 2.417]$ ($0/36$) & --- & 2.423 ($6/36$) \\
\bottomrule
\end{tabular}}

\vspace{6pt}
\resizebox{\textwidth}{!}{%
\begin{tabular}{@{}lccccccc@{}}
\toprule
\multicolumn{8}{@{}l}{\textbf{Panel B: Five-Metric Coverage Ledger for Feasible Plug-In (\texttt{Stoch}, $\alpha{=}0.05$) \& Oracle-$P^\star$ SDP Tightness}} \\
\midrule
& \multicolumn{2}{c}{Uninflated \texttt{Poly}} & \multicolumn{4}{c}{Feasible Plug-In \texttt{Stoch} ($\alpha{=}0.05$)} & Oracle-$P^\star$ SDP \\
\cmidrule(lr){2-3}\cmidrule(lr){4-7}\cmidrule(lr){8-8}
Regime (\texttt{Exp-Pz}, true $\beta_{\mathrm{spill}}=2.0$) & fin\&cov & medW & cov\_all & fin & \textbf{fin\&cov} & medW ($\times |\beta|$) & fin\&cov (SDP/In, Box/SDP) \\
\midrule
Design A: MCAR $\varphi{=}0.50$, leader & 0.972 & 3.81 & 1.000 & 0.028 & \textbf{0.028} & 26.50 ($13.2\times$) & --- \\
Design A: MCAR $\varphi{=}0.75$, leader & 0.457 & 0.95 & 1.000 & 0.657 & \textbf{0.657} & 10.16 ($5.1\times$) & --- \\
Design A: SelDeg $\varphi{=}0.50$, leader & 0.639 & 1.46 & 1.000 & 0.250 & \textbf{0.250} & 15.75 ($7.9\times$) & --- \\
Design A: MCAR $\varphi{=}0.50$, i.i.d.\ $25\%$ & 0.917 & 3.01 & 1.000 & 0.028 & \textbf{0.028} & 38.05 ($19.0\times$) & --- \\
Design A: MCAR $\varphi{=}0.50$, caste-cl. & 0.861 & 1.84 & 1.000 & 0.306 & \textbf{0.306} & \phantom{0}9.34 ($4.7\times$) & --- \\
\addlinespace[2pt]
Design B: Baseline (MCAR logistic) & 0.639 & 1.44 & 1.000 & 0.611 & \textbf{0.611} & 13.79 ($6.9\times$) & 1.000 ($1.005\times$, $2.20\times$) \\
Design B: NonCirc-1 (Non-MCAR+Fiedler) & 0.833 & 1.55 & 1.000 & 0.667 & \textbf{0.667} & \phantom{0}9.55 ($4.8\times$) & 1.000 ($1.029\times$, $2.15\times$) \\
Design B: NonCirc-2 (2D+Cloglog) & 0.583 & 1.25 & 1.000 & 0.528 & \textbf{0.528} & 12.48 ($6.2\times$) & 1.000 ($1.073\times$, $2.22\times$) \\
\bottomrule
\end{tabular}}
\begin{flushleft}
\scriptsize\emph{Notes:} In Panel~A, \texttt{Tstrat-EIV}$^{\text{up}}$ is the upper-variance Cauchy--Schwarz EIV ratio (an asymptotic ratio-of-expectations upper-magnitude endpoint, Proposition~\ref{prop:tstrat}(ii) and Remark~\ref{rem:eiv_scope}), explaining why its cross-village medians sit above $2.0$. In Panel~B, single-village \texttt{Stoch} has $0.000$ right-sign and wrong-sign zero-exclusion rates across all cells above, whereas Oracle-$P^\star$ \texttt{FastShorSDP+PermRefine} achieves $1.000$ right-sign exclusion and $0.000$ wrong-sign exclusion across all three Design~B modes. Complete \texttt{WC}, \texttt{FH\_Box}, low-SNR ($\beta=0.015$), and synthetic $10\times$-hub ledgers appear in Appendix~\ref{app:tables}.
\end{flushleft}
\end{table}

\subsection{Real Microfinance Adoption: Two Core Empirical Findings}
\label{sec:real_mf}

We now examine real microfinance adoption (\texttt{MF}) regressed on household leader exposure $E_i^{\mathrm{obs}} = \sum_{j\neq i} A_{ij} z_j$ across three pooled village-fixed-effects specifications: Spec~1 (village FEs + own \texttt{leader}), Spec~2 (adding caste/electricity, roof type, and rooms), and Spec~3 (adding control-neighbor degree $\log(D_{i,0}+1)$). Because \texttt{leader} status is not randomly assigned, all slopes are interpreted strictly as associational linear-projection targets $\theta^\star$ rather than causal parameters. Standard errors use village-clustered CR1 variance with small-cluster Student-$t_{G-1}$ reference distributions, verified against null-imposed (WCR) Rademacher wild cluster bootstrap $p$-values ($B = 4{,}999$, \citealp{cameron2008bootstrap}).

The two findings below answer distinct empirical questions on the same regressions: Finding~1 asks whether non-respondent relabeling under a fitted link model leaves the sign of the leader-spillover slope ambiguous even out of sample; Finding~2 asks why the actual hybrid-ARD estimate sits well below the released union-graph benchmark.

\subsubsection{Finding 1: Held-Out Sign Ambiguity Under the Fitted Link Model}

Because our low-SNR benchmark shows that searching over the pooled village permutation group $\prod_{v=1}^{21} \Pi_{\Nset,v}$ on noisy binary outcomes can produce sign straddling in sample even when $\ThetaI(P^\star) > 0$, we lead with an out-of-sample cross-fitting design under the fitted degree-conditioned logistic model $\widehat{P}$ (holding the transferred non-respondent activity distribution fixed). Focusing on Spec~1 across the $21$ clean-caste MF villages ($n_{\Rset} = 2{,}240$; complete diagnostics in Table~\ref{tab:karnataka_bounds} and Appendix Table~\ref{tab:app_finding1_diag}):
\begin{enumerate}[label=(\alph*),leftmargin=1.8em]
  \item \emph{Within-village 50/50 held-out cross-fitting:} We stratify respondents $50/50$ within each village, select the minimizing and maximizing permutations $(\hat{\pi}_{\min}, \hat{\pi}_{\max})$ strictly on Fold~A outcomes, evaluate the resulting OLS slopes on held-out Fold~B outcomes (and symmetrically $B \to A$), and average across the two fold directions over $10$ random splits. On permuted null outcomes, the mean held-out slopes sit near zero ($+0.00078$ at the training minimizer and $-0.00086$ at the training maximizer), consistent with negligible held-out extremum bias. On real \texttt{MF}, the training minimizer and maximizer yield \textbf{opposite-signed held-out slopes across all $10/10$ random splits}---a check of split stability on the same $21$ villages rather than ten independent replications---averaging \textbf{$-0.00894$} (range $[-0.01256, -0.00436]$) and \textbf{$+0.01981$} (range $[+0.01622, +0.02479]$; mean separation $+0.02875$, range $+0.02229$--$+0.03735$; median $-0.00864$ and $+0.01934$; deterministic split $-0.00195$ and $+0.01449$). Individual half-sample slopes retain substantial village-clustered sampling uncertainty (average half-sample $\mathrm{SE}_{\mathrm{CR1}} = 0.0130$ at $\hat{\pi}_{\min}$ and $0.0122$ at $\hat{\pi}_{\max}$). Splitting villages ($10$ vs.\ $11$) yields opposite-signed in-sample witnesses in both sub-pools, whereas fitting a single shared cross-village rank-2 SVD alignment direction on training villages does not transfer sign reversal to disjoint held-out villages (held-out slopes $+0.00570$ at the training minimizer and $+0.00189$ at the training maximizer), indicating that village-specific non-respondent alignment drives the cross-fit contrast.
  \item \emph{In-sample feasible witnesses, outer certification, and placebo width benchmark:} Across all six pooled specifications on both the $21$ clean-caste villages and all $49$ MF villages (Table~\ref{tab:karnataka_bounds}, Panel~A; Figure~\ref{fig:karnataka}(a)), constructive 2-opt search finds explicit feasible permutations witnessing opposite-signed slopes---for example, $\{-0.01809, +0.03018\} \subseteq \ThetaIhat$ in 21-village Spec~1 (certified outer span $[-0.01809, +0.03020]$, $1.0003\times$ of the inner span and $7.54\times$ narrower than the exact village-FE rearrangement box; full specifications in Table~\ref{tab:karnataka_bounds}). As a width benchmark, randomly permuting \texttt{MF} within villages ($20$ unstratified and $8$ leader-stratified draws, true null slope $\beta_{\mathrm{pop}}\equiv 0$) yields median outer widths of $0.02571$ and $0.02436$---roughly half ($50.4\%$--$53.2\%$) of the observed $0.04829$ width---underscoring why in-sample width alone overstates structural ambiguity without the held-out check in~(a).
  \item \emph{Within-cell heterogeneity shrinkage (in-sample vs.\ out-of-sample):} Shrinking fitted within-cell heterogeneity continuously toward its cell-mean matrix $\bar{P}_{\Rset,\Nset}$ via $P_{\Rset,\Nset}(\lambda) = (1-\lambda)\bar{P}_{\Rset,\Nset} + \lambda \widehat{P}_{\Rset,\Nset}$ ($\lambda \in [0,1]$) collapses $\ThetaIhat$ to the point $+0.00512$ at $\lambda = 0$ (Proposition~\ref{prop:orbit_set}(iii)). In sample, the minimum numerator hits zero at $\lambda^\star_{\mathrm{in}} = 18.9\%$ (with strictly negative feasible slopes appearing above $18.9\%$); however, in-sample null placebos that cross zero at $\lambda=1$ themselves flip at median $\lambda = 25.5\%$ (unstratified) and $39.6\%$ (leader-stratified) because their $\lambda=0$ numerator is centered near zero. Evaluating the same linear shrinkage path \emph{out of sample} across the ten 50/50 splits shifts the breakdown threshold to \textbf{median $\lambda^\star_{\mathrm{OOS}} = 32.1\%$} (mean $33.6\%$, range $25.4\%$--$49.9\%$, with $10/10$ splits crossing zero below $50\%$; deterministic split $59.3\%$). Thus roughly one-third of fitted within-cell probability heterogeneity suffices for held-out sign reversal.
\end{enumerate}

Panel~B of Table~\ref{tab:karnataka_bounds} reports the separate single-village diagnostic comparing expected-exposure bounds against village-level released union-graph OLS $\hat{\beta}_{\mathrm{obs}}$. Because this comparison mixes realized-versus-expected exposure noise with $\widehat{P}$ estimation error, uninflated single-village \texttt{FastShorSDP} commits $5/21$ ($23.8\%$) wrong-sign exclusions in $G=21$, while inflated single-village \texttt{Stoch} removes false exclusions at the cost of $57.1\%$ finite rate ($49.0\%$ in $G=49$) and wide intervals ($50.3\times$--$76.3\times |\hat{\beta}_{\mathrm{obs}}|$; worst-case \texttt{WC} reaches $84.6\times$ with $38.1\%$ finite rate in $G=21$).

\begin{table}[t]
\centering
\caption{Model-implied partial-identification bounds and held-out cross-fit slopes on real Karnataka microfinance adoption (\texttt{MF}) under the fitted degree-conditioned logistic model $\widehat{P}$. Panel~A reports pooled village-fixed-effects endpoint witnesses and certified outer envelopes across all three specifications on both the $21$ clean-caste villages and all $49$ microfinance villages (plus the within-village 50/50 held-out cross-fit for Spec~1). Panel~B reports the single-village five-metric coverage ledger benchmarked against village-level released union-graph OLS $\hat{\beta}_{\mathrm{obs}}$.}
\label{tab:karnataka_bounds}
\scriptsize
\setlength{\tabcolsep}{3.8pt}
\resizebox{\textwidth}{!}{%
\begin{tabular}{@{}lccccc@{}}
\toprule
\multicolumn{6}{@{}l}{\textbf{Panel A: Pooled Real Karnataka \texttt{MF} — Feasible Endpoint Witnesses, Held-Out Slopes \& Certified Outer Envelopes (Fitted $\widehat{P}$)}} \\
\midrule
Sample \& Specification & Constructive 2-opt Witnesses & Opposite Signs? & Certified Pooled Perm-QP & Outer/Inner & Exact VilFE Box (Box/Outer) \\
\midrule
\multicolumn{6}{@{}l}{\emph{21 Clean-Caste MF Villages (\texttt{caste} $\times$ \texttt{roof}, $G=21$, $n_{\Rset}=2{,}240$)}} \\
Spec~1: VilFE + leader & $\{-0.01809,\; +0.03018\}$ & \textbf{Yes} & $[-0.01809,\; +0.03020]$ & $1.0003\times$ & $[-0.16765,\; +0.19652]$ ($7.54\times$) \\
\quad\emph{Spec~1 Held-Out 50/50 Cross-Fit (10 splits)} & $-0.00894$ / $+0.01981$ & \textbf{Yes} ($10/10$) & \multicolumn{3}{l}{\emph{Placebo held-out: $+0.00078$ / $-0.00086$; \; OOS $\lambda^\star_{\mathrm{OOS}} = 32.1\%$ (in-samp.\ $18.9\%$)}} \\
Spec~2: + demographics & $\{-0.01022,\; +0.03717\}$ & \textbf{Yes} & $[-0.01022,\; +0.03718]$ & $1.0001\times$ & $[-0.16724,\; +0.22026]$ ($8.18\times$) \\
Spec~3: + $\log(D_0{+}1)$ & $\{-0.05416,\; +0.06562\}$ & \textbf{Yes} & $[-0.05418,\; +0.06567]$ & $1.0006\times$ & $[-0.60550,\; +0.59687]$ ($10.03\times$) \\
\addlinespace[3pt]
\multicolumn{6}{@{}l}{\emph{ALL 49 MF Villages (\texttt{elec} $\times$ \texttt{roof}, $G=49$, $n_{\Rset}=4{,}891$)}} \\
Spec~1: VilFE + leader & $\{-0.02702,\; +0.03031\}$ & \textbf{Yes} & $[-0.02702,\; +0.03031]$ & $1.0001\times$ & $[-0.20804,\; +0.21380]$ ($7.36\times$) \\
Spec~2: + demographics & $\{-0.02110,\; +0.03656\}$ & \textbf{Yes} & $[-0.02111,\; +0.03657]$ & $1.0001\times$ & $[-0.20961,\; +0.23303]$ ($7.67\times$) \\
Spec~3: + $\log(D_0{+}1)$ & $\{-0.07949,\; +0.06582\}$ & \textbf{Yes} & $[-0.07966,\; +0.06602]$ & $1.0025\times$ & $[-0.70763,\; +0.63794]$ ($9.24\times$) \\
\bottomrule
\end{tabular}}

\vspace{6pt}
\begin{tabular}{@{}lcccccc@{}}
\toprule
\multicolumn{7}{@{}l}{\textbf{Panel B: Single-Village Real \texttt{MF} Five-Metric Coverage Ledger (Benchmarked Against Released Union-Graph $\hat{\beta}_{\mathrm{obs}}$)}} \\
\midrule
Method (Sample) & cov\_all & fin & \textbf{fin\&cov} & medW ($\times |\hat{\beta}_{\mathrm{obs}}|$) & ex0 Right & ex0 Wrong \\
\midrule
Uninflated \texttt{FastShorSDP} ($G{=}21$) & 0.476 & 1.000 & \textbf{0.476} & 0.057 ($1.78\times$) & 0.429 & \textbf{0.238} ($5/21$) \\
Uninflated Rearr.\ Box ($G{=}21$) & 1.000 & 1.000 & \textbf{1.000} & 0.363 ($17.1\times$) & 0.000 & 0.000 \\
Feasible \texttt{Stoch} ($G{=}21$) & 1.000 & 0.571 & \textbf{0.571} & 1.092 ($50.3\times$) & 0.000 & 0.000 \\
Worst-case \texttt{WC} ($G{=}21$) & 1.000 & 0.381 & \textbf{0.381} & 0.915 ($84.6\times$) & 0.000 & 0.000 \\
\addlinespace[2pt]
Uninflated \texttt{FastShorSDP} ($G{=}49$) & 0.878 & 1.000 & \textbf{0.878} & 0.118 ($5.72\times$) & 0.163 & 0.020 ($1/49$) \\
Feasible \texttt{Stoch} ($G{=}49$) & 1.000 & 0.490 & \textbf{0.490} & 0.942 ($76.3\times$) & 0.000 & 0.000 \\
\bottomrule
\end{tabular}
\begin{flushleft}
\scriptsize\emph{Notes:} In Panel~A, each constructive inner row reports an explicit pair of attainable slopes $\{\underline{\beta}_{\mathrm{in}}, \overline{\beta}_{\mathrm{in}}\} \subseteq \ThetaIhat$ witnessed by feasible permutations $\pi_-, \pi_+ \in \PiN$ ($14$ multi-start 2-opt restarts per direction) under the fitted degree-conditioned logistic model $\widehat{P}$; the discrete orbit $\ThetaIhat$ lies inside the certified outer envelope $[\underline{\beta}_{\mathrm{out}}, \overline{\beta}_{\mathrm{out}}]$, and \emph{Outer/Inner} compares envelope width to the inner witness span $\overline{\beta}_{\mathrm{in}} - \underline{\beta}_{\mathrm{in}}$. The held-out cross-fit row reports mean held-out slopes evaluated at the training minimizer / training maximizer (full placebo, cross-fit, and shrinkage diagnostics appear in Appendix Table~\ref{tab:app_finding1_diag}). Panel~B benchmarks expected-exposure single-village bounds against village-level released union-graph OLS $\hat{\beta}_{\mathrm{obs}}$ (mixing realization noise with $\widehat{P}$ estimation error).
\end{flushleft}
\end{table}

\subsubsection{Finding 2: Not Mechanical ARD Attenuation---Asymmetric Reporting in Partially Sampled Union Graphs}

In 21-village Spec~1, the released union-graph benchmark slope is \textbf{$+0.0141$} (Table~\ref{tab:karnataka_reporting}, $p_{t_{20}} = 0.015$, $p_{\mathrm{WCR}} = 0.0114$), mechanical ARD scaling predicts a slight \emph{upward} shift to \textbf{$+0.0147$}, yet the actual \texttt{Raw-Hyb} ARD slope drops to \textbf{$+0.0057$}---a $59\%$ reduction (to $0.41\times$; paired difference $+0.0084$, $p_{\mathrm{WCR}} = 0.0086$). Mechanical scaling cannot even explain the direction of the shift.

To see why, recall that in Design~B realized exposure $y \sim Az$ on the exact same graphs and \texttt{leader} vector (Table~\ref{tab:semisynth}, Panel~A), \texttt{Raw-Hyb} shifts upward to $2.092$ ($+4.6\%$ above $2.000$). Letting $e := \tilde{y} - \hat{\beta}_{\mathrm{obs}} \widetilde{E}^{\mathrm{obs}}$ be the benchmark residual ($\langle e, \widetilde{E}^{\mathrm{obs}} \rangle = 0$) and $\widetilde{E}^{\mathrm{RH}} := \MZ \widehat{E}^{\mathrm{raw\text{-}hyb}}$, substituting $\tilde{y} = \hat{\beta}_{\mathrm{obs}} \widetilde{E}^{\mathrm{obs}} + e$ into \texttt{Raw-Hyb} OLS gives the exact identity:
\begin{equation}
  \label{eq:kappa_decomp}
  \hat{\beta}_{\mathrm{RawHyb}}
  \;=\;
  \hat{\beta}_{\mathrm{obs}} \cdot \underbrace{\frac{\langle \widetilde{E}^{\mathrm{obs}},\, \widetilde{E}^{\mathrm{RH}} \rangle}{\|\widetilde{E}^{\mathrm{RH}}\|_2^2}}_{\kappa\text{ (mechanical scaling)}}
  \;+\;
  \underbrace{\frac{\langle e,\, \widetilde{E}^{\mathrm{RH}} \rangle}{\|\widetilde{E}^{\mathrm{RH}}\|_2^2}}_{\Delta_{\mathrm{res}}\text{ (residual correlation)}}.
\end{equation}

\begin{figure}[t]
\centering
\resizebox{\textwidth}{!}{%
\begin{tikzpicture}
  \begin{axis}[
    name=axK1,
    width=0.47\textwidth,
    height=6.3cm,
    xmin=-0.24, xmax=0.26,
    ymin=0.3, ymax=6.7,
    ytick={1,2,3,4,5,6},
    yticklabels={
      {49 MF: S3}, {49 MF: S2}, {49 MF: S1},
      {21 Clean: S3}, {21 Clean: S2}, {21 Clean: S1}
    },
    xlabel={Spillover slope $\theta$ (pooled village-FE)},
    title={\textbf{(a) Pooled \texttt{MF} Bounds: Inner vs.\ Outer vs.\ Box}},
    title style={font=\small},
    label style={font=\small},
    tick label style={font=\scriptsize},
    xmajorgrids=true,
    grid style={dashed, gray!30},
    legend style={at={(0.5,-0.25)}, anchor=north, legend columns=2, font=\scriptsize, draw=black!30, column sep=4pt}
  ]
    \addplot[black!60, dashed, thick, forget plot] coordinates {(0, 0.3) (0, 6.7)};

    \addplot[gray!55, line width=3.5pt] coordinates {(-0.16765, 6) (0.19652, 6)};
    \addlegendentry{Exact VilFE Box}
    \addplot[gray!55, line width=3.5pt, forget plot] coordinates {(-0.16724, 5) (0.22026, 5)};
    \addplot[gray!55, line width=3.5pt, forget plot] coordinates {(-0.235, 4) (0.255, 4)};
    \addplot[gray!55, line width=3.5pt, forget plot] coordinates {(-0.20804, 3) (0.21380, 3)};
    \addplot[gray!55, line width=3.5pt, forget plot] coordinates {(-0.20961, 2) (0.23303, 2)};
    \addplot[gray!55, line width=3.5pt, forget plot] coordinates {(-0.235, 1) (0.255, 1)};

    \addplot[blue!75!black, line width=4.2pt] coordinates {(-0.01809, 6) (0.03020, 6)};
    \addlegendentry{Perm-QP Outer}
    \addplot[blue!75!black, line width=4.2pt, forget plot] coordinates {(-0.01022, 5) (0.03718, 5)};
    \addplot[blue!75!black, line width=4.2pt, forget plot] coordinates {(-0.05418, 4) (0.06567, 4)};
    \addplot[blue!75!black, line width=4.2pt, forget plot] coordinates {(-0.02702, 3) (0.03031, 3)};
    \addplot[blue!75!black, line width=4.2pt, forget plot] coordinates {(-0.02111, 2) (0.03657, 2)};
    \addplot[blue!75!black, line width=4.2pt, forget plot] coordinates {(-0.07966, 1) (0.06602, 1)};

    \addplot[only marks, mark=*, mark size=1.8pt, red!80!black] coordinates {
      (-0.01809, 6) (0.03018, 6)
      (-0.01022, 5) (0.03717, 5)
      (-0.05416, 4) (0.06562, 4)
      (-0.02702, 3) (0.03031, 3)
      (-0.02110, 2) (0.03656, 2)
      (-0.07949, 1) (0.06582, 1)
    };
    \addlegendentry{2-opt Inner}

    \addplot[only marks, mark=diamond*, mark size=2.4pt, black] coordinates {
      (0.0141, 6) (0.0179, 5) (0.0166, 4)
      (0.0031, 3) (0.0070, 2) (0.0049, 1)
    };
    \addlegendentry{Observed $\hat{\beta}_{\mathrm{obs}}$}
  \end{axis}

  \begin{axis}[
    name=axK2,
    at={($(axK1.east)+(2.15cm,0)$)},
    anchor=west,
    width=0.47\textwidth,
    height=6.3cm,
    ymin=-0.020, ymax=0.076,
    xmin=0.4, xmax=5.6,
    scaled y ticks=false,
    ytick={-0.02, 0.00, 0.02, 0.04, 0.06},
    yticklabels={$-0.02$, $0.00$, $0.02$, $0.04$, $0.06$},
    xtick={1,2,3,4,5},
    xticklabels={
      {$\hat{\beta}_{\mathrm{obs}}$},
      {$\hat{\beta}_{\mathrm{obs}}\kappa$},
      {$\hat{\beta}_{\mathrm{RH}}$},
      {$\hat{\beta}_{\Rset\Rset}$},
      {$\hat{\beta}_{\Rset\Nset}$}
    },
    ylabel={Estimated slope (Spec~1)},
    title={\textbf{(b) Not Attenuation: $\hat{\beta}_{\Rset\Nset} \gg \hat{\beta}_{\Rset\Rset}$}},
    title style={font=\small},
    label style={font=\small},
    tick label style={font=\scriptsize},
    ymajorgrids=true,
    grid style={dashed, gray!30},
    legend style={at={(0.5,-0.25)}, anchor=north, legend columns=1, font=\scriptsize, draw=black!30, cells={anchor=west}, column sep=4pt}
  ]
    \addplot[black!55, dashed, thick, forget plot] coordinates {(0.4, 0) (5.6, 0)};

    \addplot[
      only marks, mark=*, mark size=2.3pt, blue!75!black,
      error bars/.cd, y dir=both, y explicit, error bar style={thick, blue!75!black}
    ] coordinates {
      (0.88, 0.0141) +- (0, 0.0110)
      (1.88, 0.0147) +- (0, 0.0000)
      (2.88, 0.0057) +- (0, 0.0124)
      (3.88, 0.0033) +- (0, 0.0128)
      (4.88, 0.0384) +- (0, 0.0204)
    };
    \addlegendentry{21 Clean-Caste ($t_{20}$ 95\% CI)}

    \addplot[
      only marks, mark=square*, mark size=2.1pt, red!75!black,
      error bars/.cd, y dir=both, y explicit, error bar style={thick, red!75!black}
    ] coordinates {
      (1.12, 0.0031) +- (0, 0.0075)
      (2.12, 0.0032) +- (0, 0.0000)
      (3.12, -0.0017) +- (0, 0.0083)
      (4.12, -0.0054) +- (0, 0.0083)
      (5.12, 0.0222) +- (0, 0.0154)
    };
    \addlegendentry{All 49 MF ($t_{48}$ 95\% CI)}

    \node[font=\scriptsize, anchor=north west, align=left, text=blue!70!black, fill=white, fill opacity=0.85, text opacity=1, inner sep=2pt] at (axis cs:0.55, 0.074)
      {$\kappa=1.041>1$ predicts slight \emph{upward} shift;\\$\hat{\beta}_{\Rset\Nset}-\hat{\beta}_{\Rset\Rset}=+0.0351\ (p_{\mathrm{WCR}}=0.0074)$};
  \end{axis}
\end{tikzpicture}}
\caption{Two empirical findings on real Karnataka microfinance adoption (\texttt{MF}). \textbf{(a)}~Under the fitted degree-conditioned logistic model $\widehat{P}$, constructive 2-opt feasible permutations (red dots) witness opposite-signed slopes across all three pooled specifications on both the $21$ clean-caste villages and all $49$ MF villages, with Certified Pooled Perm-QP outer envelopes (blue bars) matching the inner endpoint span within $1.0001\times$--$1.0025\times$ ($7.36\times$--$10.03\times$ tighter than exact village-FE rearrangement boxes in gray, whose S3 endpoints extend to $\pm 0.60$--$0.71$). For 21 Clean S1, 50/50 held-out cross-fitting yields opposite-signed held-out slopes in $10/10$ splits (mean $-0.0089$ at the training minimizer vs.\ $+0.0198$ at the training maximizer). \textbf{(b)}~Exact decomposition of the Spec~1 drop from the released union-graph benchmark $\hat{\beta}_{\mathrm{obs}}$ ($+0.0141$) to $\hat{\beta}_{\mathrm{RH}}$ ($+0.0057$): mechanical ARD scaling $\hat{\beta}_{\mathrm{obs}}\kappa$ ($+0.0147$, plotted as a deterministic algebraic scaling of $\hat{\beta}_{\mathrm{obs}}$ without separate error bars) shifts slightly \emph{upward} ($\kappa=1.041$--$1.046$), while the joint regression reveals a large gap between ego-nominated non-respondent ties ($\hat{\beta}_{\Rset\Nset}$) and union-sampled respondent ties ($\hat{\beta}_{\Rset\Rset}$). Error bars show CR1 small-cluster $t_{G-1}$ 95\% CIs ($t_{20,0.975}=2.086$, $t_{48,0.975}=2.011$).}
\label{fig:karnataka}
\end{figure}
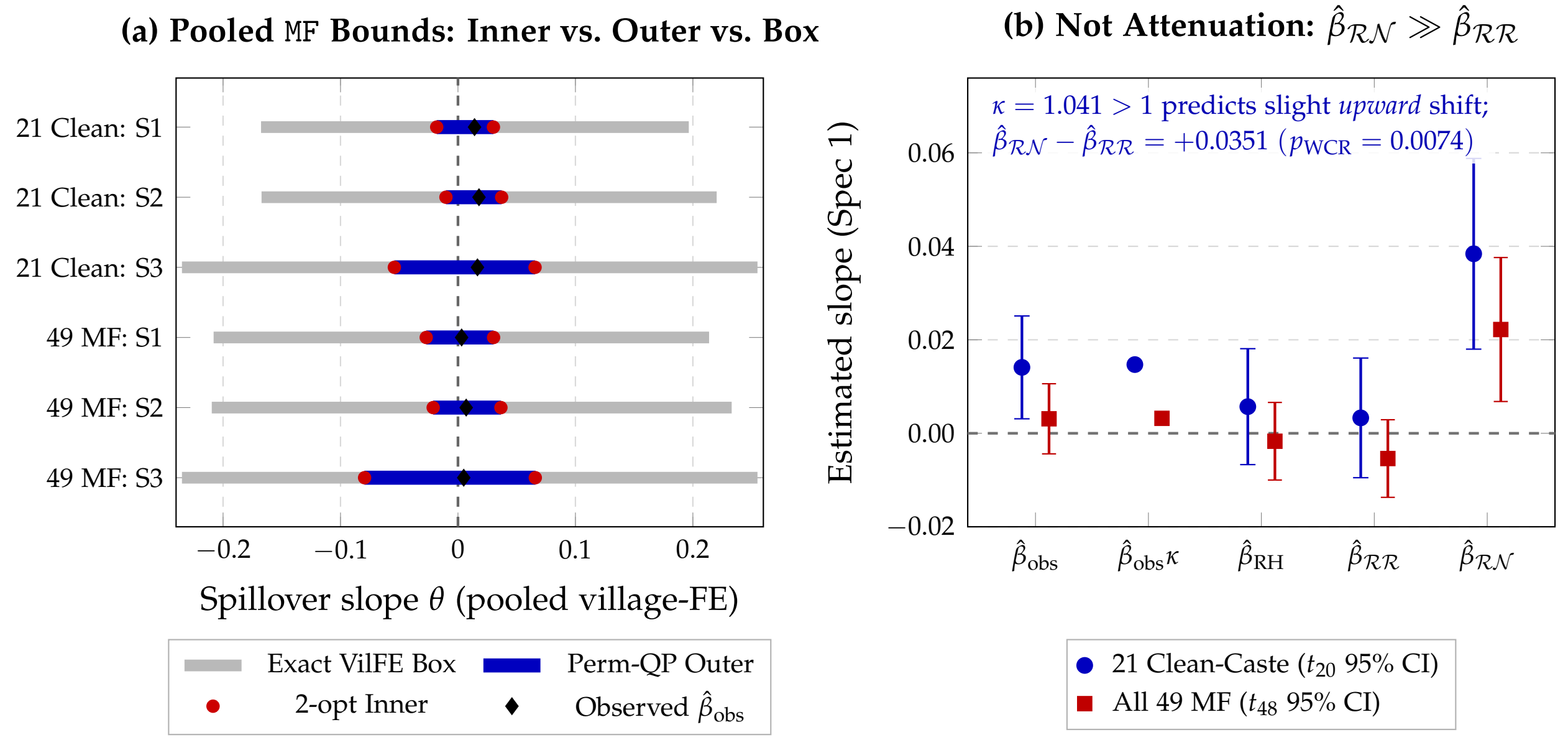

Table~\ref{tab:karnataka_reporting} and Appendix Table~\ref{tab:app_reporting} (illustrated in Figure~\ref{fig:karnataka}(b)) evaluate \eqref{eq:kappa_decomp} across all $15$ sample--specification cells (verified to machine precision $\le 5.6\times 10^{-17}$): in every row, $\kappa \in [1.0305, 1.0773] > 1$, so the entire downward shift comes from the residual correlation term ($\Delta_{\mathrm{res}} = -0.0090$ in 21-village Spec~1). Decomposing observed exposure $E_i^{\mathrm{obs}} = E_{i,\Rset\Rset} + E_{i,\Rset\Nset}$ into ties to surveyed leaders ($E_{i,\Rset\Rset} := \sum_{j\in\Rset\setminus\{i\}} A_{ij} z_j$) and ties to unsurveyed leaders ($E_{i,\Rset\Nset} := \sum_{j\in\Nset} A_{ij} z_j$) and regressing \texttt{MF} jointly on $(E_{\Rset\Rset}, E_{\Rset\Nset})$ pinpoints the source:
\begin{itemize}[leftmargin=1.5em]
  \item In 21-village Spec~1, $\hat{\beta}_{\Rset\Nset} = +0.0384$ vs.\ $\hat{\beta}_{\Rset\Rset} = +0.0033$, a contrast of \textbf{$+0.0351$} ($p_{t_{20}} = 0.0059$, wild cluster bootstrap $p_{\mathrm{WCR}} = 0.0074$).
  \item Across all $49$ MF villages, the contrast is \textbf{$+0.0276$} ($p_{\mathrm{WCR}} = 0.0016$), persisting across Specs~1--3 and the Banerjee-43 subset ($p_{\mathrm{WCR}} \le 0.0142$ across all nine rows; Table~\ref{tab:karnataka_reporting}). Because $21$ clean-caste and Banerjee-43 are overlapping subsets of the $49$ MF villages, these are not independent replications; in the complementary $28$ blank-caste villages, the point gap moves in the same positive direction ($+0.0199$--$+0.0210$) without reaching statistical significance ($p_{\mathrm{WCR}} = 0.137$--$0.177$).
\end{itemize}

\paragraph{Mechanism validation: non-leader placebo, leader balance, and directional relations.}
Because the released household adjacency matrix $A$ is a pre-symmetrized OR-union graph (Section~\ref{sec:data_hygiene}), $100\%$ of recorded ties in $A_{\Rset,\Nset}$ were personally nominated by respondent $i$, whereas $A_{\Rset,\Rset}$ combines ego-nominated ties with ties reported only by the alter $j \in \Rset$. Three empirical checks examine this salience-dilution mechanism (full results in Appendix Tables~\ref{tab:app_placebo_wcr}--\ref{tab:app_balance_dir}):
\begin{enumerate}[label=(\alph*),leftmargin=1.8em]
  \item \emph{Non-leader placebo and 4-regressor horserace:} Pure non-leader regressions on $(E^{\mathrm{nl}}_{\Rset\Rset}, E^{\mathrm{nl}}_{\Rset\Nset})$ show near-zero $\Rset\Nset$--$\Rset\Rset$ differences across every sample and specification ($+0.0004$ to $+0.0019$, $p = 0.64$--$0.95$ across the clean-caste, full-49, and Banerjee-43 subsamples; $+0.0019$ to $+0.0030$, $p = 0.56$--$0.68$ in the $28$ blank-caste villages). Entering all four exposures $(E^{\mathrm{lead}}_{\Rset\Rset}, E^{\mathrm{lead}}_{\Rset\Nset}, E^{\mathrm{nl}}_{\Rset\Rset}, E^{\mathrm{nl}}_{\Rset\Nset})$ jointly---controlling directly for ego's non-leader $\Rset\to\Nset$ and $\Rset\to\Rset$ out-degrees---leaves the leader contrast virtually unchanged (e.g., $+0.0357, p=0.0061$ in 21-village Spec~1; $+0.0278, p=0.0022$ in 49-village Spec~1), with significant difference-in-differences contrasts across all nine clean-caste, full-49, and Banerjee-43 specifications ($p = 0.008$--$0.045$). Adding household religion and census listing order to Spec~2 likewise preserves the gap ($+0.0301, p=0.0050$ in $G=21$; $+0.0253, p=0.0033$ in $G=49$).
  \item \emph{Per-dyad density ratio, uniform-thinning bound, and leader balance:} Across all $49$ MF villages, per-dyad link density in $\Rset\times\Rset$ is roughly $2\times$ that in $\Rset\times\Nset$ ($0.0910$ vs.\ $0.0405$ across all dyads, $2.25\times$; $0.1178$ vs.\ $0.0610$ across leader dyads, $1.93\times$), reflecting OR-union double-recording inside $\Rset\times\Rset$ (and any underlying density difference between surveyed and unsurveyed alters). Survey rates are mildly higher for leaders ($53.4\%$) than for non-leaders ($45.1\%$), and surveyed vs.\ unsurveyed leaders are closely balanced on caste, roof, electricity, religion, and census position, with surveyed leaders having slightly more rooms ($2.91$ vs.\ $2.59$) and slightly lower own \texttt{MF} adoption ($0.205$ vs.\ $0.246$). Under pure uniform binomial thinning of true exposure $E^{\mathrm{true}}$ with retention rate $q \in (0,1)$ ($E^{\mathrm{obs}} \mid E^{\mathrm{true}} \sim \operatorname{Bin}(E^{\mathrm{true}}, q)$), the univariate OLS slope is $\beta_{\mathrm{OLS}} = \beta / [q + (1-q)\E[E^{\mathrm{true}}]/\Var(E^{\mathrm{true}})]$, which under degree overdispersion ($\Var(E^{\mathrm{true}}) > \E[E^{\mathrm{true}}]$) generates an \emph{upward} augmentation bias bounded strictly above by $1/q$ (consistent with Proposition~1 of \citealp{lewbel2026estimating}, while in a joint regression under positive degree correlation the omitted $(1-q)E_{\Rset\Nset}^{\mathrm{true}}$ also leaks positively into $\hat{\beta}_{\Rset\Rset}$). Separating the model bound $1/q$ from the empirical $1.93\times$ leader-dyad ratio: if surveyed and unsurveyed leaders share comparable true link densities (or if surveyed leaders are slightly more connected, which makes $1/q < 1.93\times$), then pure uniform thinning can amplify $\hat{\beta}_{\Rset\Nset}$ by at most $1/q \lesssim 1.93\times$ while strictly preserving sign. By contrast, Table~\ref{tab:karnataka_reporting} shows observed slope ratios $\hat{\beta}_{\Rset\Nset}/\hat{\beta}_{\Rset\Rset}$ of $4.3\times$--$11.6\times$ across the $21$ clean-caste specifications ($+0.0384$ vs.\ $+0.0033$ in Spec~1) and opposite-signed point estimates in the full-$49$ and Banerjee-43 pools ($\hat{\beta}_{\Rset\Rset} \in [-0.0056, -0.0009]$ vs.\ $\hat{\beta}_{\Rset\Nset} \in [+0.0214, +0.0245]$). Pure uniform under-recording therefore cannot explain the magnitude or sign contrast, pointing instead to tie-strength dilution from alter-only links inside $\Rset\times\Rset$, potentially reinforced by unobserved selection between surveyed and unsurveyed leaders.
  \item \emph{Paired directional relation matrices:} The four paired relation matrices in \texttt{datav4.0} are distinct ($8{,}458$--$14{,}349$ differing dyads across $75$ villages), and on ego-only $\Rset\to\mathrm{Leader}_{\Nset}$ links upward reliance is denser (\texttt{helpdecision}/\texttt{giveadvice} $= 1.27\times$; \texttt{borrowmoney}/\texttt{lendmoney} $= 1.26\times$). Across relations (Appendix Table~\ref{tab:app_balance_dir}, Panel~B), the $\Rset\Nset$--$\Rset\Rset$ contrast is larger for advice sought (\texttt{helpdecision}: $+0.0714, p=0.0157$ in $G=21$; $+0.0412, p=0.0265$ in $G=49$) than for advice given (\texttt{giveadvice}: $+0.0244, p=0.250$; $+0.0243, p=0.109$), with similar ordering for \texttt{visitgo} vs.\ \texttt{visitcome} and \texttt{keroricego} vs.\ \texttt{keroricecome}. Because each matrix in \texttt{datav4.0} is pre-symmetrized and cannot identify edge-by-edge who nominated whom inside $\Rset\times\Rset$ (and we do not test the cross-relation difference of differences), we treat these point comparisons as directional evidence consistent with respondents nominating more salient ties.
\end{enumerate}

Because \texttt{Raw-Hyb} replaces the ego-nominated component $E_{\Rset\Nset}$ with cell-averaged counts while retaining the union-sampled component $E_{\Rset\Rset}$ verbatim, it drives $\Delta_{\mathrm{res}} < 0$ and pulls the slope downward. This mechanism is independently corroborated by directed-roster audits of the same $75$-village survey: \citet{ready2021measuring} find only $9\%$--$11\%$ dyadic concordance among surveyed pairs (causing OR-union density to nearly double, with $41\%$ repeat-alter nomination in advice versus $73\%$ in visits), and \citet{debacco2023latent} report the lowest reciprocity in advice ($0.296$) alongside hub under-reporting under four-alter nomination caps. It also connects directly to unilateral/bilateral reporting disagreement \citep{comola2014testing,comola2017missing} and sampled-network measurement error \citep{chandrasekhar2016econometrics,griffith2022name}---complementing the directed Add Health evidence of \citet{boucher2025estimating} on missing and unmatched nominations with the mirror-image distortion induced by undirected OR-union symmetrization---while cautioning against empirical designs on \texttt{datav4.0} that assume homogeneous edge observation across the $\Rset/\Nset$ boundary \citep{sun2026flexible}, conditional independence across paired visit layers \citep{lewbel2026estimating}, or equal spillover slopes across observed and missing ties \citep{marray2025estimating}. Benchmarks on partially sampled union graphs should therefore test $H_0 : \beta_{\Rset\Nset} = \beta_{\Rset\Rset}$ before attributing discrepancies to mechanical ARD attenuation.

\begin{table}[t]
\centering
\caption{Asymmetric-reporting decomposition and village-clustered inference on real Karnataka microfinance adoption (\texttt{MF}). Left columns verify the exact algebraic identity $\hat{\beta}_{\mathrm{RawHyb}} = \hat{\beta}_{\mathrm{obs}}\kappa + \Delta_{\mathrm{res}}$ ($|\text{err}| \le 5.6\times 10^{-17}$); right columns report joint village-FE regressions on $E_{\Rset\Rset} = A_{\Rset,\Rset}z_{\Rset}$ and $E_{\Rset\Nset} = A_{\Rset,\Nset}z_{\Nset}$ with CR1 standard errors, small-cluster $t_{G-1}$ $p$-values, and null-imposed Rademacher wild cluster bootstrap $p$-values ($p_{\mathrm{WCR}}$, $B=4{,}999$).}
\label{tab:karnataka_reporting}
\scriptsize
\setlength{\tabcolsep}{3.0pt}
\resizebox{\textwidth}{!}{%
\begin{tabular}{@{}lcccccccccc@{}}
\toprule
& \multicolumn{4}{c}{Exact Decomposition ($\hat{\beta}_{\mathrm{RH}} = \hat{\beta}_{\mathrm{obs}}\kappa + \Delta_{\mathrm{res}}$)} & \multicolumn{6}{c}{Joint Village-FE Regression on $(E_{\Rset\Rset}, E_{\Rset\Nset})$} \\
\cmidrule(lr){2-5}\cmidrule(lr){6-11}
Sample / Spec & $\hat{\beta}_{\mathrm{obs}}$ ($p_{t}$) & $\kappa$ & $\hat{\beta}_{\mathrm{obs}}\kappa$ & $\Delta_{\mathrm{res}} \to \hat{\beta}_{\mathrm{RH}}$ & $\hat{\beta}_{\Rset\Rset}$ ($p_{t}$) & $\hat{\beta}_{\Rset\Nset}$ ($p_{t}$) & $\hat{\beta}_{\Rset\Nset}-\hat{\beta}_{\Rset\Rset}$ & $t_{\mathrm{CR1}}$ & $p_{t_{G-1}}$ & $p_{\mathrm{WCR}}$ \\
\midrule
\multicolumn{11}{@{}l}{\emph{21 Clean-Caste MF (\texttt{caste} $\times$ \texttt{roof}, $G=21$, $df=20$, $n_{\Rset}=2{,}240$; subset of ALL 49)}} \\
Spec~1 (VilFE+$z$) & $+0.0141$ (.015) & 1.0409 & $+0.0147$ & $-0.0090 \to +0.0057$ & $+0.0033$ (.594) & $+0.0384$ (.0008) & $+0.0351$ & $+3.08$ & \textbf{.0059} & \textbf{.0074} \\
Spec~2 (+demog) & $+0.0179$ (.005) & 1.0490 & $+0.0188$ & $-0.0069 \to +0.0119$ & $+0.0088$ (.176) & $+0.0378$ (.0009) & $+0.0290$ & $+2.68$ & \textbf{.0145} & \textbf{.0142} \\
Spec~3 (+$\log D_0$) & $+0.0166$ (.011) & 1.0748 & $+0.0178$ & $-0.0088 \to +0.0090$ & $+0.0070$ (.284) & $+0.0368$ (.0014) & $+0.0299$ & $+2.79$ & \textbf{.0113} & \textbf{.0102} \\
\addlinespace[2pt]
\multicolumn{11}{@{}l}{\emph{ALL 49 MF Villages (\texttt{elec} $\times$ \texttt{roof}, $G=49$, $df=48$, $n_{\Rset}=4{,}891$)}} \\
Spec~1 (VilFE+$z$) & $+0.0031$ (.415) & 1.0457 & $+0.0032$ & $-0.0050 \to -0.0017$ & $-0.0054$ (.197) & $+0.0222$ (.0057) & $+0.0276$ & $+3.19$ & \textbf{.0025} & \textbf{.0016} \\
Spec~2 (+demog) & $+0.0070$ (.065) & 1.0447 & $+0.0073$ & $-0.0042 \to +0.0031$ & $-0.0009$ (.829) & $+0.0245$ (.0018) & $+0.0254$ & $+2.99$ & \textbf{.0045} & \textbf{.0052} \\
Spec~3 (+$\log D_0$) & $+0.0049$ (.214) & 1.0709 & $+0.0053$ & $-0.0060 \to -0.0007$ & $-0.0034$ (.428) & $+0.0230$ (.0039) & $+0.0264$ & $+3.12$ & \textbf{.0031} & \textbf{.0036} \\
\addlinespace[2pt]
\multicolumn{11}{@{}l}{\emph{Banerjee-43 MF Villages (\texttt{elec} $\times$ \texttt{roof}, $G=43$, $df=42$, $n_{\Rset}=4{,}413$; subset of ALL 49)}} \\
Spec~1 (VilFE+$z$) & $+0.0027$ (.504) & 1.0497 & $+0.0028$ & $-0.0049 \to -0.0020$ & $-0.0056$ (.203) & $+0.0214$ (.0134) & $+0.0270$ & $+2.93$ & \textbf{.0054} & \textbf{.0040} \\
Spec~2 (+demog) & $+0.0066$ (.099) & 1.0490 & $+0.0070$ & $-0.0040 \to +0.0029$ & $-0.0010$ (.826) & $+0.0237$ (.0052) & $+0.0246$ & $+2.72$ & \textbf{.0095} & \textbf{.0072} \\
Spec~3 (+$\log D_0$) & $+0.0045$ (.287) & 1.0772 & $+0.0048$ & $-0.0059 \to -0.0011$ & $-0.0036$ (.429) & $+0.0221$ (.0098) & $+0.0257$ & $+2.84$ & \textbf{.0069} & \textbf{.0078} \\
\addlinespace[2pt]
\multicolumn{11}{@{}l}{\emph{28 Blank-Caste MF Villages (\texttt{elec} $\times$ \texttt{roof}, $G=28$, $df=27$, $n_{\Rset}=2{,}652$; complement of 21 Clean-Caste)}} \\
Spec~1 (VilFE+$z$) & $-0.0053$ (.271) & 1.0578 & $-0.0057$ & $-0.0024 \to -0.0080$ & $-0.0115$ (.054$^\dagger$) & $+0.0084$ (.465) & $+0.0199$ & $+1.46$ & .1561 & .1768 \\
Spec~2 (+demog) & $-0.0012$ (.804) & 1.0522 & $-0.0012$ & $-0.0022 \to -0.0034$ & $-0.0074$ (.199) & $+0.0129$ (.254) & $+0.0203$ & $+1.51$ & .1425 & .1488 \\
Spec~3 (+$\log D_0$) & $-0.0032$ (.535) & 1.0773 & $-0.0034$ & $-0.0036 \to -0.0071$ & $-0.0097$ (.113) & $+0.0113$ (.329) & $+0.0210$ & $+1.57$ & .1286 & .1372 \\
\bottomrule
\end{tabular}}
\begin{flushleft}
\scriptsize\emph{Notes:} Bold $p_{t_{G-1}}$ and $p_{\mathrm{WCR}}$ mark $p < 0.05$ under small-cluster $t_{G-1}$ and null-imposed Rademacher wild cluster bootstrap ($B=4{,}999$; \citealp{cameron2008bootstrap}); in 21-village Spec~1, $\hat{\beta}_{\mathrm{obs}} = +0.0141$ has $p_{\mathrm{WCR}} = 0.0114$. $^\dagger$Indicates a borderline significance flip relative to normal $\pm 1.96$ ($t = -2.014$, $p_z = 0.044 \to p_{t_{27}} = 0.0541$); similarly, paired \texttt{Obs - RawHyb} in \texttt{ALL 49 Spec 2} ($t=+1.978$, $p_{t_{48}}=0.0537$, $p_{\mathrm{WCR}}=0.0544$) and \texttt{21 Clean (elec x roof) Spec 2} ($t=+2.043$, $p_{t_{20}}=0.0544$) move above $0.05$. Non-leader placebos, 4-regressor horseraces, leader balance, and paired directional relation regressions appear in Appendix Tables~\ref{tab:app_placebo_wcr}--\ref{tab:app_balance_dir}.
\end{flushleft}
\end{table}

\FloatBarrier

\section{Discussion, Practical Guidance, and Open Boundaries}
\label{sec:discussion}

Aggregated Relational Data remain a cost-effective tool for measuring social networks in the field. Separating what ARD identifies at the level of group-invariant network statistics \citep{breza2020using,breza2023consistently} from what happens when non-respondent links enter individual-level spillover regressions under partial participation ($\varphi < 1$) yields concrete workflow choices for both retrospective analysis and prospective survey design.

\subsection{A Practical Decision Guide for Applied Researchers}

Table~\ref{tab:decision_guide} distills our theoretical and empirical results into concrete workflow choices depending on whether data collection is already complete (retrospective analysis) or still in design (prospective survey).

\begin{table}[htbp]
\centering
\caption{Practical workflow guide for spillover inference under partial-participation ARD.}
\label{tab:decision_guide}
\small
\begin{tabular}{@{} >{\raggedright\arraybackslash}p{3.4cm} >{\raggedright\arraybackslash}p{5.5cm} >{\raggedright\arraybackslash}p{5.8cm} @{}}
\toprule
Setting & Recommended Method & Key Diagnostic / Boundary to Report \\
\midrule
\textbf{Retrospective $K$-group ARD} (already collected, $\varphi < 1$) & Run the exact $O(r_{\mathrm{r}}r_{\mathrm{n}} + \sum m_{\mathrm{n},k}\log m_{\mathrm{n},k})$ sign test (Theorem~\ref{thm:sign_test}); compute both constructive 2-opt inner witnesses and \texttt{FastShorSDP+PermRefine} / Pooled \texttt{Perm-QP} outer bounds (Theorem~\ref{thm:certifiers}). & Always distinguish inner witnesses from outer bounds (Remark~\ref{rem:inner_vs_outer}). On noisy real outcomes, pair pooled bounds with null-outcome placebos and out-of-sample cross-fitting (Section~\ref{sec:real_mf}). When adding plug-in uncertainty (Section~\ref{sec:feasible_plugin}), report the full five-metric ledger (\texttt{fin\&cov}, \texttt{medW}), never overall coverage alone. \\
\addlinespace[4pt]
\textbf{Expected exposure $y \sim Pz$} (high SNR, heterogeneous degrees, no $\widehat{P}$) & Use the model-free \texttt{Raw-Pure} Cauchy--Schwarz EIV bracket \eqref{eq:eiv_bracket} (Proposition~\ref{prop:raw_eiv}). & Verify positive debiased denominator ($0$ dropouts) and do \emph{not} apply EIV variance subtraction when the structural exposure is realized ($y \sim Az$). \\
\addlinespace[4pt]
\textbf{Prospective survey design} (salient binary $z_j$, realized $y\sim Az$ or $\omega_{ik}\neq 0$) & Use $(K{+}1)$ Augmented ARD for a scalar spillover slope under a tight question budget; upgrade to $2K$ $\Tstrat$ when group-heterogeneous spillovers or disjoint untreated control functions under expected exposure + homophily are needed (Propositions~\ref{prop:tstrat} and \ref{prop:budget}). & Under a strict question budget $\Budget = r_{\mathrm{r}} \Qitems$ with sparse single-graph $y \sim Pz$ and rare randomized $z$, check Proposition~\ref{prop:budget}(ii): standard $K$-ARD + EIV (or $(K{+}1)$ falling back to its $K$ counts) outperforms direct $E_{\mathrm{real}}$-EIV. \\
\addlinespace[4pt]
\textbf{Partially sampled undirected union graph} (e.g., field benchmarks) & Run the exact $\kappa$ decomposition \eqref{eq:kappa_decomp}, joint $(E_{\Rset\Rset}, E_{\Rset\Nset})$ regression, and non-leader placebo with small-cluster $t_{G-1}$ and wild cluster bootstrap inference (Section~\ref{sec:real_mf}). & Test $H_0: \beta_{\Rset\Nset} = \beta_{\Rset\Rset}$ (and compare against non-leader exposures) before attributing discrepancies between released union-graph and ARD regressions to mechanical attenuation. Rejecting $H_0$ diagnoses reporting asymmetry, but failing to reject $H_0$ does \emph{not} certify the absence of heterogeneous reporting bias or exposure misspecification \citep{gao2026impossibility}. \\
\bottomrule
\end{tabular}
\end{table}

\subsection{Explicit Open Boundaries}

In the spirit of transparent scope reporting, four genuine limitations and open questions remain beyond the scope of the present paper:
\begin{enumerate}[label=(\arabic*),leftmargin=2em]
  \item \emph{Single-village plug-in power and pooled uncertainty quantification:} Although our feasible plug-in perturbation certifier corrects rank-1 cell-error projection, hub-kurtosis effective degrees of freedom, and wrong-sign exclusions, single-village inflated intervals remain wide (heuristic \texttt{Stoch} spans $4.7\times$--$76.3\times$ the effect size and returns $(-\infty,+\infty)$ in $33\%$--$97\%$ of sparse single villages across Tables~\ref{tab:semisynth}--\ref{tab:karnataka_bounds}, with $17.5\%$--$27.0\%$ non-finite in Appendix Table~\ref{tab:app_low_snr}, while worst-case \texttt{WC} reaches $84.6\times$ and up to $100\%$ non-finite in sparse subgraphs). Unlike unlinked i.i.d.\ linear projections whose 1D optimal-transport endpoints admit $\sqrt{n}$ Wasserstein-2 central limit theorems \citep{dhaultfoeuille2025linear}, graph-coupled quadratic-denominator orbits do not inherit standard i.i.d.\ endpoint asymptotics. Developing formal multi-village pooled plug-in perturbation bounds---averaging independent node-activity estimation errors across $G$ villages while jointly addressing finite-sample extremum inflation---is an important direction for future work.
  \item \emph{Non-MCAR non-respondent transfer under unobserved $\Nset\times\Nset$:} Although our NonCircular-1 and NonCircular-2 benchmarks show that Oracle-$P^\star$ \texttt{FastShorSDP+PermRefine} stays tight across non-logistic link kernels $P^\star$ and feasible \texttt{Stoch} maintains overall coverage under moderate non-MCAR shifts ($\delta_{\Nset}=+0.40$), severe unobserved non-response selection (e.g., if only network hubs opt out of the survey entirely) cannot be identified from $\mathcal{D}_{\mathrm{ARD}}$ without auxiliary census degree or roster covariates.
  \item \emph{Field measurement error in $(K{+}1)$ and $2K$ $\Tstrat$:} Our survey-design results treat reported counts $Y_{i,\mathrm{tr}}$ and $Y_{i,(k,1)}$ as error-free static sums; quantifying optimal $K$ under empirically calibrated imperfect transmission/visibility, stigma/response bias, size-asymmetric recall and rounding-to-five heaping, cognitive load, and per-interview fixed costs \citep{mccormick2010how,laga2021thirty} requires experimental field validation.
  \item \emph{Direct validation of ego vs.\ alter nominations:} Because the standard pre-symmetrized \texttt{datav4.0} adjacency-matrix bundle pre-symmetrizes all $2{,}100$ matrices, our finding $\hat{\beta}_{\Rset\Nset} \gg \hat{\beta}_{\Rset\Rset}$ on \texttt{datav4.0} is consistent with reporting asymmetry across the surveyed/unsurveyed boundary (with larger point contrasts for advice sought than advice given in the paired directional matrices), though it is not separable from unobserved selection between surveyed and unsurveyed leaders from \texttt{datav4.0} alone. Replicating this diagnostic directly on the separate Harvard Dataverse raw directed individual nomination rosters (\texttt{doi:10.7910/DVN/U3BIHX}; \citealp{ready2021measuring,debacco2023latent})---where ego-only, alter-only, and mutual ties inside $\Rset\times\Rset$ can be separated explicitly \citep{comola2014testing,comola2017missing}---is a high-priority follow-up to isolate tie-strength dilution from leader selection.
\end{enumerate}

\subsection*{Statement on AI use}
An AI assistant based on Gemini, accessed through Antigravity, contributed substantially to formulating the theoretical framework, developing the proofs and polynomial-time certification algorithms, writing the standalone Python simulation and Karnataka empirical-audit scripts, verifying primary literature, and drafting and revising the manuscript. AI assistants based on OpenAI Codex and Claude also contributed independent adversarial reviews, finite-sample counterexample checks, and verification of empirical scope, citations, and exposition. The author directed the research agenda, scope decisions, and empirical design, reviewed the analyses and manuscript, and takes full responsibility for all mathematical statements, code, empirical results, references, and conclusions.

\clearpage
\bibliographystyle{plainnat}
\bibliography{refs}

\newpage
\appendix
\section{Auxiliary Proofs and Diagnostics for Section~\ref{sec:orbit}}
\label{app:orbit}

\subsection{Complete Proof of Proposition~\ref{prop:orbit_set}}

\begin{proof}[Proof of Proposition~\ref{prop:orbit_set}]
Consider the population replication thought experiment in which $(m, P_{\Rset,\Rset}, \{P_{\Rset,G_k^{\mathrm{n}}}\}_{k=1}^K \bmod \PiN)$ is observed, where $m = (\E[y_i \mid \xi, z, W_{\Rset}])_{i\in\Rset}$ and the column-profile oracle supplies the unordered multiset of non-respondent column profiles $\{(p_{ij})_{i\in\Rset} : j \in G_k^{\mathrm{n}}\}$ per trait group $k$.

\emph{Part~(i) (Sharpness for the column-profile oracle under unrestricted $m$):}
For any $\pi \in \PiN$, construct the alternative column-labeled structure $(\xi^\pi, m)$ keeping $m$ and the noise distribution of $y_{\Rset} - m$ identical. By Proposition~\ref{prop:swap_invariance}, $\mathcal{L}_{\xi^\pi}(A_{\Rset,\Rset}, Y_{\Rset,\cdot}) = \mathcal{L}_\xi(A_{\Rset,\Rset}, Y_{\Rset,\cdot})$, and the column-profile oracle is invariant to $\pi \in \PiN$ by definition, while the target linear-projection estimand shifts from $\theta(\pi^\star)$ to $\theta(\pi)$. Conversely, any column-aligned structure compatible with the column-profile oracle differs from the true column ordering only by some $\pi \in \PiN$, so its projection estimand belongs to $\ThetaI$. Hence $\ThetaI$ is sharp relative to the column-profile oracle and unrestricted $m$. (Without cross-row column alignment, Proposition~\ref{prop:swap_invariance} shows that arbitrary row-specific permutations within each $(i,k)$ preserve $\mathcal{L}_\xi$, expanding the general-$\psi$ identified set strictly beyond $\ThetaI$.)

\emph{Part~(ii) (Outer validity under linear structural models):}
Under Assumption~\ref{assump:causal}, $m_i = \beta_0 + \beta_1 z_i + \beta_{\mathrm{spill}} E_i(\pi^\star) + W_i^\top \vartheta_W$, so applying the orthogonal projector $\MZ$ gives $\tilde{m} = \beta_{\mathrm{spill}} \widetilde{E}(\pi^\star)$. Assuming $\widetilde{E}(\pi^\star) \neq 0$, substituting into \eqref{eq:orbit_set} yields $\theta(\pi^\star) = \beta_{\mathrm{spill}} \in \ThetaI$. When $\dim(\operatorname{Im}\MZ) \ge 2$, $\ThetaI$ does not enforce $\tilde{m} \in \spanop\{\widetilde{E}(\pi)\}$ and is generically strictly wider than the exact zero-residual structural conditional-mean set $\{\beta : \exists \pi \in \PiN,\, \tilde{m} = \beta \widetilde{E}(\pi)\}$.

\emph{Part~(iii) (Collapse-to-point characterization):}
\emph{(a)} Suppose $E_{\Rset}(\pi) = E_{\Rset}(\pi')$ for all $\pi,\pi' \in \PiN$. Fix any group $k \in \{1,\dots,K\}$, any respondent $i \in \Rset$, and any two distinct non-respondents $j, j' \in G_k^{\mathrm{n}}$. Comparing the identity permutation with the transposition $\sigma_{j,j'} = (j \leftrightarrow j') \in \PiN$ gives
\[
  0 \;=\; E_i(\mathrm{id}) - E_i(\sigma_{j,j'})
  \;=\; (p_{ij} z_j + p_{ij'} z_{j'}) - (p_{ij} z_{j'} + p_{ij'} z_j)
  \;=\; (p_{ij} - p_{ij'})(z_j - z_{j'}).
\]
If $z^{(k)}$ is not constant on $G_k^{\mathrm{n}}$, pick $j_0, j_1 \in G_k^{\mathrm{n}}$ with $z_{j_0} \neq z_{j_1}$; then $p_{i j_0} = p_{i j_1}$, and for any third index $j \in G_k^{\mathrm{n}}$, either $z_j \neq z_{j_0}$ or $z_j \neq z_{j_1}$, forcing $p_{ij} = p_{i j_0}$. Thus $p_{ij}$ is constant across $j \in G_k^{\mathrm{n}}$ for every $i \in \Rset$; the converse is immediate. Under $0 < p_{ij} < 1$, constancy in $j \in G_k^{\mathrm{n}}$ is the degenerate special case $\logit(p_{ij}) = \gamma_{i,k}$ (zero non-respondent effect $\gamma_{j, k(i)} \equiv 0$) of the $k$-beta additive log-odds family $\logit(p_{ij}) = \gamma_{i,k(j)} + \gamma_{j,k(i)}$ in Proposition~1 of \citet{tseng2026countfibres}.

\emph{(b)} For the projected exposure $\widetilde{E}(\pi) = \MZ E_{\Rset}(\pi)$, comparing the identity with any transposition $\sigma_{j,j'}$ within $G_k^{\mathrm{n}}$ yields
\[
  \widetilde{E}(\mathrm{id}) - \widetilde{E}(\sigma_{j,j'})
  \;=\; (z_j - z_{j'})\, \MZ P_{\Rset, G_k^{\mathrm{n}}}(e_j - e_{j'}).
\]
Because transpositions generate $\PiN$, $\widetilde{E}(\pi)$ is constant across $\PiN$ if and only if \eqref{eq:exact_collapse_iff} holds. Since $\theta(\pi) = \langle \tilde{m}, \widetilde{E}(\pi)\rangle / \|\widetilde{E}(\pi)\|_2^2$ is constant across $\PiN$ for every $\tilde{m} \in \operatorname{Im}(\MZ)$ if and only if $\widetilde{E}(\pi)$ is constant across $\PiN$ (as $u \mapsto u/\|u\|_2^2$ is injective on $\R^{r_{\mathrm{r}}}\setminus\{0\}$), \eqref{eq:exact_collapse_iff} is necessary and sufficient for $\ThetaI$ to collapse to a point for all $m$.
\end{proof}

\begin{remark}[Numerical Contrast: Orbit $\ThetaI$ vs.\ Structural Conditional-Mean Filtering]
\label{rem:orbit_vs_cm}
On a small degree-heterogeneous test network with true $\beta_{\mathrm{spill}} = 2.0$, the discrete column-oracle Orbit Identified Set under unrestricted $m$ has endpoint range $[1.1918, 3.8372]$ (diameter $2.6453$), whereas imposing exact zero-residual linear collinearity $\tilde{m} = \beta \widetilde{E}(\pi)$ collapses to the singleton $\{2.0000\}$. In finite samples with outcome noise or nonlinearities in $m$, zero-residual filtering fails immediately, whereas $\ThetaI$ remains misspecification-robust. We also verified on the same testbed that generic residualized basis vectors $(\MZ u_0, \MZ u_1)$ are non-orthogonal ($|\Corr(\MZ u_0, \MZ u_1)| = 0.8563 \neq 0$, while remaining linearly independent with Gram $\lambda_{\min} = 0.0792 > 0$), so none of our proofs assume artificial orthogonality after FWL projection.
\end{remark}

\subsection{Complete Proof of Theorem~\ref{thm:certifiers} and Pooled Matrix-Free Perm-QP}

\begin{proof}[Proof of Theorem~\ref{thm:certifiers}]
Let $\mathcal{O} = \prod_{k=1}^K \mathcal{O}_k \subseteq \mathcal{P} = \conv(\mathcal{O})$.

\emph{Tier~1 (\texttt{LP-Secant}) validity:}
Define $N(x) := \|\tilde{e}_0 + \widetilde{P} x\|_2^2$, which is convex quadratic on the compact polytope $\mathcal{P}$. At every iterate $x^{(t)} \in \mathcal{P}$ of Pairwise Frank--Wolfe \citep{frank1956algorithm}, convexity yields $N(x) \ge N(x^{(t)}) + \nabla N(x^{(t)})^\top (x - x^{(t)})$ for all $x \in \mathcal{P}$. Minimizing the right-hand side over $x \in \mathcal{P}$ by cell-wise sorting gives a valid lower bound $N_{\mathrm{LB}}^{(t)} \le \min_{x\in\mathcal{P}} N(x) \le \min_{x\in\mathcal{O}} N(x)$, so $N_{\mathrm{LB}} := \max_t N_{\mathrm{LB}}^{(t)} > 0$ unconditionally certifies $\widetilde{E}(x) \neq 0$ on all of $\mathcal{P}$. On the convex side (say $\beta > 0$ upper bound), $\hat{\theta}(x) \le \overline{L} / N_{\mathrm{LB}}$. On the non-convex side (say $\beta > 0$ lower bound), compute exact row extrema $\ell_i := \min_{x\in\mathcal{P}} \widetilde{E}_i(x)$ and $u_i := \max_{x\in\mathcal{P}} \widetilde{E}_i(x)$ by sorting row $i$ of $\widetilde{P}$. By convexity of $u \mapsto u^2$ on $[\ell_i, u_i]$, $\widetilde{E}_i(x)^2 \le (\ell_i + u_i)\widetilde{E}_i(x) - \ell_i u_i$ holds pointwise for every $x \in \mathcal{P}$. Hence for $\beta > 0$,
\[
  Q_\beta(x) \;=\; \tilde{y}^\top \widetilde{E}(x) - \beta \sum_{i=1}^{r_{\mathrm{r}}} \widetilde{E}_i(x)^2
  \;\ge\;
  \tilde{y}^\top \widetilde{E}(x) - \beta \sum_{i=1}^{r_{\mathrm{r}}} \Big((\ell_i + u_i)\widetilde{E}_i(x) - \ell_i u_i\Big)
  \;=\; Q_\beta^{\mathrm{sec}}(x),
\]
whose exact minimum over $\mathcal{P}$ is computed in one sorting pass. Bisecting $\beta$ until $\min_{\mathcal{P}} Q_\beta^{\mathrm{sec}} > 0$ certifies $\underline{\beta}_{\mathrm{LP}} \le \min_{\mathcal{O}} \hat{\theta}(x)$ by Lemma~\ref{lem:level_set}(i).

\emph{Tier~2 (\texttt{Perm-QP}) validity and matrix-free village-FE extension:}
On the convex side ($\max_{\mathcal{P}} Q_\beta$ for $\beta > 0$, or $\min_{\mathcal{P}} Q_\beta$ for $\beta < 0$), minimizing the convex quadratic $-Q_\beta(x)$ over $\mathcal{P}$ by Pairwise Frank--Wolfe produces a certified dual lower bound $L_{\mathrm{FW}}(-\beta) \le \min_{x\in\mathcal{P}} (-Q_\beta(x)) = -\max_{x\in\mathcal{P}} Q_\beta(x)$. Whenever $L_{\mathrm{FW}}(-\beta) > 0$, Lemma~\ref{lem:level_set}(i) certifies $\max_{x\in\mathcal{O}} \hat{\theta}(x) < \beta$. Intersecting with Tier~1 guarantees $[\underline{\beta}_{\mathrm{QP}}, \overline{\beta}_{\mathrm{QP}}] \subseteq [\underline{\beta}_{\mathrm{LP}}, \overline{\beta}_{\mathrm{LP}}]$ unconditionally even if Frank--Wolfe stops early. Because every step needs only matrix--vector evaluations $x \mapsto \widetilde{P} x$ and $u \mapsto \widetilde{P}^\top u$ plus per-cell sorting, the exact same algorithm extends matrix-free to $G$ pooled villages with $\mathcal{P}_{\mathrm{pool}} = \prod_{v=1}^G \mathcal{P}_v$ and exact village-demeaned FWL projector $\MZ = M_{[V, z, W]}$.

\emph{Tier~3 (\texttt{FastShorSDP+PermRefine}) Cholesky certification and scale equivariance:}
Every $x \in \mathcal{O}$ satisfies four exact orbit invariants:
\begin{equation}
  \label{eq:orbit_invariants}
  x \in \mathcal{P},
  \qquad
  \mathbf{1}_{m_{\mathrm{n},k}}^\top x^{(k)} = S_k,
  \qquad
  \|x^{(k)}\|_2^2 = Q_k := \sum_{j \in G_k^{\mathrm{n}}} z_j^2,
  \qquad
  (x_\ell - a_k^-)(x_\ell - b_k^+) \le 0,
\end{equation}
where $a_k^- := \min_{G_k^{\mathrm{n}}} z$ and $b_k^+ := \max_{G_k^{\mathrm{n}}} z$ (with equality $x_\ell^2 - x_\ell = 0$ on binary cells). Introducing dual multipliers $\lambda \in \R^K$ and $\nu \in \R^{r_{\mathrm{n}}}$ ($\nu_\ell \ge 0$ on non-binary cells) yields the Shor Lagrangian
\begin{equation}
  \label{eq:shor_lagrangian}
  \begin{aligned}
    \mathcal{L}_\beta(x; \lambda, \nu)
    &\;:=\;
    Q_\beta(x)
    + \sum_{k=1}^K \lambda_k \big(\|x^{(k)}\|_2^2 - Q_k\big) \\
    &\qquad
    + \sum_{\ell=1}^{r_{\mathrm{n}}} \nu_\ell \big(x_\ell^2 - (a_{k(\ell)}^- + b_{k(\ell)}^+) x_\ell + a_{k(\ell)}^- b_{k(\ell)}^+\big)
    \;\le\; Q_\beta(x)
  \end{aligned}
\end{equation}
for every $x \in \mathcal{O}$. Writing $x = \bar{x} + Q_V w$ where $\bar{x}_\ell = S_{k(\ell)}/m_{\mathrm{n},k(\ell)}$ and $Q_V = \blkdiag(H_1,\dots,H_K)$ spans $\ker([\mathbf{1}_{m_{\mathrm{n},1}}^\top, \dots, \mathbf{1}_{m_{\mathrm{n},K}}^\top])$, we obtain
\[
  \mathcal{L}_\beta(\bar{x} + Q_V w; \lambda, \nu)
  \;=\;
  w^\top H(\lambda,\nu) w + 2 h(\lambda,\nu)^\top w + c_0(\lambda,\nu),
\]
with $H(\lambda,\nu) = -\beta Q_V^\top M_0 Q_V + \blkdiag_k(\lambda_k I_{m_{\mathrm{n},k}-1} + H_k^\top \diag(\nu^{(k)}) H_k)$. If Cholesky factorization verifies $H(\lambda,\nu) = R^\top R \succ 0$ (with an explicit floating-point diagonal safety shift if rigorous machine-arithmetic certification is desired; \citealp{neumaier2004safe,rump2006verification}), then two valid lower bounds follow simultaneously:
\begin{enumerate}[label=(\alph*),leftmargin=1.8em]
  \item Completing the square over all $w \in \R^{r_{\mathrm{n}}-K}$ gives $\mathcal{L}_\beta(\bar{x}+Q_V w;\lambda,\nu) = \|R w + R^{-\top} h\|_2^2 + c_0 - \|R^{-\top} h\|_2^2 \ge c_0 - h^\top H^{-1} h$.
  \item Because $H(\lambda,\nu) \succ 0$, the restriction of $\mathcal{L}_\beta(x;\lambda,\nu)$ to the affine hull $\{\bar{x} + Q_V w : w \in \R^{r_{\mathrm{n}}-K}\} \supseteq \mathcal{P}$ is strictly convex. Therefore, at \emph{any} feasible iterate $\hat{x} \in \mathcal{P}$ produced by Pairwise Frank--Wolfe on $\min_{x\in\mathcal{P}} \mathcal{L}_\beta(x;\lambda,\nu)$ (\texttt{PermRefine}), first-order convexity yields the exact certified linear dual lower bound
  \begin{equation}
    \label{eq:cholesky_cert}
    \begin{aligned}
      \min_{x \in \mathcal{O}} Q_\beta(x)
      \;\ge\;
      \max\Big\{
        &c_0(\lambda,\nu) - h(\lambda,\nu)^\top H(\lambda,\nu)^{-1} h(\lambda,\nu), \\
        &\mathcal{L}_\beta(\hat{x};\lambda,\nu) + \min_{s\in\mathcal{P}} \nabla_x \mathcal{L}_\beta(\hat{x};\lambda,\nu)^\top (s - \hat{x})
      \Big\}.
    \end{aligned}
  \end{equation}
\end{enumerate}
Intersecting Tier~3 with Tier~2 preserves nesting unconditionally. Finally, under rescaling $y \mapsto \kappa_y y$ ($\kappa_y > 0$) and $z \mapsto \kappa_z z$ ($\kappa_z > 0$), we have $\hat{\theta}(x) \mapsto (\kappa_y / \kappa_z)\hat{\theta}(x)$, and rescaling $\beta \mapsto (\kappa_y/\kappa_z)\beta$ together with $(\lambda_k, \nu_\ell) \mapsto (\kappa_y/\kappa_z)(\lambda_k, \nu_\ell)$ rescales $\mathcal{L}_\beta$ homogeneously by $\kappa_y \kappa_z$, proving exact scale equivariance of the population relaxations (up to finite-step bisection tolerances and discrete two-valued-vs.-binary tolerance checks in numerical implementations).
\end{proof}

\section{Auxiliary Proofs, Calibration Formulas, and Diagnostics for Section~\ref{sec:plugin_homophily}}
\label{app:plugin}

\subsection{Proof of Lemma~\ref{lem:perturbation}}

\begin{proof}[Proof of Lemma~\ref{lem:perturbation}]
Let $\widetilde{E}^\star := \MZ E(\pi^\star)$, $\widehat{\widetilde{E}} := \MZ \widehat{E}(\tilde{\pi})$, and $\delta_E := \widehat{\widetilde{E}} - \widetilde{E}^\star$. By assumption, $\|\delta_E\|_2 / \sqrt{r_{\mathrm{r}}} \le \eta_{\mathrm{plug}}$ and $\|\widehat{\widetilde{E}}\|_2 / \sqrt{r_{\mathrm{r}}} \ge \underline{\sigma}_{\widehat{E}} > 0$, so $\|\delta_E\|_2 \le \varrho \|\widehat{\widetilde{E}}\|_2$ with $\varrho < 1$. By the reverse triangle inequality, $\|\widetilde{E}^\star\|_2 \ge (1-\varrho)\|\widehat{\widetilde{E}}\|_2 > 0$.

Write $\tilde{y} = \tilde{m} + \tilde{\varepsilon}$ ($\tilde{\varepsilon} := \MZ \varepsilon$) and decompose $\tilde{m} = \theta^\star \widetilde{E}^\star + r^\star$, where $\langle r^\star, \widetilde{E}^\star \rangle = 0$ and $\|r^\star\|_2 = \sqrt{r_{\mathrm{r}}}\,\sigma_{r^\star}$. Expanding $\hat{\theta}(\tilde{\pi}) = \langle \tilde{m} + \tilde{\varepsilon}, \widehat{\widetilde{E}} \rangle / \|\widehat{\widetilde{E}}\|_2^2$ around $\theta^\star$ gives
\[
  \hat{\theta}(\tilde{\pi}) - \theta^\star
  \;=\; \frac{\langle \theta^\star \widetilde{E}^\star + r^\star + \tilde{\varepsilon},\, \widehat{\widetilde{E}} \rangle - \theta^\star \|\widehat{\widetilde{E}}\|_2^2}{\|\widehat{\widetilde{E}}\|_2^2}
  \;=\; \frac{-\theta^\star \langle \delta_E, \widehat{\widetilde{E}} \rangle + \langle r^\star, \delta_E \rangle + \langle \tilde{\varepsilon}, \widehat{\widetilde{E}} \rangle}{\|\widehat{\widetilde{E}}\|_2^2},
\]
where we used $\langle r^\star, \widehat{\widetilde{E}} \rangle = \langle r^\star, \widetilde{E}^\star + \delta_E \rangle = \langle r^\star, \delta_E \rangle$. Applying Cauchy--Schwarz and the hypothesis $|\langle \tilde{\varepsilon}, \widehat{\widetilde{E}} \rangle| / (r_{\mathrm{r}} \underline{\sigma}_{\widehat{E}}^2) \le \mathrm{se}_{\max}$ yields
\[
  |\hat{\theta}(\tilde{\pi}) - \theta^\star|
  \;\le\; |\theta^\star| \frac{\|\delta_E\|_2}{\|\widehat{\widetilde{E}}\|_2}
  + \frac{\|r^\star\|_2 \|\delta_E\|_2}{\|\widehat{\widetilde{E}}\|_2^2}
  + \frac{|\langle \tilde{\varepsilon}, \widehat{\widetilde{E}} \rangle|}{\|\widehat{\widetilde{E}}\|_2^2}
  \;\le\; \varrho |\theta^\star| + \varrho \frac{\sigma_{r^\star}}{\underline{\sigma}_{\widehat{E}}} + \mathrm{se}_{\max}.
\]
Using $|\theta^\star| \le |\hat{\theta}(\tilde{\pi})| + |\hat{\theta}(\tilde{\pi}) - \theta^\star| \le b_{\max}^{\mathrm{abs}} + |\hat{\theta}(\tilde{\pi}) - \theta^\star|$ and solving for $|\hat{\theta}(\tilde{\pi}) - \theta^\star|$ gives
\[
  |\hat{\theta}(\tilde{\pi}) - \theta^\star|
  \;\le\; \frac{\varrho\, b_{\max}^{\mathrm{abs}} + \varrho(\sigma_{r^\star}/\underline{\sigma}_{\widehat{E}}) + \mathrm{se}_{\max}}{1 - \varrho}.
\]
Because $\tilde{\pi} \in \PiN$ implies $\hat{\theta}(\tilde{\pi}) \in \ThetaIhat(\widehat{P}) \subseteq [\hat{b}_{\min}, \hat{b}_{\max}]$, it follows unconditionally that $\theta^\star \in \widehat{\mathcal{I}}(\varrho, \mathrm{se}_{\max}, \sigma_{r^\star})$ in \eqref{eq:perturbation_iv}.
\end{proof}

\subsection{Explicit Plug-In Calibration Formulas (Worst-Case and Stochastic-EIV Heuristic)}

Given the joint degree-conditioned logistic fit $(\hat{p}_{ij}, \hat{\alpha}_{\Rset}, \hat{\zeta}_{\Rset})$ and transferred non-respondent multiset $\widehat{P}_{\Rset, G_k^{\mathrm{n}}}$, write $\hat{w}_{ij} := \hat{p}_{ij}(1-\hat{p}_{ij})$ and $\Rset_s := \{i \in \Rset : \hat{\zeta}_i = s\}$ for $s \in \{+1,-1\}$. We compute three variance components:
\begin{enumerate}[label=(\arabic*),leftmargin=2em]
  \item \emph{Node-activity component ($S_{\mathrm{node}}$):}
  \begin{align*}
    v_{\mathrm{node},i}
    &\;:=\;
    \frac{\Big(\sum_{j\in\Rset\setminus\{i\}} \hat{w}_{ij} z_j + \sum_{k=1}^K \bar{z}_{\Nset,k}\sum_{j\in G_k^{\mathrm{n}}} \hat{w}_{ij}\Big)^2}{\max\!\big(\sum_{j\neq i} \hat{w}_{ij},\, 2\big)},\\
    q_i &\;:=\; (1 - h_{ii}) v_{\mathrm{node},i},
    \qquad
    S_{\mathrm{node}} \;:=\; \sum_{i=1}^{r_{\mathrm{r}}} q_i.
  \end{align*}
  \item \emph{Exact rank-1 FWL-projected cell-profile component ($S_{\mathrm{cell}}$):} Let $a_i^{\mathrm{eff}} := \sum_{k=1}^K \sum_{j\in G_k^{\mathrm{n}}} \hat{w}_{ij}$ (or $a_i$ in DCSBM) and $v_s := (a_i^{\mathrm{eff}} \ind{\hat{\zeta}_i = s})_{i\in\Rset}$. Projecting $v_s$ directly through $\MZ$ gives
  \begin{equation}
    \label{eq:rank1_cell}
    S_{\mathrm{cell},s}
    \;:=\;
    \|\MZ v_s\|_2^2 \cdot \frac{\sum_{k=1}^K \bar{z}_{\Nset,k}^2 \sum_{u \in \Rset_s}\sum_{j\in G_k^{\mathrm{n}}} \hat{w}_{uj}}{\big(\sum_{u \in \Rset_s} a_u^{\mathrm{eff}}\big)^2},
    \qquad
    S_{\mathrm{cell}} := S_{\mathrm{cell},+1} + S_{\mathrm{cell},-1}.
  \end{equation}
  \item \emph{Within-cell spread component ($S_{\mathrm{spr}}$):}
  \[
    S_{\mathrm{spr}}
    \;:=\;
    \sum_{i=1}^{r_{\mathrm{r}}} (1 - h_{ii}) \sum_{k=1}^K m_{\mathrm{n},k}\, \widehat{\Var}_{G_k^{\mathrm{n}}}(\hat{p}_{i\cdot})\, \bar{z}_{\Nset,k}(1 - \bar{z}_{\Nset,k}).
  \]
\end{enumerate}

Defining Satterthwaite--Welch \citep{satterthwaite1946approximate} effective degrees of freedom
\begin{equation}
  \label{eq:reff_sw}
  r_{\mathrm{eff}} \;:=\; \frac{S_{\mathrm{node}}^2}{\sum_{i=1}^{r_{\mathrm{r}}} q_i^2},
  \qquad
  r_{\mathrm{eff,tot}} \;:=\; \frac{(S_{\mathrm{node}} + S_{\mathrm{cell}})^2}{\sum_{i=1}^{r_{\mathrm{r}}} q_i^2 + S_{\mathrm{cell},+1}^2 + S_{\mathrm{cell},-1}^2},
\end{equation}
and the cube-root Wilson--Hilferty upper multiplier $c_{\mathrm{WH}}(df, z) := \max\!\big(1, (1 - \frac{2}{9\,df} + z\sqrt{\frac{2}{9\,df}})^{3/2}\big)$, the worst-case perturbation ratio entering Theorem~\ref{thm:plugin_asymp} at one-sided tail $\alpha^{(1)}$ is
\begin{equation}
  \label{eq:rho_wc}
  \varrho_{\mathrm{WC}}(\alpha^{(1)})
  \;:=\;
  \frac{c_{\mathrm{WH}}(r_{\mathrm{eff,tot}}, z_{1-\alpha^{(1)}}) \sqrt{(S_{\mathrm{node}}+S_{\mathrm{cell}})/r_{\mathrm{r}}} + \sqrt{S_{\mathrm{spr}}/r_{\mathrm{r}}}}{\underline{\sigma}_{\widehat{E}}}.
\end{equation}
The tighter \emph{stochastic-EIV heuristic ratio} reported in empirical tables (calibrated at nominal $\alpha = 0.05$, explicitly labeled heuristic because $\varrho_{\mathrm{stoch}}$ approximates a slope perturbation rather than an $\ell_2$ exposure-norm bound) is
\begin{equation}
  \label{eq:rho_stoch}
  \varrho_{\mathrm{stoch}}
  \;:=\;
  \varrho_{\mathrm{node}}^2\, c_{\mathrm{WH}}(r_{\mathrm{eff}}, z_{1-\alpha})
  \;+\; z_{1-\alpha/2}\sqrt{\sigma_{\mathrm{lin,node}}^2 + \varrho_{\mathrm{cell}}^2}
  \;+\; \varrho_{\mathrm{spr}},
\end{equation}
where $\varrho_{\bullet} := \sqrt{S_{\bullet}/r_{\mathrm{r}}} / \underline{\sigma}_{\widehat{E}}$ and $\sigma_{\mathrm{lin,node}} := \max\!\Big(\frac{\sqrt{\sum_i (\widehat{\widetilde{E}}_i^{\mathrm{mean}})^2 v_{\mathrm{node},i}}}{\max(\sqrt{\|\widehat{\widetilde{E}}^{\mathrm{mean}}\|_2^2 \, r_{\mathrm{r}}\underline{\sigma}_{\widehat{E}}^2},\, r_{\mathrm{r}}\underline{\sigma}_{\widehat{E}}^2)},\; \frac{\varrho_{\mathrm{node}}}{\sqrt{r_{\mathrm{eff}}}}\Big)$. Both procedures return $(-\infty,+\infty)$ whenever $\varrho \ge 0.88$.

\subsection{Proof Sketch of Theorem~\ref{thm:plugin_asymp}}

\begin{proof}[Proof sketch of Theorem~\ref{thm:plugin_asymp}]
\emph{Part~(i):} Under \eqref{eq:logistic_model} with sparsity offset $\rho_r$, bounded $|\alpha_i| \le M_\alpha$, and $d_{\mathrm{avg}} \asymp r e^{\rho_r} \gg \log r$, the Fisher information matrix $\mathcal{I} \in \R^{r_{\mathrm{r}}\times r_{\mathrm{r}}}$ for respondent activities $\alpha_{\Rset}$ (profiling out the finite-dimensional cell parameters) has diagonal entries $\mathcal{I}_{ii} = \sum_{j\neq i} w_{ij} \asymp d_{\mathrm{avg}}$ and off-diagonal entries $\mathcal{I}_{ii'} = w_{ii'} \asymp d_{\mathrm{avg}}/r$ for $i \neq i'$. Using the inverse-Fisher approximation for undirected $\beta$-models \citep{chatterjee2011random,yan2013central,graham2017econometric}, write $\mathcal{I} = D + W_{\mathrm{off}}$ with $D = \diag(\mathcal{I}_{ii})$. Under MCAR partial participation $\varphi < 1$, $\|D^{-1/2} W_{\mathrm{off}} D^{-1/2}\|_{\mathrm{op}} \le \varphi + o(1) < 1$ (since $W_{\mathrm{off}}\mathbf{1} \approx \varphi D \mathbf{1}$), so the Neumann series $\mathcal{I}^{-1} = \sum_{m=0}^\infty (-1)^m D^{-1}(W_{\mathrm{off}} D^{-1})^m$ converges absolutely. Note that every higher term $(D^{-1}W_{\mathrm{off}})^m D^{-1}$ for $m \ge 1$ has entries of order $O((r d_{\mathrm{avg}})^{-1})$ with geometric decay factor $\varphi^m$: hence $(\mathcal{I}^{-1})_{ii'} = O((r d_{\mathrm{avg}})^{-1})$ for $i \neq i'$ (predominantly negative under partial participation $\varphi < 1$, with rare positive pairs arising from logistic hub saturation as verified in Table~\ref{tab:block1_diag}(c)), while $(\mathcal{I}^{-1})_{ii} = \mathcal{I}_{ii}^{-1}(1 + O(r^{-1}))$. Consequently, for any exposure Jacobian $J$ whose normalized rows $J/d_{\mathrm{avg}}$ have uniformly bounded norms (with $J_{ii} = O(d_{\mathrm{avg}})$ and $J_{ii'} = O(d_{\mathrm{avg}}/r)$), bounded-leverage controls $(\MZ)_{ii'} = O(r_{\mathrm{r}}^{-1})$, and non-degenerate trace $\tr(\MZ J D^{-1} J^\top) \asymp r_{\mathrm{r}} d_{\mathrm{avg}}$, the off-diagonal contribution satisfies $|\tr(\MZ J (\mathcal{I}^{-1} - D^{-1}) J^\top)| / \tr(\MZ J D^{-1} J^\top) = O(r^{-1})$.

\emph{Part~(ii):} Decompose $\MZ(\widehat{E}(\tilde{\pi}) - E(\pi^\star)) = \MZ \Delta_{\mathrm{node}} + \MZ \Delta_{\mathrm{cell}} + \MZ \Delta_{\mathrm{spr}} + o_p(d_{\mathrm{avg}}^{-1/2}\sqrt{r_{\mathrm{r}}})$. Under the asymptotically Gaussian score expansion $\hat{\alpha}_i - \alpha_i = \mathcal{I}_{ii}^{-1}\sum_{j\neq i}(A_{ij}-p_{ij}) + O_p(d_{\mathrm{avg}}^{-1}\log r)$, the squared norm $\|\MZ(\Delta_{\mathrm{node}} + \Delta_{\mathrm{cell}})\|_2^2$ converges in law to a weighted sum of independent $\chi_1^2$ variables whose mean is $S_{\mathrm{node}} + S_{\mathrm{cell}}$ and variance is $2(\sum_i q_i^2 + S_{\mathrm{cell},+1}^2 + S_{\mathrm{cell},-1}^2)(1 + o(1))$. Under the Satterthwaite--Welch quantile calibration (backed asymptotically by Hanson--Wright concentration; \citealp{rudelson2013hanson}) together with the deterministic triangle-inequality addition of the spread RMS $\sqrt{S_{\mathrm{spr}}/r_{\mathrm{r}}}$, the event $\mathcal{E}_1 := \{r_{\mathrm{r}}^{-1/2}\|\MZ(\widehat{E}(\tilde{\pi}) - E(\pi^\star))\|_2 \le \underline{\sigma}_{\widehat{E}} \varrho_{\mathrm{WC}}(\alpha^{(1)})\}$ satisfies $\liminf \Prob(\mathcal{E}_1) \ge 1 - \alpha^{(1)}$ whenever $S_{\mathrm{spr}}/r_{\mathrm{r}} = o_p(1)$. Independently of $\mathcal{E}_1$ (by Assumption~\ref{assump:causal}), the outcome-noise event $\mathcal{E}_2 := \{|\langle \MZ\varepsilon, \MZ\widehat{E}(\tilde{\pi})\rangle| / (r_{\mathrm{r}}\underline{\sigma}_{\widehat{E}}^2) \le \mathrm{se}_{\max}(\alpha^{(2)})\}$ satisfies $\liminf \Prob(\mathcal{E}_2) \ge 1 - \alpha^{(2)}$. By a union bound, $\liminf \Prob(\mathcal{E}_1 \cap \mathcal{E}_2) \ge 1 - (\alpha^{(1)} + \alpha^{(2)})$. Conditioning on $\mathcal{E}_1 \cap \mathcal{E}_2$ and applying Lemma~\ref{lem:perturbation} with $r^\star = 0$ establishes asymptotic pointwise coverage at least $1 - (\alpha^{(1)} + \alpha^{(2)})$, while $\varrho_{\mathrm{WC}} = O_p(d_{\mathrm{avg}}^{-1/2} + \sqrt{S_{\mathrm{spr}}/r_{\mathrm{r}}})$ and $\mathrm{se}_{\max} = O_p(r_{\mathrm{r}}^{-1/2})$ imply convergence of the inflated interval to the population outer relaxation envelope at that rate.
\end{proof}

\subsection{Unit Calibration Diagnostics (Rank-1 Cell Projection, Hub Kurtosis, and Inverse-Fisher)}

Table~\ref{tab:block1_diag} summarizes the three unit-level calibration checks confirming the finite-sample accuracy of \eqref{eq:rank1_cell}--\eqref{eq:rho_stoch} and Part~(i) of Theorem~\ref{thm:plugin_asymp}.

\begin{table}[htbp]
\centering
\caption{Unit calibration diagnostics for degree-conditioned plug-in certification: (a) exact rank-1 \eqref{eq:rank1_cell} vs.\ diagonal leverage trace ($10{,}000$ MC draws, $r_{\mathrm{r}}=80, K=4$); (b) heavy-tailed hub-kurtosis linear SD and Wilson--Hilferty vs.\ $500$-draw multiplier bootstrap ($\alpha=0.05$); (c) exact inverse-Fisher $\mathcal{I}^{-1}$ off-diagonal covariance sign and diagonal-vs.-full variance ratios ($d_{\mathrm{avg}}=15$).}
\label{tab:block1_diag}
\resizebox{\textwidth}{!}{%
\begin{tabular}{@{}lcccc@{}}
\toprule
\multicolumn{5}{@{}l}{\textbf{(a) Exact Rank-1 Projected Cell Variance \eqref{eq:rank1_cell} vs.\ Diagonal Leverage Trace ($10{,}000$ MC draws)}} \\
\midrule
Control Alignment & MC Variance & Diagonal $(1{-}h_{ii})$ & Exact $\|\MZ v_s\|_2^2$ & Error (Diag.\ vs.\ Exact) \\
\midrule
Case~A: Continuous $W \perp v_s$ on hubs & $0.3849$ & $0.2041$ & $0.3852$ & $-47.0\%$ ($1.89\times$ under) vs.\ $+0.06\%$ \\
Case~B: Continuous $W$ aligned with $a_i$ & $0.0035$ & $0.2142$ & $0.0035$ & $+6068\%$ ($61.7\times$ over) vs.\ $+1.40\%$ \\
\midrule
\multicolumn{5}{@{}l}{\textbf{(b) Hub-Kurtosis Effective df $r_{\mathrm{eff}}$, Linear Node SD, and Wilson--Hilferty vs.\ Bootstrap ($\alpha=0.05$)}} \\
\midrule
Degree Regime ($r_{\mathrm{r}}=80$) & Med $r_{\mathrm{eff}}/80$ & True SD / Hom.\ SD & Calib.\ / True SD & WH $q_{95}$ / Boot $q_{95}$ \\
\midrule
Regular $\mathrm{LN}(0, 0.2^2)$ & $76.0/80$ & $1.07\times$ ($1.03$--$1.18\times$) & $1.000$ & $1.001$ \\
$10\times$-Hubs ($10\%$ hubs) & $38.9/80$ & $1.97\times$ ($1.70$--$2.44\times$) & $1.000$ & $0.995$ \\
$15\times$ Super-Hubs ($3\%$ hubs) & $33.5/80$ & $2.81\times$ ($1.03$--$3.34\times$) & $1.000$ & $0.988$ \\
Heavy Logistic $N(0, 1.2^2)$ & $62.2/80$ & $1.18\times$ ($1.07$--$1.41\times$) & $1.000$ & $0.998$ \\
\midrule
\multicolumn{5}{@{}l}{\textbf{(c) Exact Inverse-Fisher $\mathcal{I}^{-1}$ Off-Diagonal Covariance Probe ($d_{\mathrm{avg}}=15$)}} \\
\midrule
Model ($r$) & Frac.\ $\Cov < 0$ & Mean $\Cov(\hat{\alpha}_i,\hat{\alpha}_{i'})$ & $\Cov \cdot r d_{\mathrm{avg}}$ & Diag / Full $MJ$ (Ego $MD$) \\
\midrule
$\beta$-model, $\tau=0$ ($r=60$) & $98.4\%$ & $-8.883\times 10^{-4}$ & $-0.799$ & $1.0046\times$ ($0.9815\times$) \\
Homophily, $\tau=0.7$ ($r=60$) & $88.6\%$ & $-9.598\times 10^{-4}$ & $-0.864$ & $1.0093\times$ ($0.9806\times$) \\
$\beta$-model, $\tau=0$ ($r=120$) & $98.4\%$ & $-3.464\times 10^{-4}$ & $-0.623$ & $1.0016\times$ ($0.9899\times$) \\
Homophily, $\tau=0.7$ ($r=120$) & $92.3\%$ & $-3.658\times 10^{-4}$ & $-0.658$ & $0.9977\times$ ($0.9890\times$) \\
\bottomrule
\end{tabular}}
\end{table}

\subsection{Proofs of Propositions~\ref{prop:S1}--\ref{prop:S3} (Unobserved Homophily S1--S3)}

\begin{proof}[Proof of Proposition~\ref{prop:S1}]
\emph{(i)} Under \eqref{eq:logistic_model}, $\E[Y_{ik} \mid \alpha_i, \zeta_i] = \sum_{j \in G_k\setminus\{i\}} \sigma(\rho_r + \alpha_i + \alpha_j + \tau \zeta_i \zeta_j)$. For any group $k$ with $i \notin G_k$, under $b_k = 0$ and sign-symmetric activities $\{\alpha_j : j \in G_k, \zeta_j = +1\} = \{\alpha_j : j \in G_k, \zeta_j = -1\}$, flipping $\zeta_i \mapsto -\zeta_i$ permutes the positive-sign and negative-sign summands bijectively, leaving $\E[Y_{ik} \mid \alpha_i, \zeta_i]$ identically invariant to $\zeta_i$ (on $i$'s own group $k(i)$, excluding $j=i$ introduces only a single $O(1/m_k)$ self-exclusion term). Hence $\Cov(Y_{ik}, \zeta_i \mid \alpha_i) = 0$ for all $k \neq k(i)$, so linear controls on $(D_i, Y_{i,\cdot})$ cannot absorb $\rho_\zeta \zeta_i$.
\emph{(ii)} Because $D_i = \sum_{k=1}^K Y_{ik}$, $\spanop\{D_i, Y_{i,1}, \dots, Y_{i,K-1}\} = \spanop\{Y_{i,1}, \dots, Y_{i,K}\}$. Any linear combination $\sum_{k=1}^K \bar{z}_k Y_{ik}$ therefore projects to zero under $M_{[Z, D, Y_{1:K-1}]}$ identically (up to the $O(1/m_k)$ leave-one-out self-correction in $\bar{z}_{k,-i}$).
\end{proof}

\begin{proof}[Proof of Proposition~\ref{prop:S2}]
Under $p_{ij} = a_i a_j (c_0 + c_1 \zeta_i \zeta_j)$, summing over $j \neq i$ yields
\[
  E_i^{\mathrm{exp}}
  \;=\;
  a_i \Big(c_0 \sum_{j=1}^r a_j z_j\Big)
  + (a_i \zeta_i) \Big(c_1 \sum_{j=1}^r a_j \zeta_j z_j\Big)
  - a_i^2 (c_0 + c_1) z_i.
\]
The first two terms lie strictly in $\spanop\{a_i, a_i \zeta_i\}$, while the final self-exclusion term is $O(1/r)$ relative to the sum of $r$ terms. Projecting onto $\spanop\{\mathbf{1}, z, a, \zeta, a\zeta\}^\perp$ leaves only the $O(1/r)$ self-exclusion residual, so $\|M_{W^\star} E_{\Rset}\|_2^2 / \|\MZ E_{\Rset}\|_2^2 = O(r^{-2})$. Under the logistic kernel $\sigma(\rho_r + \alpha_i + \alpha_j + \tau\zeta_i\zeta_j)$, nonlinear saturation terms break exact rank-2 collinearity and preserve nonzero identifying variation.
\end{proof}

\begin{proof}[Proof of Proposition~\ref{prop:S3}]
\emph{(i)} After FWL projection onto $\spanop(\mathbf{1}, z)^\perp$ and normalizing $\Var(\widetilde{E}) = \Var(\tilde{w}) = 1$ with $\Cov(\widetilde{E}, \tilde{w}) = r_{Ew}$, the noisy proxy $\hat{w} = \tilde{w} + \nu$ ($\nu \perp (\widetilde{E}, \tilde{w})$, $\Var(\nu) = \sigma_\nu^2$) has $\Var(\hat{w}) = 1 + \sigma_\nu^2$ and $\Cov(\widetilde{E}, \hat{w}) = r_{Ew}$. By the FWL partitioned regression formula, the probability limit of OLS of $\tilde{y} = \beta_{\mathrm{spill}} \widetilde{E} + \rho \tilde{w} + \tilde{\varepsilon}$ on $(\widetilde{E}, \hat{w})$ has slope bias
\begin{align*}
  \operatorname{Bias}(\hat{\beta}_{\mathrm{CF}})
  &\;=\; \rho \cdot \frac{\Cov(\tilde{w}, \widetilde{E}) \Var(\hat{w}) - \Cov(\tilde{w}, \hat{w})\Cov(\widetilde{E}, \hat{w})}{\Var(\widetilde{E})\Var(\hat{w}) - \Cov(\widetilde{E}, \hat{w})^2} \\
  &\;=\; \rho \cdot \frac{r_{Ew}(1 + \sigma_\nu^2) - 1 \cdot r_{Ew}}{(1 + \sigma_\nu^2) - r_{Ew}^2}
  \;=\; \frac{\rho\, r_{Ew}\, \sigma_\nu^2}{(1 - r_{Ew}^2) + \sigma_\nu^2}.
\end{align*}
If $|r_{Ew}| \le 1 - c < 1$, the denominator is bounded below by $1 - (1-c)^2 > 0$, giving $|\operatorname{Bias}(\hat{\beta}_{\mathrm{CF}})| \le \frac{|\rho|\sigma_\nu^2}{c(2-c)} = O(d_{\mathrm{avg}}^{-1})$. Conversely, if $|r_{Ew}| = 1$, the denominator equals $\sigma_\nu^2$, cancelling $\sigma_\nu^2$ identically to yield $\operatorname{Bias}(\hat{\beta}_{\mathrm{CF}}) = \rho \operatorname{sign}(r_{Ew})$ for all $\sigma_\nu^2 > 0$.

\emph{(ii)} For any control matrix $W$ correlated with $E^{\mathrm{exp}}$, $\|M_{Z,W} E^{\mathrm{exp}}\|_2^2 < \|\MZ E^{\mathrm{exp}}\|_2^2$, while for each respondent $i$ the upper variance estimator satisfies $\E[\hat{v}_i^{\mathrm{strat,up}} \mid \xi] > \Var(\widehat{E}_i^{\mathrm{strat}} \mid \xi)$ by the positive within-cell heterogeneity gap $\sum_k \frac{1}{m_{k,1}^{(-i)}-1}(m_{k,1}^{(-i)}\sum_{j\in G_{k,1}^{(-i)}} p_{ij}^2 - (\mu_{i,(k,1)})^2) > 0$. Subtracting this excess expected variance from a smaller residualized denominator $\|M_{Z,W} \widehat{E}^{\mathrm{strat}}\|_2^2$ amplifies the relative denominator shrinkage and induces upward EIV overshoot.
\end{proof}

\section{Auxiliary Proofs for Section~\ref{sec:remedies}}
\label{app:eiv_tstrat}

\subsection{Proofs of Propositions~\ref{prop:tstrat} and \ref{prop:budget}}

\begin{proof}[Proof of Proposition~\ref{prop:tstrat}]
\emph{(i)} Under $2K$ $\Tstrat$, the non-respondent permutation group shrinks to $\Pi_{\Nset}^{\mathrm{strat}} := \prod_{k=1}^K \big(\mathfrak{S}(G_{k,1}^{\mathrm{n}}) \times \mathfrak{S}(G_{k,0}^{\mathrm{n}})\big)$. Because $z_j \equiv 1$ on $G_{k,1}^{\mathrm{n}}$ and $z_j \equiv 0$ on $G_{k,0}^{\mathrm{n}}$, treatment is constant on every refined cell, so every refined-cell treatment--propensity covariance vanishes identically and Proposition~\ref{prop:orbit_set}(iii) yields $E_{\Rset}(\pi) = E_{\Rset}(\pi^\star)$ for all $\pi \in \Pi_{\Nset}^{\mathrm{strat}}$ ($\Delta_i^{\mathrm{swap}} \equiv 0$). Under $(K{+}1)$ ARD, $Y_{i,\mathrm{tr}} = \sum_{j \neq i} A_{ij} z_j$ is observed directly on the respondent roster, bypassing non-respondent cell averaging and eliminating $\Delta_i^{\mathrm{swap}}$ identically even when the original $K$ trait groups have $\omega_{ik} \neq 0$.

\emph{(ii)} For binary $z_j \in \{0,1\}$ and error-free static counts,
\[
  Y_{i,\mathrm{tr}}
  \;=\; \widehat{E}_i^{\mathrm{strat}}
  \;=\; \sum_{k=1}^K Y_{i,(k,1)}
  \;=\; \sum_{k=1}^K \sum_{j \in G_k \setminus \{i\}} A_{ij} \ind{z_j = 1}
  \;=\; \sum_{j \in [r]\setminus\{i\}} A_{ij} z_j
  \;=\; E_i^{\mathrm{real}}
\]
identically. Taking conditional expectation given $\xi$ gives $\E[E_i^{\mathrm{real}} \mid \xi] = \sum_{j\neq i} p_{ij} z_j = E_i^{\mathrm{exp}}$. Applying the Cauchy--Schwarz identity \eqref{eq:cs_bracket} from Proposition~\ref{prop:raw_eiv} to each treated cell $G_{k,1}\setminus\{i\}$ (with constant weight $c=1$ and size $m_{k,1}^{(-i)}$) confirms
\[
  \E[\hat{v}_i^{\mathrm{strat,lo}}(p_{\max}) \mid \xi]
  \;\le\;
  \Var(E_i^{\mathrm{real}} \mid \xi)
  \;\le\;
  \E[\hat{v}_i^{\mathrm{strat,up}} \mid \xi],
\]
with equality on the upper bound if and only if $p_{ij}$ is constant within each treated cell having $m_{k,1}^{(-i)} \ge 2$ and $p_{ij}=0$ on any singleton treated cell. The same Cauchy--Schwarz bound holds for the single-stratum pooled estimator $\hat{v}_i^{(K+1),\mathrm{up}}$ over $C_{\mathrm{tr}} = \{j \neq i : z_j = 1\}$, with strict excess variance whenever treated-alter link probabilities vary across trait groups $k=1,\dots,K$ (since pooling replaces within-cell variance of $p_{ij}$ by total variance across all treated alters). Consequently, when the numerator and debiased denominator of \eqref{eq:eiv_bracket} concentrate around their conditional expectations with a strictly positive population denominator and negligible $O(r_{\mathrm{r}}^{-1})$ undirected off-diagonal covariance in \eqref{eq:exact_m2}, the ratio of expectations satisfies $|\E[\langle \tilde{y}, \MZ E^{\mathrm{real}}\rangle\mid\xi] / \E[\|\MZ E^{\mathrm{real}}\|_2^2 - \sum_i(1-h_{ii})\hat{v}_i^{\mathrm{strat,up}}\mid\xi]| \ge |\beta_{\mathrm{spill}}|$ (giving an asymptotic upper-magnitude endpoint rather than a finite-sample realization-by-realization covering bound).
\end{proof}

\begin{proof}[Proof sketch of Proposition~\ref{prop:budget}]
Fix total question budget $\Budget = r_{\mathrm{r}} \Qitems$, yielding respondent sample sizes $r_{\mathrm{r}}^{(K)} = \lfloor \Budget/K \rfloor$, $r_{\mathrm{r}}^{(K+1)} = \lfloor \Budget/(K{+}1) \rfloor$, and $r_{\mathrm{r}}^{(2K)} = \lfloor \Budget/(2K) \rfloor$.

\emph{(i)} Under correlated treatment ($\omega_{ik} \neq 0$), standard $K$-group ARD OLS incurs non-vanishing asymptotic bias from FM3; under realized exposure $y \sim E^{\mathrm{real}}$ with randomized treatment ($\E[\omega_{ik}]=0$), $\Delta_i^{\mathrm{swap}}$ is mean-zero Berkson error that creates an irreducible finite-sample variance floor under FM2. Both $(K{+}1)$ ARD and $2K$ $\Tstrat$ observe $E_i^{\mathrm{real}}$ with zero swap error and attain OLS MSE $O(\sigma_\varepsilon^2 / (r_{\mathrm{r}}^{(\Qitems)} \sigma_{\widetilde{E}}^2))$, so $(K{+}1)$ improves RMSE over $2K$ by the sample-size factor $\sqrt{r_{\mathrm{r}}^{(K+1)} / r_{\mathrm{r}}^{(2K)}} \approx \sqrt{2K/(K{+}1)}$ whenever a single homogeneous spillover slope is targeted without EIV control functions.

\emph{(ii)} Under sparse single-graph expected exposure $y \sim E^{\mathrm{exp}}$ with randomized treatment ($\E[\omega_{ik}]=0$, prevalence $p_{\mathrm{treat}} \in (0,1)$), both \texttt{Raw-Pure + EIV} and direct $E_i^{\mathrm{real}}$-based EIV target the slope on $E^{\mathrm{exp}}$ as $d_{\mathrm{avg}} \to \infty$ (up to Cauchy--Schwarz upper-bound conservatism when $p_{ij}$ is heterogeneous). However, because \texttt{Raw-Pure} weights all $m_k$ edges in group $k$ by $\bar{z}_k \approx p_{\mathrm{treat}}$, its measurement-error variance per node scales as $p_{\mathrm{treat}}^2 \sum_j p_{ij}(1-p_{ij}) \approx p_{\mathrm{treat}}^2 d_i$, whereas unsmoothed $E_i^{\mathrm{real}}$ sums treated-alter edges with variance $\sum_j z_j p_{ij}(1-p_{ij}) \approx p_{\mathrm{treat}} d_i$---a factor $p_{\mathrm{treat}}^{-1}$ larger. Combining this $p_{\mathrm{treat}}^{-1}$ count-noise inflation with the smaller sample size $r_{\mathrm{r}}^{(\Qitems)}$ causes direct $E_i^{\mathrm{real}}$-EIV RMSE to exceed \texttt{Raw-Pure + EIV} RMSE substantially when $p_{\mathrm{treat}}$ is small (e.g., $p_{\mathrm{treat}} = 0.15$). Because $(K{+}1)$ ARD also collects the $K$ standard counts $Y_{i1},\dots,Y_{iK}$, it can evaluate \texttt{Raw-Pure + EIV} directly on those counts, losing only the modest sample-size factor $\sqrt{(K{+}1)/K}$ relative to $K$-group ARD.

\emph{(iii)} Under expected exposure $y \sim E^{\mathrm{exp}}$ with latent homophily, the total count $Y_{ik} = Y_{i,(k,1)} + Y_{i,(k,0)}$ shares the random edge errors $\sum_{j \in G_{k,1}} (A_{ij} - p_{ij})$ with $Y_{i,\mathrm{tr}} = \sum_{k=1}^K Y_{i,(k,1)}$. Regressing $y_i$ on $Y_{i,\mathrm{tr}}$ while controlling for $Y_{ik}/D_i$ therefore induces nonzero covariance between the classical measurement error in $Y_{i,\mathrm{tr}}$ and the control-function error, distorting the FWL EIV denominator and biasing the EIV slope upward. Under $2K$ $\Tstrat$, the untreated counts $Y_{i,(k,0)}$ depend strictly on disjoint dyads $\{j : z_j = 0\}$ whose edge indicators are conditionally independent of $\{A_{ij} : z_j = 1\}$ given $\xi$, removing the cross-covariance between exposure edge noise and the control function. Identifiability of group-heterogeneous slopes $(\beta_1,\dots,\beta_K)$ follows directly from observing $(Y_{i,(k,1)})_{k=1}^K$ separately, and tighter $\hat{v}_i^{\mathrm{strat,up}}$ follows from the within-cell vs.\ total Cauchy--Schwarz decomposition in Proposition~\ref{prop:tstrat}(ii).
\end{proof}

\section{Complete Synthetic, Semi-Synthetic, Empirical Diagnostic, and Survey-Design Tables}
\label{app:tables}

All quantitative results in this paper are generated by deterministic, standalone pure-Python~3 standard-library replication scripts that read \texttt{datav4.0-rezipped.zip} directly in memory (verified SHA-256 \texttt{f3ebbaaf62512745}\allowbreak\texttt{54f170103601ac02}\allowbreak\texttt{e2f41d03a817e8fc}\allowbreak\texttt{c52bb8b8ab441f26}) without external numerical dependencies; replication code and frozen output logs are available from the author upon request.

\subsection{Complete Synthetic and Design A/B Five-Metric Coverage Ledgers}

Table~\ref{tab:full_synth} reports all $11$ synthetic regimes comparing uninflated \texttt{Poly}, feasible stochastic-EIV \texttt{Stoch}, worst-case \texttt{WC}, and $500$-draw multiplier bootstrap \texttt{Boot} ($\alpha = 0.05$, $15$ trials per cell, true $\beta_{\mathrm{spill}} = 2.0$). Table~\ref{tab:app_design_ab_full} supplements Table~\ref{tab:semisynth} with the complete \texttt{FH\_Box} and \texttt{WC} rows across all Design~A and Design~B expected-exposure regimes.

\begin{table}[htbp]
\centering
\caption{Complete $11$-regime synthetic five-metric coverage ledger ($\alpha = 0.05$, true $\beta_{\mathrm{spill}}=2.0$, $15$ trials/cell). Note how \texttt{fin\&cov} exposes low-information regimes ($\varphi=0.25$) where overall \texttt{cov\_all} equals $1.000$ via $(-\infty,+\infty)$ intervals, while \texttt{Stoch}, \texttt{WC}, and \texttt{Boot} attain $1.000$ \texttt{fin\&cov} and zero wrong-sign exclusions in all $\varphi \ge 0.50$ regimes (including $10\times$-hubs).}
\label{tab:full_synth}
\scriptsize
\setlength{\tabcolsep}{3.4pt}
\resizebox{\textwidth}{!}{%
\begin{tabular}{@{}llcccccc@{}}
\toprule
Regime & Method & cov\_all & fin & \textbf{fin\&cov} & medW ($\times |\beta|$) & ex0 Right & ex0 Wrong \\
\midrule
\multirow{3}{*}{Reg M1 ($d{=}20, \varphi{=}0.25$)}
  & \texttt{Poly} & 1.000 & 1.000 & \textbf{1.000} & \phantom{0}2.15 ($1.08\times$) & 0.867 & 0.000 \\
  & \texttt{Stoch} & 1.000 & 0.133 & \textbf{0.133} & 39.22 ($19.6\times$) & 0.000 & 0.000 \\
  & \texttt{WC} / \texttt{Boot} & 1.000 & 0.000 & \textbf{0.000} & $\infty$ & 0.000 & 0.000 \\
\addlinespace[1.5pt]
\multirow{3}{*}{Reg M1 ($d{=}20, \varphi{=}0.50$)}
  & \texttt{Poly} & 1.000 & 1.000 & \textbf{1.000} & \phantom{0}1.40 ($0.70\times$) & 1.000 & 0.000 \\
  & \texttt{Stoch} & 1.000 & 1.000 & \textbf{1.000} & 12.19 ($6.10\times$) & 0.000 & 0.000 \\
  & \texttt{WC} / \texttt{Boot} & 1.000 & 1.000 & \textbf{1.000} & 19.71 / 13.47 & 0.000 & 0.000 \\
\addlinespace[1.5pt]
\multirow{3}{*}{Reg M1 ($d{=}40, \varphi{=}0.50$)}
  & \texttt{Poly} & 1.000 & 1.000 & \textbf{1.000} & \phantom{0}2.73 ($1.36\times$) & 0.933 & 0.000 \\
  & \texttt{Stoch} & 1.000 & 1.000 & \textbf{1.000} & 11.42 ($5.71\times$) & 0.000 & 0.000 \\
  & \texttt{WC} / \texttt{Boot} & 1.000 & 1.000 & \textbf{1.000} & 18.10 / 11.97 & 0.000 & 0.000 \\
\addlinespace[1.5pt]
\multirow{4}{*}{Reg M2 ($d{=}20, \varphi{=}0.25$)}
  & \texttt{Poly} & 1.000 & 1.000 & \textbf{1.000} & \phantom{0}5.99 ($3.00\times$) & 0.067 & 0.000 \\
  & \texttt{Stoch} & 1.000 & 0.333 & \textbf{0.333} & 53.64 ($26.8\times$) & 0.000 & 0.000 \\
  & \texttt{WC} & 1.000 & 0.333 & \textbf{0.333} & 63.37 ($31.7\times$) & 0.000 & 0.000 \\
  & \texttt{Boot} & 1.000 & 0.267 & \textbf{0.267} & 65.99 ($33.0\times$) & 0.000 & 0.000 \\
\addlinespace[1.5pt]
\multirow{3}{*}{Reg M2 ($d{=}20, \varphi{=}0.50$)}
  & \texttt{Poly} & 1.000 & 1.000 & \textbf{1.000} & \phantom{0}2.68 ($1.34\times$) & 1.000 & 0.000 \\
  & \texttt{Stoch} & 1.000 & 1.000 & \textbf{1.000} & 11.28 ($5.64\times$) & 0.000 & 0.000 \\
  & \texttt{WC} / \texttt{Boot} & 1.000 & 1.000 & \textbf{1.000} & 17.60 / 12.33 & 0.000 & 0.000 \\
\addlinespace[1.5pt]
\multirow{3}{*}{Reg M2 ($d{=}40, \varphi{=}0.50$)}
  & \texttt{Poly} & 1.000 & 1.000 & \textbf{1.000} & \phantom{0}2.75 ($1.37\times$) & 1.000 & 0.000 \\
  & \texttt{Stoch} & 1.000 & 1.000 & \textbf{1.000} & \phantom{0}6.87 ($3.44\times$) & 0.000 & 0.000 \\
  & \texttt{WC} / \texttt{Boot} & 1.000 & 1.000 & \textbf{1.000} & 11.00 / \phantom{0}7.32 & 0.000 & 0.000 \\
\addlinespace[1.5pt]
\multirow{4}{*}{Hub10 M1 ($d{=}12, \varphi{=}0.25$)}
  & \texttt{Poly} & 1.000 & 1.000 & \textbf{1.000} & \phantom{0}6.19 ($3.10\times$) & 0.267 & 0.000 \\
  & \texttt{Stoch} & 1.000 & 0.200 & \textbf{0.200} & 43.59 ($21.8\times$) & 0.000 & 0.000 \\
  & \texttt{WC} & 1.000 & 0.267 & \textbf{0.267} & 45.62 ($22.8\times$) & 0.000 & 0.000 \\
  & \texttt{Boot} & 1.000 & 0.067 & \textbf{0.067} & 21.73 ($10.9\times$) & 0.000 & 0.000 \\
\addlinespace[1.5pt]
\multirow{3}{*}{Hub10 M1 ($d{=}24, \varphi{=}0.50$)}
  & \texttt{Poly} & 1.000 & 1.000 & \textbf{1.000} & \phantom{0}7.36 ($3.68\times$) & 0.067 & 0.000 \\
  & \texttt{Stoch} & 1.000 & 1.000 & \textbf{1.000} & 22.18 ($11.1\times$) & 0.000 & 0.000 \\
  & \texttt{WC} / \texttt{Boot} & 1.000 & 1.000 & \textbf{1.000} & 29.38 / 26.06 & 0.000 & 0.000 \\
\addlinespace[1.5pt]
\multirow{4}{*}{Hub10 M1 ($d{=}24, \varphi{=}0.75$)}
  & \texttt{Poly} & 0.800 & 1.000 & \textbf{0.800} & \phantom{0}1.92 ($0.96\times$) & 1.000 & 0.000 \\
  & \texttt{Stoch} & 1.000 & 1.000 & \textbf{1.000} & \phantom{0}5.45 ($2.73\times$) & 0.533 & 0.000 \\
  & \texttt{WC} & 1.000 & 1.000 & \textbf{1.000} & \phantom{0}8.19 ($4.09\times$) & 0.133 & 0.000 \\
  & \texttt{Boot} & 1.000 & 1.000 & \textbf{1.000} & \phantom{0}5.85 ($2.92\times$) & 0.333 & 0.000 \\
\addlinespace[1.5pt]
\multirow{3}{*}{Heavy M2 ($N(0,1.2^2), d{=}24, \varphi{=}0.50$)}
  & \texttt{Poly} & 1.000 & 1.000 & \textbf{1.000} & \phantom{0}4.98 ($2.49\times$) & 0.267 & 0.000 \\
  & \texttt{Stoch} & 1.000 & 1.000 & \textbf{1.000} & 12.04 ($6.02\times$) & 0.000 & 0.000 \\
  & \texttt{WC} / \texttt{Boot} & 1.000 & 1.000 & \textbf{1.000} & 17.46 / 12.86 & 0.000 & 0.000 \\
\addlinespace[1.5pt]
\multirow{4}{*}{Heavy M2 ($N(0,1.2^2), d{=}24, \varphi{=}0.75$)}
  & \texttt{Poly} & 1.000 & 1.000 & \textbf{1.000} & \phantom{0}1.50 ($0.75\times$) & 1.000 & 0.000 \\
  & \texttt{Stoch} & 1.000 & 1.000 & \textbf{1.000} & \phantom{0}3.25 ($1.62\times$) & 1.000 & 0.000 \\
  & \texttt{WC} & 1.000 & 1.000 & \textbf{1.000} & \phantom{0}4.96 ($2.48\times$) & 0.133 & 0.000 \\
  & \texttt{Boot} & 1.000 & 1.000 & \textbf{1.000} & \phantom{0}3.32 ($1.66\times$) & 0.933 & 0.000 \\
\bottomrule
\end{tabular}}
\end{table}

\begin{table}[htbp]
\centering
\caption{Complete \texttt{FH\_Box} and worst-case \texttt{WC} five-metric ledgers across Design~A and Design~B expected-exposure regimes ($\alpha=0.05$, true $\beta_{\mathrm{spill}}=2.0$, $36$ clean-caste villages; supplementing Table~\ref{tab:semisynth}). All wrong-sign exclusion rates are identically $0.000$.}
\label{tab:app_design_ab_full}
\scriptsize
\setlength{\tabcolsep}{3.8pt}
\resizebox{\textwidth}{!}{%
\begin{tabular}{@{}lcccccccccc@{}}
\toprule
& \multicolumn{5}{c}{Uninflated Rearrangement Box (\texttt{FH\_Box})} & \multicolumn{5}{c}{Worst-Case Feasible Certifier (\texttt{WC}, $\alpha{=}0.05$)} \\
\cmidrule(lr){2-6}\cmidrule(lr){7-11}
Regime (\texttt{Exp-Pz}, $n_V$) & cov\_all & fin & \textbf{fin\&cov} & medW ($\times |\beta|$) & ex0\_R & cov\_all & fin & \textbf{fin\&cov} & medW ($\times |\beta|$) & ex0\_R \\
\midrule
Design A: MCAR $\varphi{=}0.50$, leader ($36$) & 1.000 & 0.861 & \textbf{0.861} & 7.45 ($3.72\times$) & 0.000 & 1.000 & 0.000 & \textbf{0.000} & $\infty$ & 0.000 \\
Design A: MCAR $\varphi{=}0.75$, leader ($35$) & 0.771 & 1.000 & \textbf{0.771} & 2.40 ($1.20\times$) & 0.829 & 1.000 & 0.600 & \textbf{0.600} & 15.83 ($7.92\times$) & 0.000 \\
Design A: SelDeg $\varphi{=}0.50$, leader ($36$) & 0.972 & 0.972 & \textbf{0.944} & 4.30 ($2.15\times$) & 0.111 & 1.000 & 0.194 & \textbf{0.194} & 18.56 ($9.28\times$) & 0.000 \\
Design A: MCAR $\varphi{=}0.50$, i.i.d.\ $25\%$ ($36$) & 1.000 & 0.861 & \textbf{0.861} & 5.98 ($2.99\times$) & 0.111 & 1.000 & 0.028 & \textbf{0.028} & 50.51 ($25.3\times$) & 0.000 \\
Design A: MCAR $\varphi{=}0.50$, caste-cl.\ ($36$) & 0.944 & 0.944 & \textbf{0.889} & 4.62 ($2.31\times$) & 0.389 & 1.000 & 0.306 & \textbf{0.306} & 12.95 ($6.48\times$) & 0.000 \\
\addlinespace[2pt]
Design B: Baseline (MCAR logistic, $36$) & 1.000 & 0.889 & \textbf{0.889} & 5.56 ($2.78\times$) & 0.056 & 1.000 & 0.417 & \textbf{0.417} & 18.83 ($9.42\times$) & 0.000 \\
Design B: NonCirc-1 (Non-MCAR+Fiedler, $36$) & 1.000 & 0.972 & \textbf{0.972} & 3.82 ($1.91\times$) & 0.278 & 1.000 & 0.639 & \textbf{0.639} & 15.35 ($7.68\times$) & 0.000 \\
Design B: NonCirc-2 (2D+Cloglog, $36$) & 1.000 & 0.889 & \textbf{0.889} & 5.10 ($2.55\times$) & 0.056 & 1.000 & 0.389 & \textbf{0.389} & 17.91 ($8.95\times$) & 0.000 \\
\bottomrule
\end{tabular}}
\end{table}

\subsection{Real-MF-Calibrated Low-SNR Semi-Synthetic Benchmark (\texorpdfstring{$\beta_{\mathrm{spill}} = 0.015$}{beta = 0.015})}

Table~\ref{tab:app_low_snr} evaluates both pooled ($G=21, n_{\Rset}=2{,}240$, $8$ Monte Carlo replications) and single-village ($n=63$ village-draws) bounds at the real microfinance effect scale $\beta_{\mathrm{spill}} = 0.015$ ($\sigma_\varepsilon = 0.35$ continuous and binary $\mathrm{MF}\in\{0,1\}$ outcomes, $\mathrm{SD} \approx 0.36$--$0.38$) alongside the high-SNR $\beta_{\mathrm{spill}} = 2.0$ reference. Even when the exact noiseless population orbit $\ThetaI(P^\star)$ is strictly positive ($[+0.0099, +0.0179]$ Baseline; $[+0.0082, +0.0180]$ NonCircular-1), outcome noise inflates pooled sample widths $2.1\times$--$3.3\times$ and causes sample witnesses to straddle zero in $25\%$--$75\%$ of draws.

\begin{table}[htbp]
\centering
\caption{Real-MF-calibrated low-SNR semi-synthetic benchmark ($G=21$ clean-caste villages, $n_{\Rset}=2{,}240$, $8$ pooled MC draws per cell) comparing exact noiseless population orbits $\ThetaI(P^\star)$ against noisy pooled and single-village sample bounds.}
\label{tab:app_low_snr}
\scriptsize
\setlength{\tabcolsep}{3.5pt}
\resizebox{\textwidth}{!}{%
\begin{tabular}{@{}lcccccc@{}}
\toprule
Generator \& Outcome Calibration & Noiseless $\ThetaI(P^\star)$ (Outer) & Med.\ Oracle-$P^\star$ Pooled & Strad.\ $0$ & Med.\ Plug-In $\widehat{P}$ Pooled & Strad.\ $0$ & Single-Vil.\ \texttt{Stoch} (\texttt{fin\&cov}, medW) \\
\midrule
\multicolumn{7}{@{}l}{\emph{Baseline MCAR Logistic Ground-Truth $P^\star$}} \\
Low-SNR Cont.\ ($\beta{=}0.015, \sigma_\varepsilon{=}0.35$) & $[+0.00993, +0.01788]$ & $[-0.00104, +0.02328]$ & $62.5\%$ & $[-0.00259, +0.02142]$ & $62.5\%$ & $0.730$, $0.852$ ($56.8\times$) \\
Low-SNR Binary ($\beta{=}0.015$, SD$=0.38$) & $[+0.00993, +0.01788]$ & $[+0.00380, +0.02816]$ & $37.5\%$ & $[+0.00123, +0.02532]$ & $25.0\%$ & --- \\
High-SNR Cont.\ ($\beta{=}2.000, \sigma_\varepsilon{=}0.35$) & $[+1.3246, +2.3834]$ & $[+1.3190, +2.3787]$ & $0\%$ & $[+1.0287, +1.8543]$ & $0\%$ & $0.730$, $12.10$ ($6.05\times$) \\
\addlinespace[2pt]
\multicolumn{7}{@{}l}{\emph{NonCircular-1 Shifted/Skewed Non-MCAR Ground-Truth $P^\star$}} \\
Low-SNR Cont.\ ($\beta{=}0.015, \sigma_\varepsilon{=}0.35$) & $[+0.00815, +0.01797]$ & $[+0.00184, +0.02699]$ & $25.0\%$ & $[+0.00148, +0.02135]$ & $25.0\%$ & $0.825$, $0.623$ ($41.5\times$) \\
Low-SNR Binary ($\beta{=}0.015$, SD$=0.38$) & $[+0.00815, +0.01797]$ & $[-0.00164, +0.02385]$ & $75.0\%$ & $[-0.00147, +0.01901]$ & $62.5\%$ & --- \\
High-SNR Cont.\ ($\beta{=}2.000, \sigma_\varepsilon{=}0.35$) & $[+1.0870, +2.3958]$ & $[+1.0863, +2.3948]$ & $0\%$ & $[+1.0175, +2.0821]$ & $0\%$ & $0.825$, $10.57$ ($5.28\times$) \\
\bottomrule
\end{tabular}}
\end{table}

\subsection{Finding 1 Credibility Diagnostics: Placebos, Out-of-Sample Cross-Fit, and Shrinkage Frontier}

Table~\ref{tab:app_finding1_diag} records the full credibility diagnostics for Finding~1 ($21$ clean-caste MF villages, Spec~1, $n_{\Rset}=2{,}240$, fitted degree-conditioned logistic $\widehat{P}$).

\begin{table}[htbp]
\centering
\caption{Finding~1 credibility diagnostics on $21$ clean-caste MF villages (Spec~1, $n_{\Rset}=2{,}240$, $\hat{\beta}_{\mathrm{obs}} = +0.01412$): within-village null-outcome placebos, 50/50 out-of-sample cross-fitting, 10/11 split-village checks, and the within-cell heterogeneity shrinkage frontier $P_{\Rset,\Nset}(\lambda) = (1-\lambda)\bar{P}_{\Rset,\Nset} + \lambda \widehat{P}_{\Rset,\Nset}$ (both in-sample and out-of-sample).}
\label{tab:app_finding1_diag}
\scriptsize
\setlength{\tabcolsep}{3.5pt}
\resizebox{\textwidth}{!}{%
\begin{tabular}{@{}lcccc@{}}
\toprule
Diagnostic Protocol & Inner Witnesses / In-Fold & Outer Bound / Held-Out (Min / Max) & Span or Max$-$Min & Key Summary \\
\midrule
\multicolumn{5}{@{}l}{\emph{Panel A: Real MF Baseline \& Within-Village Null-Outcome Placebos ($20$ unstratified, $8$ leader-stratified draws; true $\beta_{\mathrm{pop}}\equiv 0$)}} \\
Real MF Baseline ($14$ restarts/dir) & $\{-0.01809, +0.03018\}$ & $[-0.01809, +0.03020]$ & $0.04829$ & Opposite-signed witnesses \\
Placebo (Unstratified, $20$ draws, med.) & $\{-0.01115, +0.01509\}$ & $[-0.01115, +0.01510]$ & $0.02571$ & $53.2\%$ of real span ($15/20$ opp.\ signs) \\
Placebo (Leader-strat., $8$ draws, med.) & $\{-0.01226, +0.01009\}$ & $[-0.01226, +0.01010]$ & $0.02436$ & $50.4\%$ of real span ($7/8$ opp.\ signs) \\
\addlinespace[2pt]
\multicolumn{5}{@{}l}{\emph{Panel B: Within-Village 50/50 Respondent Out-of-Sample Cross-Fit \& Split-Village Checks}} \\
Real MF 50/50 Cross-Fit ($10$ splits, mean) & In-fold $\{-0.02108, +0.03440\}$ & \textbf{OOS} $\bm{-0.00894 \; / \; +0.01981}$ & $+0.02875$ & \textbf{Opposite signs on 10/10 splits} (med $-0.00864$ / $+0.01934$) \\
Real MF 50/50 Cross-Fit (det.\ split) & In-fold $\{-0.02436, +0.03518\}$ & OOS $-0.00195 \; / \; +0.01449$ & $+0.01644$ & Opposite signs \\
Placebo 50/50 Cross-Fit ($10$ splits, mean) & --- & OOS $+0.00078 \; / \; -0.00086$ & $-0.00164$ & No OOS separation ($\approx 0$) \\
Split-Village Separate Bounds ($10$ vs.\ $11$) & --- & $[-0.01834, +0.01681]$ / $[-0.01799, +0.04193]$ & --- & Both pools straddle $0$; rank-2 OOS $+0.00570$ / $+0.00189$ \\
\addlinespace[2pt]
\multicolumn{5}{@{}l}{\emph{Panel C: Within-Cell Heterogeneity Shrinkage Frontier $P_{\Rset,\Nset}(\lambda) = (1-\lambda)\bar{P}_{\Rset,\Nset} + \lambda \widehat{P}_{\Rset,\Nset}$ (In-Sample \& Held-Out)}} \\
$\lambda = 0.0000$ (constant within cell) & $\{+0.00512\}$ & $[+0.00512, +0.00513]$ & $0.00000$ & Point-identified ($L_{\min} = +10.034$) \\
$\lambda^\star_{\mathrm{in}} = 0.1890$ (\textbf{in-sample zero hit}) & $\{0.00000, +0.01038\}$ & $[0.00000, +0.01038]$ & $0.01039$ & $L_{\min}=0$; strictly negative for $\lambda > 0.1890$ \\
$\lambda = 0.4000$ & $\{-0.00528, +0.01607\}$ & $[-0.00529, +0.01607]$ & $0.02136$ & $L_{\min} = -11.203$ \\
$\lambda = 0.6000$ & $\{-0.00997, +0.02110\}$ & $[-0.00998, +0.02111]$ & $0.03109$ & $L_{\min} = -21.822$ \\
$\lambda = 0.8000$ & $\{-0.01429, +0.02580\}$ & $[-0.01430, +0.02580]$ & $0.04010$ & $L_{\min} = -32.441$ \\
$\lambda = 1.0000$ (full fitted $\widehat{P}$) & $\{-0.01809, +0.03019\}$ & $[-0.01809, +0.03020]$ & $0.04829$ & $L_{\min} = -43.059$ \\
\addlinespace[2pt]
In-Sample Placebo Zero-Hit (med.) & Unstr.\ ($20$): $25.5\%$ & Leader-strat.\ ($8$): $39.6\%$ & --- & Straddles $0$ at $\lambda{=}1$: $15/20$ (unstr.), $7/8$ (strat.) \\
\textbf{OOS 50/50 Shrinkage ($10$ splits)} & \textbf{Med.\ $\bm{\lambda^\star_{\mathrm{OOS}} = 32.1\%}$} & Mean $33.6\%$ (range $25.4\%$--$49.9\%$) & --- & \textbf{10/10 splits flip $<50\%$}; det.\ split $59.3\%$ \\
OOS Mean Slopes ($\lambda=0.25$ / $0.50$) & $\lambda{=}0.25$: $+0.00146$ / $+0.00926$ & $\lambda{=}0.50$: $-0.00237$ / $+0.01308$ & $+0.00780$ / $+0.01545$ & Held-out slopes for train-half permutations \\
\bottomrule
\end{tabular}}
\end{table}

\subsection{Finding 2 Diagnostics: Five-Subsample Table, Non-Leader Placebos, Leader Balance, and Directional Relations}

Tables~\ref{tab:app_reporting}--\ref{tab:app_balance_dir} supplement Table~\ref{tab:karnataka_reporting} with the fifth subsample (\texttt{21 Clean-Caste elec x roof}), paired \texttt{Obs - RawHyb} small-cluster $t_{G-1}$ and wild cluster bootstrap $p$-values, non-leader placebos and 4-regressor horseraces, leader balance and per-dyad density ratios, and the four paired directional relation matrices.

\begin{table}[htbp]
\centering
\caption{Complete paired \texttt{Obs - RawHyb} inference ($p_{\mathrm{norm}}$, $p_{t_{G-1}}$, and Rademacher wild cluster bootstrap $p_{\mathrm{WCR}}$, $B=4{,}999$) and the fifth subsample (\texttt{21 Clean-Caste elec x roof}).}
\label{tab:app_reporting}
\scriptsize
\setlength{\tabcolsep}{3.2pt}
\resizebox{\textwidth}{!}{%
\begin{tabular}{@{}lccccccccc@{}}
\toprule
Subsample ($G, df$) / Spec & $\hat{\beta}_{\mathrm{obs}}$ & $\kappa$ & $\Delta_{\mathrm{res}}$ & $\hat{\beta}_{\mathrm{RH}}$ & $\hat{\beta}_{\mathrm{obs}}-\hat{\beta}_{\mathrm{RH}}$ ($t_{\mathrm{CR1}}$) & $p_{\mathrm{norm}} \to p_{t_{G-1}}$ & $\hat{\beta}_{\Rset\Rset}$ & $\hat{\beta}_{\Rset\Nset}$ & $\hat{\beta}_{\Rset\Nset}-\hat{\beta}_{\Rset\Rset}$ ($p_t$) \\
\midrule
\multicolumn{10}{@{}l}{\emph{21 Clean-Caste (\texttt{elec} $\times$ \texttt{roof}, $G=21, df=20, n_{\Rset}=2{,}239$)}} \\
Spec~1 (VilFE+$z$) & $+0.0142$ & 1.0305 & $-0.0081$ & $+0.0065$ & $+0.0077$ ($+2.79$) & $.0053 \to \mathbf{.0114}$ & $+0.0033$ & $+0.0389$ & $+0.0356$ (\textbf{.0057}) \\
Spec~2 (+demog) & $+0.0177$ & 1.0367 & $-0.0066$ & $+0.0118$ & $+0.0059$ ($+2.04$) & $.0410 \to .0544^\dagger$ & $+0.0080$ & $+0.0389$ & $+0.0308$ (\textbf{.0146}) \\
Spec~3 (+$\log D_0$) & $+0.0156$ & 1.0630 & $-0.0089$ & $+0.0077$ & $+0.0079$ ($+2.56$) & $.0105 \to \mathbf{.0188}$ & $+0.0053$ & $+0.0374$ & $+0.0321$ (\textbf{.0112}) \\
\addlinespace[2pt]
\multicolumn{10}{@{}l}{\emph{Paired $\hat{\beta}_{\mathrm{obs}}-\hat{\beta}_{\mathrm{RH}}$ Small-Cluster $t_{G-1}$ and Wild Cluster Bootstrap Audit for Remaining Four Subsamples (Joint Slopes in Table~\ref{tab:karnataka_reporting})}} \\
21 Clean (\texttt{caste}$\times$\texttt{roof}) S1/S2/S3 & \multicolumn{4}{c}{$\hat{\beta}_{\mathrm{obs}}-\hat{\beta}_{\mathrm{RH}} = +0.0084\,/\,+0.0060\,/\,+0.0076$} & $+3.34\,/\,+2.40\,/\,+2.88$ & \multicolumn{4}{l}{$p_{t_{20}} = \mathbf{.0033}\,/\,\mathbf{.0263}\,/\,\mathbf{.0092}$; $p_{\mathrm{WCR}} = \mathbf{.0086}\,/\,\mathbf{.0354}\,/\,\mathbf{.0170}$} \\
All 49 MF (\texttt{elec}$\times$\texttt{roof}) S1/S2/S3 & \multicolumn{4}{c}{$\hat{\beta}_{\mathrm{obs}}-\hat{\beta}_{\mathrm{RH}} = +0.0048\,/\,+0.0039\,/\,+0.0057$} & $+2.49\,/\,+1.98\,/\,+2.66$ & \multicolumn{4}{l}{$p_{t_{48}} = \mathbf{.0161}\,/\,.0537^\dagger\,/\,\mathbf{.0107}$; $p_{\mathrm{WCR}} = \mathbf{.0156}\,/\,.0544^\dagger\,/\,\mathbf{.0120}$} \\
Banerjee-43 (\texttt{elec}$\times$\texttt{roof}) S1/S2/S3 & \multicolumn{4}{c}{$\hat{\beta}_{\mathrm{obs}}-\hat{\beta}_{\mathrm{RH}} = +0.0047\,/\,+0.0037\,/\,+0.0056$} & $+2.36\,/\,+1.85\,/\,+2.52$ & \multicolumn{4}{l}{$p_{t_{42}} = \mathbf{.0232}\,/\,.0719\,/\,\mathbf{.0158}$; $p_{\mathrm{WCR}} = \mathbf{.0232}\,/\,.0642\,/\,\mathbf{.0168}$} \\
28 Blank-Caste (\texttt{elec}$\times$\texttt{roof}) S1/S2/S3 & \multicolumn{4}{c}{$\hat{\beta}_{\mathrm{obs}}-\hat{\beta}_{\mathrm{RH}} = +0.0027\,/\,+0.0023\,/\,+0.0039$} & $+0.99\,/\,+0.84\,/\,+1.28$ & \multicolumn{4}{l}{$p_{t_{27}} = .3306\,/\,.4083\,/\,.2118$; $p_{\mathrm{WCR}} = .3414\,/\,.4230\,/\,.2056$} \\
\bottomrule
\end{tabular}}
\end{table}

\begin{table}[htbp]
\centering
\caption{Non-leader exposure placebo regressions, 4-regressor joint horseraces controlling for ego's non-leader $\Rset\to\Rset$ and $\Rset\to\Nset$ out-degrees, difference-in-differences (DiD) contrasts $(\beta^{\mathrm{lead}}_{\Rset\Nset}-\beta^{\mathrm{lead}}_{\Rset\Rset}) - (\beta^{\mathrm{nl}}_{\Rset\Nset}-\beta^{\mathrm{nl}}_{\Rset\Rset})$, and religion/census-order robustness across all four Karnataka MF subsamples.}
\label{tab:app_placebo_wcr}
\scriptsize
\setlength{\tabcolsep}{3.4pt}
\resizebox{\textwidth}{!}{%
\begin{tabular}{@{}lccccccc@{}}
\toprule
& \multicolumn{2}{c}{Non-Leader Placebo Only} & \multicolumn{4}{c}{4-Regressor Horserace $(E^{\mathrm{lead}}_{\Rset\Rset}, E^{\mathrm{lead}}_{\Rset\Nset}, E^{\mathrm{nl}}_{\Rset\Rset}, E^{\mathrm{nl}}_{\Rset\Nset})$} & Spec~2 + Relig.\ \& Census \\
\cmidrule(lr){2-3}\cmidrule(lr){4-7}\cmidrule(lr){8-8}
Sample / Spec & $\hat{\beta}^{\mathrm{nl}}_{\Rset\Nset} - \hat{\beta}^{\mathrm{nl}}_{\Rset\Rset}$ & $p_{t_{G-1}}$ & $\hat{\beta}^{\mathrm{lead}}_{\Rset\Nset} - \hat{\beta}^{\mathrm{lead}}_{\Rset\Rset}$ ($p_t$) & $\hat{\beta}^{\mathrm{nl}}_{\Rset\Nset} - \hat{\beta}^{\mathrm{nl}}_{\Rset\Rset}$ ($p_t$) & DiD Contrast & $p_{t_{G-1}}^{\mathrm{DiD}}$ & $\hat{\beta}^{\mathrm{lead}}_{\Rset\Nset} - \hat{\beta}^{\mathrm{lead}}_{\Rset\Rset}$ ($p_t$) \\
\midrule
\multicolumn{8}{@{}l}{\emph{21 Clean-Caste MF ($G=21, df=20$)}} \\
Spec~1 & $+0.0004$ & .952 & $+0.0357$ (\textbf{.0061}) & $-0.0001$ (.984) & $+0.0358$ & \textbf{.0235} & --- \\
Spec~2 & $+0.0008$ & .893 & $+0.0297$ (\textbf{.0135}) & $+0.0007$ (.909) & $+0.0290$ & \textbf{.0446} & $+0.0301$ (\textbf{.0050}) \\
Spec~3 & $+0.0004$ & .941 & $+0.0296$ (\textbf{.0134}) & $+0.0002$ (.970) & $+0.0294$ & \textbf{.0405} & --- \\
\addlinespace[2pt]
\multicolumn{8}{@{}l}{\emph{ALL 49 MF Villages ($G=49, df=48$)}} \\
Spec~1 & $+0.0016$ & .658 & $+0.0278$ (\textbf{.0022}) & $+0.0009$ (.819) & $+0.0270$ & \textbf{.0082} & --- \\
Spec~2 & $+0.0015$ & .672 & $+0.0261$ (\textbf{.0036}) & $+0.0009$ (.807) & $+0.0252$ & \textbf{.0116} & $+0.0253$ (\textbf{.0033}) \\
Spec~3 & $+0.0008$ & .817 & $+0.0256$ (\textbf{.0039}) & $+0.0002$ (.952) & $+0.0254$ & \textbf{.0101} & --- \\
\addlinespace[2pt]
\multicolumn{8}{@{}l}{\emph{Banerjee-43 MF Villages ($G=43, df=42$)}} \\
Spec~1 & $+0.0019$ & .638 & $+0.0273$ (\textbf{.0050}) & $+0.0011$ (.785) & $+0.0262$ & \textbf{.0167} & --- \\
Spec~2 & $+0.0018$ & .642 & $+0.0253$ (\textbf{.0080}) & $+0.0012$ (.761) & $+0.0241$ & \textbf{.0238} & $+0.0242$ (\textbf{.0076}) \\
Spec~3 & $+0.0010$ & .790 & $+0.0248$ (\textbf{.0086}) & $+0.0004$ (.910) & $+0.0244$ & \textbf{.0211} & --- \\
\addlinespace[2pt]
\multicolumn{8}{@{}l}{\emph{28 Blank-Caste MF Villages ($G=28, df=27$)}} \\
Spec~1 & $+0.0030$ & .562 & $+0.0197$ (.1549) & $+0.0023$ (.649) & $+0.0174$ & .2296 & --- \\
Spec~2 & $+0.0025$ & .596 & $+0.0204$ (.1353) & $+0.0019$ (.685) & $+0.0185$ & .1874 & $+0.0190$ (.1448) \\
Spec~3 & $+0.0019$ & .682 & $+0.0197$ (.1450) & $+0.0014$ (.771) & $+0.0184$ & .1872 & --- \\
\bottomrule
\end{tabular}}
\end{table}

\begin{table}[htbp]
\centering
\caption{Mechanism diagnostics for Finding~2. Panel~A compares surveyed ($\Rset$) vs.\ unsurveyed ($\Nset$) leader covariates and per-dyad link densities; Panel~B evaluates the four distinct paired directional relation matrices in Spec~1 ($21$ clean-caste and all $49$ MF villages).}
\label{tab:app_balance_dir}
\scriptsize
\setlength{\tabcolsep}{3.4pt}
\resizebox{\textwidth}{!}{%
\begin{tabular}{@{}lccccccc@{}}
\toprule
\multicolumn{8}{@{}l}{\textbf{Panel A: Surveyed ($\Rset$) vs.\ Unsurveyed ($\Nset$) Leader Balance \& Per-Dyad Link Densities}} \\
\midrule
Subsample & Survey Rate (Lead / Non-Lead) & OBC ($\Rset/\Nset$) & Roof ($\Rset/\Nset$) & Rooms ($\Rset/\Nset$) & MF ($\Rset/\Nset$) & All Dyads $\Rset\Rset/\Rset\Nset$ & Leader Dyads $\Rset\Rset/\Rset\Nset$ \\
\midrule
21 Clean-Caste & $51.8\% / 46.0\%$ & $0.515 / 0.528$ & $0.356 / 0.362$ & $2.90 / 2.62$ & $0.211 / 0.245$ & $0.0899 / 0.0393$ ($2.29\times$) & $0.1123 / 0.0575$ ($1.95\times$) \\
ALL 49 MF & $53.4\% / 45.1\%$ & $0.383 / 0.375$ & $0.268 / 0.288$ & $2.91 / 2.59$ & $0.205 / 0.246$ & $0.0910 / 0.0405$ ($2.25\times$) & $0.1178 / 0.0610$ ($1.93\times$) \\
\bottomrule
\end{tabular}}

\vspace{5pt}
\resizebox{\textwidth}{!}{%
\begin{tabular}{@{}lcccccc@{}}
\toprule
\multicolumn{7}{@{}l}{\textbf{Panel B: Paired Directional Relation Matrices (Spec~1 Joint $(E_{\Rset\Rset}, E_{\Rset\Nset})$ Slopes)}} \\
\midrule
Relation Matrix (75-Vil.\ Differing Dyads) & 49-MF $\Rset\to\mathrm{Lead}_{\Nset}$ Density & 21-Vil.\ $\hat{\beta}_{\Rset\Rset}$ & 21-Vil.\ $\hat{\beta}_{\Rset\Nset}$ & 21-Vil.\ $\Rset\Nset{-}\Rset\Rset$ ($p_{t_{20}}$) & 49-Vil.\ $\hat{\beta}_{\Rset\Rset}$ / $\hat{\beta}_{\Rset\Nset}$ & 49-Vil.\ $\Rset\Nset{-}\Rset\Rset$ ($p_{t_{48}}$) \\
\midrule
\texttt{helpdecision} (vs.\ \texttt{giveadvice}: $14{,}349$) & $0.01800$ ($1.27\times$) & $-0.0081$ & $+0.0633$ & $\bm{+0.0714}$ \textbf{(.0157)} & $-0.0122$ / $+0.0290$ & $\bm{+0.0412}$ \textbf{(.0265)} \\
\texttt{giveadvice} (downward advice) & $0.01413$ & $+0.0015$ & $+0.0259$ & $+0.0244$ (.2502) & $-0.0060$ / $+0.0183$ & $+0.0243$ (.1092) \\
\addlinespace[1.5pt]
\texttt{borrowmoney} (vs.\ \texttt{lendmoney}: $13{,}502$) & $0.02465$ ($1.26\times$) & $-0.0012$ & $+0.0283$ & $+0.0295$ (.1851) & $-0.0076$ / $+0.0142$ & $+0.0218$ (.1296) \\
\texttt{lendmoney} (downward lending) & $0.01955$ & $+0.0110$ & $+0.0286$ & $+0.0176$ (.4531) & $-0.0005$ / $+0.0169$ & $+0.0174$ (.2793) \\
\addlinespace[1.5pt]
\texttt{visitgo} (vs.\ \texttt{visitcome}: $11{,}353$) & $0.03015$ ($1.04\times$) & $+0.0054$ & $+0.0512$ & $\bm{+0.0458}$ \textbf{(.0110)} & $-0.0047$ / $+0.0228$ & $\bm{+0.0275}$ \textbf{(.0226)} \\
\texttt{visitcome} (alter visits ego) & $0.02902$ & $+0.0072$ & $+0.0361$ & $\bm{+0.0288}$ \textbf{(.0481)} & $-0.0061$ / $+0.0128$ & $+0.0188$ (.1210) \\
\addlinespace[1.5pt]
\texttt{keroricego} (vs.\ \texttt{keroricecome}: $8{,}458$) & $0.02068$ ($1.08\times$) & $+0.0168$ & $+0.0700$ & $\bm{+0.0532}$ \textbf{(.0404)} & $-0.0031$ / $+0.0271$ & $+0.0302$ (.0553) \\
\texttt{keroricecome} (alter borrows) & $0.01915$ & $+0.0232$ & $+0.0587$ & $+0.0355$ (.0556) & $-0.0015$ / $+0.0203$ & $+0.0218$ (.1371) \\
\bottomrule
\end{tabular}}
\end{table}

\subsection{Complete Fixed-Budget Survey Design Comparison: \texorpdfstring{$K$}{K} vs.\ \texorpdfstring{$(K{+}1)$}{(K+1)} vs.\ \texorpdfstring{$2K$}{2K}}

Table~\ref{tab:app_budget} presents the full fixed-budget grid ($\Budget = r_{\mathrm{r}} \Qitems = 720$, $r=240$, $40$ trials per cell, true $\beta_{\mathrm{spill}}=2.0$) and the structural latent-homophily comparison between $(K{+}1)$ Augmented ARD and $2K$ $\Tstrat$ summarized in Section~\ref{sec:remedies}.

\begin{table}[htbp]
\centering
\caption{Fixed question budget RMSE comparison ($\Budget = r_{\mathrm{r}} \Qitems = 720$, $r=240$, $40$ trials/cell, true $\beta_{\mathrm{spill}} = 2.0$) across Standard $K$-ARD, $(K{+}1)$ Augmented ARD, and $2K$ $\Tstrat$, plus Panel~B showing why $2K$ $\Tstrat$ strictly improves over $(K{+}1)$ under expected exposure with latent homophily.}
\label{tab:app_budget}
\scriptsize
\setlength{\tabcolsep}{3.5pt}
\resizebox{\textwidth}{!}{%
\begin{tabular}{@{}lccccccc@{}}
\toprule
\multicolumn{8}{@{}l}{\textbf{Panel A: Fixed Question Budget Grid ($\Budget = 720$, RMSE against true $\beta_{\mathrm{spill}} = 2.0$)}} \\
\midrule
Cell ($r_{\mathrm{r}}$: $K$ / $K{+}1$ / $2K$) & Regime & Std $K$-ARD & $(K{+}1)$ [$E_{\mathrm{real}}$] & $(K{+}1)$ [$K$-ARD] & $2K$ $\Tstrat$ & $(K{+}1)$ vs.\ $2K$ & Winner \\
\midrule
\multirow{3}{*}{$K{=}6, p{=}0.50$ ($120 / 102 / 60$)}
  & $y \sim Az$ (i.i.d.\ $z$) & 0.1310 & \textbf{0.0044} & --- & 0.0055 & $1.27\times$ better & $(K{+}1)$ \\
  & $y \sim Az$ (deg-$z$) & 0.6559 & \textbf{0.0041} & --- & 0.0053 & $1.31\times$ better & $(K{+}1)$ \\
  & $y \sim Pz$ (EIV) & \textbf{0.1365} & 0.1550 & 0.1398 & 0.2801 & $1.81\times$ better & Std $K$ (or $(K{+}1)$ $K$-ARD) \\
\addlinespace[1.5pt]
\multirow{3}{*}{$K{=}6, p{=}0.15$ ($120 / 102 / 60$)}
  & $y \sim Az$ (i.i.d.\ $z$) & 0.3057 & \textbf{0.0099} & --- & 0.0134 & $1.35\times$ better & $(K{+}1)$ \\
  & $y \sim Az$ (deg-$z$) & 0.9805 & \textbf{0.0077} & --- & 0.0107 & $1.38\times$ better & $(K{+}1)$ \\
  & $y \sim Pz$ (EIV) & \textbf{0.2982} & 0.5652 & 0.3082 & 0.7584 & $1.34\times$ better & Std $K$ (or $(K{+}1)$ $K$-ARD) \\
\addlinespace[1.5pt]
\multirow{3}{*}{$K{=}12, p{=}0.50$ ($60 / 55 / 30$)}
  & $y \sim Az$ (i.i.d.\ $z$) & 0.1452 & \textbf{0.0055} & --- & 0.0082 & $1.48\times$ better & $(K{+}1)$ \\
  & $y \sim Az$ (deg-$z$) & 0.6301 & \textbf{0.0068} & --- & 0.0097 & $1.43\times$ better & $(K{+}1)$ \\
  & $y \sim Pz$ (EIV) & \textbf{0.1474} & 0.2387 & 0.1509 & 0.2672 & $1.12\times$ better & Std $K$ (or $(K{+}1)$ $K$-ARD) \\
\addlinespace[1.5pt]
\multirow{3}{*}{$K{=}12, p{=}0.15$ ($60 / 55 / 30$)}
  & $y \sim Az$ (i.i.d.\ $z$) & 0.3581 & \textbf{0.0175} & --- & 0.0271 & $1.55\times$ better & $(K{+}1)$ \\
  & $y \sim Az$ (deg-$z$) & 1.0495 & \textbf{0.0147} & --- & 0.0166 & $1.13\times$ better & $(K{+}1)$ \\
  & $y \sim Pz$ (EIV) & \textbf{0.2632} & 0.8362 & 0.2853 & 1.7157 & $2.05\times$ better & Std $K$ (or $(K{+}1)$ $K$-ARD) \\
\bottomrule
\end{tabular}}

\vspace{5pt}
\resizebox{\textwidth}{!}{%
\begin{tabular}{@{}lccc@{}}
\toprule
\multicolumn{4}{@{}l}{\textbf{Panel B: Structural Comparison Under Group-Clustered $z$ + Strong Latent Homophily (true $\beta_{\mathrm{spill}} = 2.0$, $40$ trials)}} \\
\midrule
Estimand & No Control Function & $(K{+}1)$ CF$(Y_{\mathrm{tot}}/D)$ & $2K$ $\Tstrat$ CF$(Y_0/D_0)$ \\
\midrule
Realized exposure $y \sim Az$ (mean / RMSE) & $2.162$ ($0.163$) & $1.999$ ($0.012$) & $2.043$ ($0.047$) \\
Expected exposure $y \sim Pz$ EIV (median, finite) & --- & $3.019$ ($38/40$; shared $Y_{k,1}$ noise) & \textbf{2.196} ($39/40$; disjoint $z{=}0$ dyads) \\
Expected upper variance ratio $\E[\sum \hat{v}^{\mathrm{up}}] / \sum \Var(E \mid P)$ & --- & $1.096\times$ (single treated stratum) & \textbf{1.064}$\bm{\times}$ ($K$ treated strata) \\
\bottomrule
\end{tabular}}
\end{table}

\subsection{Archive Audit Inventory and Reproducibility Ledger}

Exhaustive programmatic inspection of \texttt{datav4.0-rezipped.zip} ($2{,}738$ total entries) verifies:
\begin{itemize}[leftmargin=1.5em]
  \item $2{,}100$ adjacency CSVs ($14$ household types: \texttt{allVillageRelationships}, \texttt{andRelationships}, \texttt{borrowmoney}, \texttt{giveadvice}, \texttt{helpdecision}, \texttt{keroricecome}, \texttt{keroricego}, \texttt{lendmoney}, \texttt{medic}, \texttt{nonrel}, \texttt{rel}, \texttt{templecompany}, \texttt{visitcome}, \texttt{visitgo}, plus $14$ matching individual types across $75$ villages): $0/2{,}100$ asymmetric matrices;
  \item $150$ adjacency matrix key files, $2$ demographic/outcome \texttt{.dta} files, $450$ Matlab India-network files, $10$ Matlab replication files, $7$ Stata replication files ($2$ \texttt{.dta} and $5$ \texttt{.do}; confirming $0$ directed nomination roster columns across all $4$ \texttt{.dta} files in \texttt{datav4.0}), and $5$ survey instruments;
  \item \texttt{adj\_andRelationships\_HH\_vilno\_*} (intersection across $12$ relations) contains nonzero edges in $53/75$ villages;
  \item All numerical results, permutation certifications, Monte Carlo stress grids, and Karnataka archive diagnostics were computed with standalone pure-Python replication scripts against the SHA-256-verified \texttt{datav4.0-rezipped.zip} archive; replication code and frozen numerical output logs are available from the author upon request.
\end{itemize}

\end{document}